\documentclass[acmsmall]{acmart}
\AtBeginDocument{%
  }

\setcopyright{acmlicensed}
\copyrightyear{2018}
\acmYear{2018}
\acmDOI{XXXXXXX.XXXXXXX}
\acmConference[Conference acronym 'XX]{Make sure to enter the correct
  conference title from your rights confirmation email}{June 03--05,
  2018}{Woodstock, NY}
\acmISBN{978-1-4503-XXXX-X/2018/06}

\begin{document}


\title{Exploring Human-AI Collaboration in E-Textile Design: A Case Study on Flex Sensor Placement for Shoulder Motion Detection}


\author{Zhuchenyang Liu}
\authornote{Both authors contributed equally to this research.}
\affiliation{%
  \institution{Aalto University}
  \city{Espoo}
  \country{Finland}}
\email{zhuchenyang.liu@aalto.fi}

\author{Yao Zhang}
\authornotemark[1]
\affiliation{%
  \institution{Aalto University}
  \city{Espoo}
  \country{Finland}}

\author{Yalan He}
\affiliation{%
  \institution{Fudan University}
  \city{Shanghai}
  \country{China}}

\author{Hilla Paasio}
\affiliation{%
  \institution{Aalto University}
  \city{Espoo}
  \country{Finland}}

\author{Changyi Li}
\affiliation{%
  \institution{Aalto University}
  \city{Espoo}
  \country{Finland}}
  
\author{Guna Semjonova}
\affiliation{%
  \institution{Riga Stradins University}
  \city{Riga}
  \country{Latvia}}

\author{Yu Xiao}
\affiliation{%
  \institution{Aalto University}
  \city{Espoo}
  \country{Finland}}

\begin{abstract}

Flex sensors are widely used in e-textiles for detecting joint motions and, subsequently, full-body movements. A critical initial step in utilizing these sensors is determining the optimal placement on the body to accurately capture human motions. This task requires a combination of expertise in fields such as anatomy, biomechanics, and textile design, which is seldom found in a single practitioner. Generative AI, such as Large Language Models (LLMs), has recently shown promise in facilitating design. However, to our knowledge, the extent to which LLMs can aid in the e-textile design process remains largely unexplored in the literature. To address this open question, we conducted a case study focusing on shoulder motion detection using flex sensors. We enlisted three human designers to participate in an experiment involving human-AI collaborative design. We examined design efficiency across three scenarios: designs produced by LLMs alone, by humans alone, and through collaboration between LLMs and human designers. Our quantitative and qualitative analyses revealed an intriguing relationship between expertise and outcomes: the least experienced human designer achieved continuous improvement through collaboration, ultimately matching the best performance achieved by humans alone, whereas the most experienced human designer experienced a decline in performance. Additionally, the effectiveness of human-AI collaboration is affected by the granularity of feedback - incremental adjustments outperformed sweeping redesigns - and the level of abstraction, with observation-oriented feedback
producing better outcomes than prescriptive anatomical directives. These findings offer valuable insights into the opportunities and challenges associated with human-AI collaborative e-textile design.

\end{abstract}



\received{20 February 2007}
\received[revised]{12 March 2009}
\received[accepted]{5 June 2009}

\maketitle

\section{Introduction}

\begin{figure}[h]
  \centering
  \includegraphics[width=\linewidth]{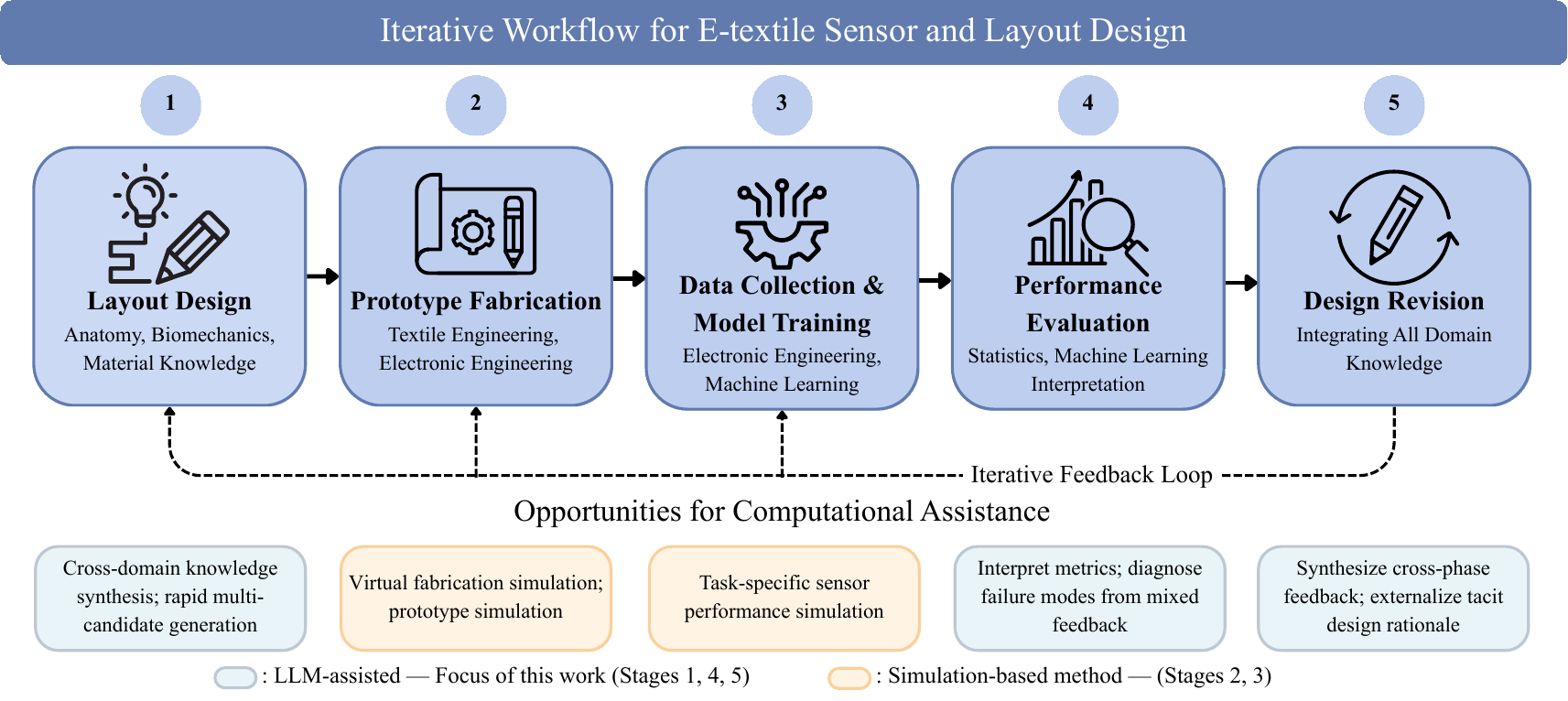}
  \caption{Iterative design process of e-textile-based motion capture systems.
  Each phase is annotated with the domain knowledge it requires.
  Below the workflow, we map opportunities for AI assistance:
  LLM-based approaches (blue) can support Phases~1, 4, and~5, while
  simulation-driven approaches (orange) primarily address Phases~2 and~3.
  This work focuses on the LLM-assisted stages.}
  \Description{A flowchart showing five phases of e-textile sensor layout design
  with iterative feedback arrows pointing back to Phases 1, 2, and 3.
  Below, blue (LLM) and orange (Simulation) annotations indicate where each
  computational approach can assist.}
  \label{fig:workflow}
\end{figure}

Electronic textiles (e-textiles) integrate sensing, computation, and communication directly into fabric, enabling unobtrusive monitoring of human physiology in everyday settings~\cite{stoppa2014wearable, komolafe2021textile, ismar2020futuristic}. Among the sensing modalities employed in e-textiles, resistive flex sensors have become a popular choice for capturing joint motions and, by extension, full-body movements: bending a flex sensor along a limb or joint produces a resistance change proportional to the deflection angle, which can be mapped to joint kinematics through machine learning models~\cite{saggio2016resistive, liu2019reconstructing, gholami2019lower}. Systems built on this principle have been used for hand gesture recognition~\cite{kim2025optisense}, gait analysis~\cite{niswander2020optimization}, and upper-limb rehabilitation monitoring~\cite{chen2020wearable}, demonstrating the breadth of applications that flex-sensor-based motion capture can address.


As shown in Fig~\ref{fig:workflow}, 
the design process for e-textile-based motion capture systems typically involves several iterative steps: designing the sensors and their layout on the human body, fabricating hardware prototypes, collecting sensor readings and corresponding motion measurements, training a prediction model, evaluating performance, and refining the design based on evaluation results \cite{liu2025sensory, frances2020wearable}. For each specific application, selecting sensor types is often straightforward. For instance, flex sensors are commonly utilized for detecting joint movements, while pressure sensors measure the force applied. Projects not focused on developing new sensor designs generally opt for commercially available or pre-evaluated sensors with clearly defined specifications such as sensitivity, sensing range, and resolution.

Given a selection of sensors, determining the optimal placement of sensors on the human body — known as \textit{sensor layout design} \cite{shirazi2024standardized, kim2025optisense} — is a complex optimization challenge. The goal is to maximize the information captured about target movements with a limited number of sensors. Effective sensor placement necessitates integrating knowledge from several fields, including anatomy (identifying which muscles and tendons are involved in the target motion), biomechanics (understanding skin deformation during movement), textile engineering (considering how fabric affects sensor-body coupling), and signal processing (mapping sensor signals to joint angles) \cite{goncalves2018wearable, haratian2022motion}. It is challenging for any designer to possess comprehensive expertise in all these areas. Typically, designers adopt a trial-and-error approach, iteratively improving the layout based on evaluation outcomes. This knowledge- and labor-intensive process has spurred efforts to introduce computational assistance to streamline and enhance the design process.



Simulation-driven approaches have made notable progress by enabling virtual prototyping through motion and/or sensor simulation, which reduces the cost of physical fabrication and evaluation~\cite{qu2025laytex, kim2025optisense, ray2025w2w, kim2024optimal}. However, in these approaches the design proposals still originate from human designers; the simulation serves primarily to evaluate candidate layouts rather than to generate or refine them. 

Large Language Models (LLMs) have recently demonstrated potential as cross-domain knowledge integrators in mechanical and physical design tasks \cite{makatura2024can, mustapha2025survey}. Their ability to synthesize information from various disciplines suggests they could be well-suited to the e-textile design workflow, particularly in sensor layout design. Nevertheless, how and to what extent LLMs can contribute to sensor layout design remains largely unexplored.


In this study, we aim to explore the opportunities and challenges associated with leveraging LLMs to assist in the e-textile design process, specifically focusing on sensor layout optimization. We selected shoulder motion detection as our case study due to the shoulder's complexity as a joint with three degrees of freedom. This choice reflects the shoulder's role as a representative and clinically significant case in previous e-textile design studies \cite{qu2025laytex, kim2025optisense, jin2020soft}. We engaged three human designers with varying levels of expertise in e-textile design — senior, intermediate, and junior — to participate in a user study aimed at addressing the following research questions:

\textbf{RQ1:} Can LLMs generate sensor layouts that match or surpass the quality of initial designs produced by human designers across different experience levels?

\textbf{RQ2:} When designs generated by LLMs are iteratively refined through human-AI collaboration, how does their quality compare to that of designs produced through a purely human-driven iterative process?

\textbf{RQ3:} How do the human designers' levels of experience, feedback strategies, and interaction patterns affect the effectiveness of human-AI collaboration in the iterative sensor layout optimisation process?

We assess design quality based on the motion detection accuracy attained by each design. The scientific contributions of this paper are threefold. First, we provide an empirical comparison of initial sensor layout designs generated by LLMs with those created by human designers, offering preliminary evidence of LLMs' capability to synthesize cross-disciplinary knowledge in physical design tasks. Second, we compare human-AI collaborative designs with those developed by human designers in subsequent design iterations. This study reveals an intriguing finding: the least experienced designer consistently improved the quality of the design through human-AI collaboration, eventually matching the highest performance achieved by participating human designers working without AI assistance. In contrast, the most experienced designer experienced a decline in final output quality when collaborating with AI. Third, through qualitative analysis, we identify two critical factors influencing the effectiveness of human-AI collaboration —\textit{granularity of feedback} (incremental adjustments outperform sweeping redesigns) and \textit{level of abstraction} (observation-oriented input yields better outcomes than prescriptive anatomical directives). These insights offer practical guidance for facilitating effective human-AI collaboration in design tasks.


\section{Background and Related Works}

This section provides the background and positions our work relative to existing literature. 

\subsection{Shoulder Motion Capture}
\label{sec:bg_shoulder}

As the most mobile and complex joint in the human body, the shoulder holds significant clinical importance, since shoulder pathologies are among the most common musculoskeletal conditions. The glenohumeral (GH) joint allows motion in three degrees of freedom through the coordinated action of four articulations~\cite{wu2005isb, carnevale2019wearable}, producing highly heterogeneous skin deformation patterns across large, overlapping muscle groups with distinct fiber orientations~\cite{jin2020soft, huang2023sensor}. 

Current measurement tools for shoulder motion are limited to subjective clinical scales or lab-bound optical tracking systems~\cite{chen2020wearable, walmsley2018measurement, zhou2024portable}. E-textile-based motion capture offers a promising path toward continuous, ambulatory shoulder motion monitoring. However, significant challenges persist in optimally placing sensors on this anatomically complex region.

Recently, the shoulder has consequently emerged as the primary evaluation site in the simulation-driven e-textile design work closely related to this paper. For instance, Laytex~\cite{qu2025laytex} validated its layout generation tool with participants performing shoulder motions, while OptiSense~\cite{kim2025optisense} optimized strain sensor placement for shoulder angle prediction.  Jin et al.~\cite{jin2020soft} and Zhang et al.~\cite{zhang2023tailored}  developed textile sensor systems specifically targeting shoulder kinematics. By selecting the same body site, our case study allows for direct methodological comparison with these prior approaches while exploring a complementary axis of innovation --LLM-assisted sensor layout design.

\subsection{LLMs in Physical Design and Sensing}
\label{sec:rw_llm}

LLMs have demonstrated emergent capabilities in reasoning about physical systems. 
Makatura et al.~\cite{makatura2024can} evaluated LLMs on engineering design tasks and found that they can generate plausible design proposals by synthesizing information across disciplines, though they noted limitations in precise geometric reasoning. Mustapha et al.~\cite{mustapha2025survey} surveyed LLM applications in engineering and highlighted their strength in integrating heterogeneous domain knowledge. In the Internet of Things domain, Xu et al.~\cite{xu2024penetrative} showed that LLMs possess considerable proficiency in employing world knowledge to interpret IoT sensor data and reason about physical tasks, demonstrating their ability to bridge the gap between raw sensor signals and meaningful physical interpretation. Kaneko and Inoue~\cite{kaneko2023toward} demonstrated that ChatGPT can suggest sensor locations for human activity recognition (HAR), achieving comparable classification accuracy with fewer sensors. Their study represents an initial exploration of LLMs for sensor placement, particularly focusing on inertial measurement unit (IMU)-based systems for classifying discrete activity categories using open datasets. In contrast, our case study extends beyond LLM-based design to evaluate the effectiveness of human-AI collaboration in the design process. Additionally, our research specifically targets the layout design of resistive flex sensors for continuous, fine-grained joint angle prediction.

\subsection{Human--AI Collaborative Design}
\label{sec:rw_hai}

A growing body of work has examined how humans and AI systems collaborate in design tasks. In digital design domains, studies have explored human--AI co-creation in code generation~\cite{zhang2023unifying}, UI prototyping~\cite{kolthoff2025guide}, and creative content production~\cite{rezwana2023designing}, identifying factors such as user trust, feedback modality, and the balance between directive and open-ended interaction as key determinants of collaboration quality. Gmeiner et al.~\cite{gmeiner2023exploring} studied AI-assisted design in engineering contexts and found that overly prescriptive human input can suppress the complementary reasoning that makes AI assistance valuable, while observation-level feedback tends to enable more synergistic outcomes. Brynjolfsson et al.~\cite{brynjolfsson2025generative} reported that generative AI productivity gains among customer support agents accrued primarily to novice workers, suggesting that AI can effectively disseminate expert-level knowledge to less experienced practitioners.

However, the majority of this work has focused on purely digital design tasks where iteration is fast and feedback is immediate. Physical design introduces embodied constraints---fabric behavior, body mechanics, sensor--surface coupling---that may be difficult for LLMs to reason about without direct physical experience~\cite{makatura2024can}. Whether insights from digital human--AI collaboration transfer to physical prototyping contexts, and how expertise level mediates the effectiveness of such collaboration, remain open questions that our study aims to address.

\section{User Study Design}

This section describes the design of our case study. We first present the
overall study protocol and explain how each phase addresses our research
questions (\S\ref{sec:protocol}). We then detail the technical
implementation that supports the implementation of the protocol (\S\ref{sec:technical}), and
finally describe participant recruitment (\S\ref{sec:participants}).

\subsection{User Study Protocol}
\label{sec:protocol}

To address our three research questions, we designed a within-subject user study in which three
human designers each completed the same design task under different
experimental conditions. The design task required placing $k{=}4$ commercially available resistive flex sensors on a garment to predict three clinical shoulder
joint angles (flexion/extension, abduction/adduction, and internal/external
rotation). 
The number of sensors was limited to $k{=}4$, as this configuration offers a balanced trade-off between prediction accuracy and garment complexity for shoulder kinematics, as established in prior studies \cite{qu2025laytex, kim2025optisense}. Keeping 
k constant also ensures a fair comparison across all conditions; designers were not allowed to use fewer or more sensors.

\begin{figure*}[t]
  \centering
  \includegraphics[width=\textwidth]{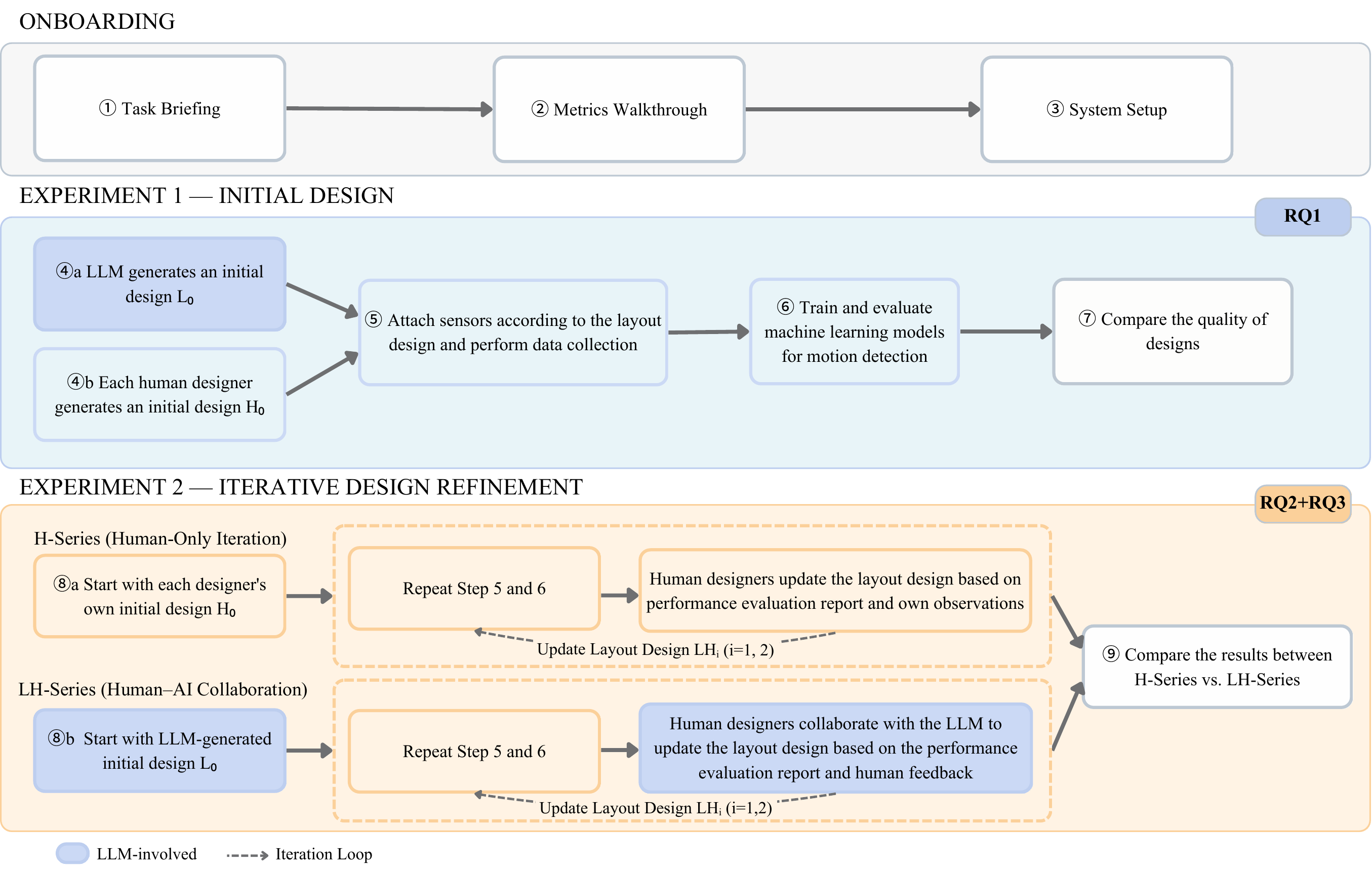}
  \caption{Overview of the user study protocol. The study proceeds through three
  phases: Onboarding, Experiment~1 (initial design comparison, addressing
  RQ1), and Experiment~2 (iterative design refinement comparison across H-Series and
  LH-Series, addressing RQ2 and RQ3). Within Experiment~2, the H-Series
  (amber) follows a human-only iteration loop, while the LH-Series (blue)
  introduces human-AI collaborative design refinements. }
  \label{fig:protocol}
\end{figure*}

The study comprised three sequential phases---\textit{Onboarding},
\textit{Experiment~1} (initial design comparison), and
\textit{Experiment~2} (iterative design refinement comparison)---structured to
progressively address the research questions (Figure~\ref{fig:protocol}). In
Experiment~1, we compare the achieved motion sensing performance of initial sensor layouts produced independently by
the LLM and by each human designer, respectively, directly
addressing \textbf{RQ1}. Experiment~2 involves two conditions for each human designer:
the \textit{H-Series} where the human designer iteratively refines their own initial design without AI assistance, and the \textit{LH-Series}, where the human designer
collaborates with the LLM to iteratively refine the LLM-generated initial design. Comparing the iteration trajectories between these two conditions addresses \textbf{RQ2}, while the records of structured feedback from human designers and human-AI interaction
logs collected during the LH-Series provide the qualitative evidence needed to address \textbf{RQ3}. 

The user study was conducted collaboratively by a researcher and a research assistant. To quickly validate the quality of the design, a test subject was recruited to immediately perform pre-defined shoulder movements while wearing a garment equipped with flex sensors arranged according to the layout design. The entire user study lasted approximately two hours per designer, encompassing all three phases.


\subsubsection{Onboarding ($\sim$20 min)}

An onboarding session was conducted for each participating human designer prior to the start of the experiments. The session consisted of three parts.
First, a \textit{task briefing} presented the design challenge---placing
$k{=}4$ flex sensors on a garment to detect three clinical shoulder joint
angles (flexion/extension, abduction/adduction, internal/external
rotation)---along with physical constraints such as pre-selected sensor dimensions and attachment mechanisms. 
Second, a \textit{metrics walkthrough} acquainted each participant with the performance dashboard, which displays the quantitative evaluation results of their designs. This includes the key performance indicator, Mean Per-Joint Angular Error (MPJAE), along with other metrics such as Pearson Correlation Coefficient and average mean percentage error. More details about these metrics are presented in Section~\ref{sec:metrics}. Human designers practiced reading a sample report of these metrics to confirm comprehension.
Third, a \textit{system calibration} step fitted the garment on a test subject, synchronized the sensor acquisition and reference tracking
systems, and established a neutral-pose baseline.

\subsubsection{Experiment 1: Initial Design Comparison ($\sim$30 min per designer)}
\label{sec:exp1_protocol}

Experiment~1 was designed to address \textbf{RQ1}: \textit{Can LLMs generate sensor layouts that match or surpass the quality of initial designs produced by human designers across different experience levels?}

We compared the quality of the four sensor layout designs:

\begin{itemize}
  \item $L_0$ (LLM-created initial design): Generated prior to the workshop
        session via a divergent-convergent prompting strategy
        (detailed in~\S\ref{sec:design_gen}), without any input from human designers.
  \item $H_0^A$, $H_0^B$, $H_0^C$ (Human designer created initial designs): Each human designer
        independently proposed a 4-sensor layout,
        drawing on their own domain knowledge. Designers received no information about $L_0$ or each other's designs. They specified sensor locations using anatomical landmarks and garment reference points. The resulting layouts ($H_0^A$, $H_0^B$, $H_0^C$) serve as initial designs from human designers in Experiment~1 and as starting points for the H-Series in Experiment~2.
\end{itemize}



\subsubsection{Experiment 2: Iterative Design Refinement and Comparison between H-Series and LH-Series ($\sim$2 hrs per human designer)}
\label{sec:exp2_protocol}

Experiment~2 was designed to address \textbf{RQ2} (\textit{When designs generated by LLMs are iteratively refined through human-AI collaboration, how does their quality compare to that of designs produced through a purely human-driven iterative process?}) and \textbf{RQ3} (\textit{How do the human designers’ levels of experience, feedback strategies, and interaction
patterns affect the effectiveness of human-AI collaboration in the iterative sensor layout optimisation
process?}).

Each human designer completed the H-Series first, and then the LH-Series in Experiment 2. This fixed ordering was a practical constraint of our
workshop format: the H-Series allowed designers to familiarize themselves
with the evaluation pipeline before introducing the LLM-mediated
component. Crucially, this ordering creates a \textit{conservative test}
for H-Series superiority: any knowledge gained during independent
iteration could carry over to benefit LH-Series performance, meaning
that if the H-Series still outperforms the LH-Series, the advantage is likely
genuine. Conversely, LH-Series improvements should be interpreted with
the caveat that designers had prior iteration experience. We return to
this point when discussing the results of the experiment (\S\ref{sec:discuss_2}).

\paragraph{H-Series (Human-Only).}
Starting from their own initial design ($H_0$) produced in Experiment~1,
each human designer independently refined the layout through up to two iteration
cycles. In each cycle, the human designer: (1)~inspected the deployed prototype and
reviewed the performance evaluation report on the feedback dashboard ($\sim$10 min);
(2)~decided what to modify based on their own analysis and physically
repositioned sensors on the garment ($\sim$5 min); and (3)~ran another
data collection and evaluation round ($\sim$10 min, covering data collection, model training, and performance metric computation).
The human designer could stop early if satisfied with the result. This condition
captures the conventional iterative design process driven entirely by human
expertise.

\paragraph{LH-Series (Human--AI Collaboration).}
Starting with the LLM-generated initial layout design ($L_0$), each human designer collaborated with LLMs (Figure~\ref{fig:architecture}) over two iteration cycles at most. In each cycle, the human designer: (1) inspected the deployed prototype and reviewed the performance evaluation report; (2) completed a structured feedback template, documenting both quantitative performance evaluation (e.g., identifying underperforming sensors) and personal observations (e.g., fabric bunching, poor fit), which was submitted to the LLM along with performance metrics ($\sim$10 min); (3) reviewed the LLM-generated refined layout design along with its explicit rationale, and then deployed and tested the refined layout ($\sim$10 min).



The layout design generated in each iteration in both Experiment 1 and 2 was evaluated following a consistent procedure which involved three steps: (1) the human designer attached sensors to the garment according to the specified layout ($\sim$ 5 minutes); (2) the test subject executed a predefined series of shoulder movements ($\sim$ 5 minutes); and (3) a machine learning model was trained from the collected sensor measurements and corresponding labels (i.e. ground-truth motion data extracted from video) to detect joint angles and the relevant performance indicators were calculated ($\sim$ 5 minutes). Each layout required roughly 30 minutes from deployment to evaluation. Since all four layouts were tested on the same subject performing identical motions using the same machine learning model architecture, any observed performance differences can be attributed to the layout design rather than external confounding factors. The technical implementation for rapid design evaluation will be presented in Section~\ref{sec:validation}.


\subsection{Technical Implementation}
\label{sec:technical}

\begin{figure*}[t]
  \centering
  \includegraphics[width=\textwidth]{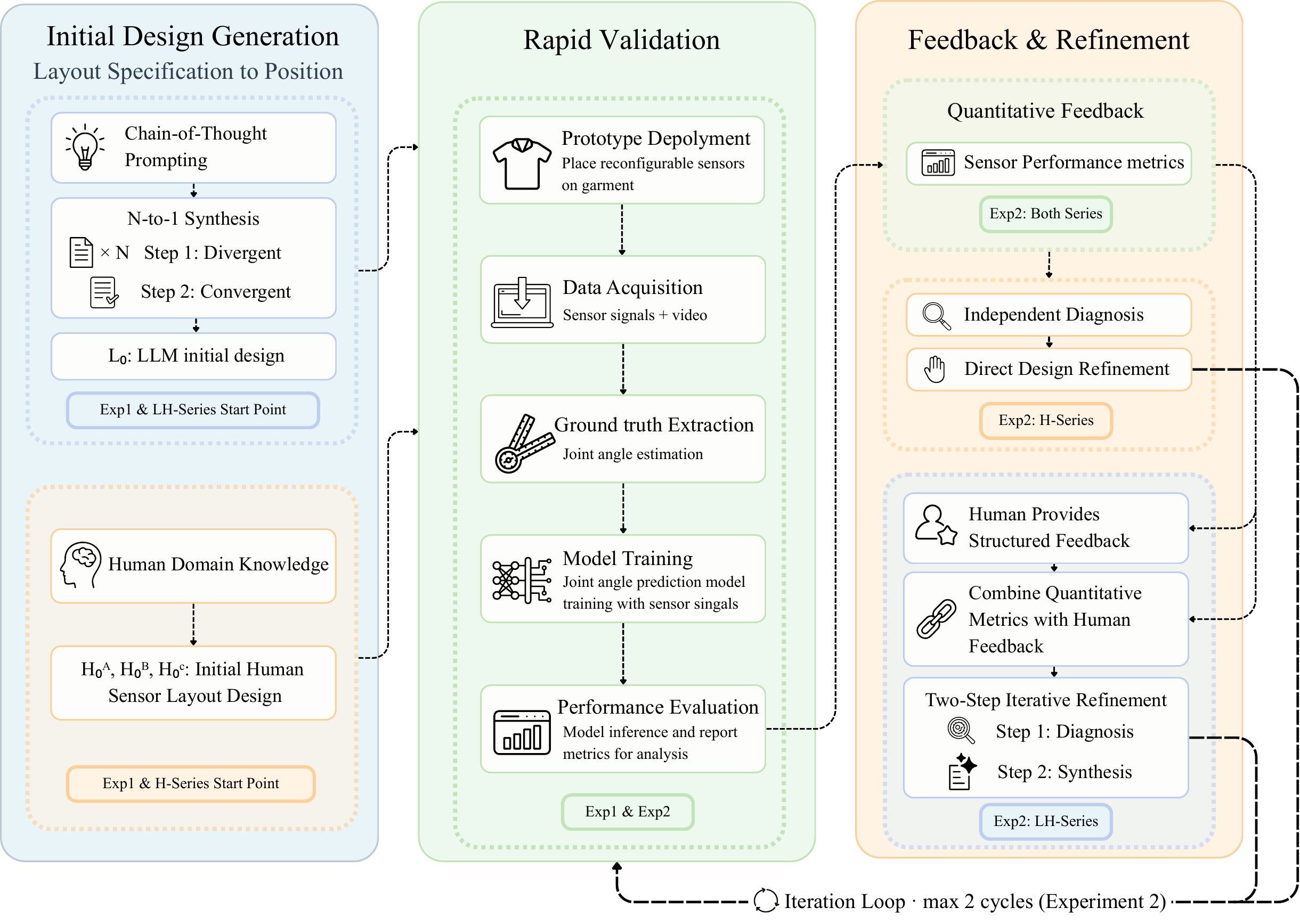}
  \caption{Overview of the technical implementation: LLM-based N-to-1 synthesis (blue) produces $L_0$ during Experiment~1. Within the Feedback \& Refinement module, there are two 
  pathways: human designer-only diagnosis and
  direct design refinement (amber; H-Series), and LLM-mediated two-step refinement informed by structured human feedback (blue;
  LH-Series). Dashed arrows indicate the iteration loop (up to two cycles). The Rapid Validation pipeline (green) is utilized for evaluating both initial designs and refined ones.}
  
  \label{fig:architecture}
\end{figure*}

Figure~\ref{fig:architecture} gives an overview of the technical implementation of the system needed for implementing the study protocol,
including: \textit{Initial Design Generation (Experiment 1)} (left),
\textit{Rapid Validation (Experiment 1 and 2)} (center) and \textit{Feedback \& Refinement (Experiment 2)}
(right).

\subsubsection{LLM-driven Initial Design Generation}
\label{sec:design_gen}
 
We applied structured prompting to generate $L_0$. The output consisted of a specification detailing the positions of four flex sensors relative to anatomical landmarks. To mitigate single-generation randomness, we employed GPT-4 as the
core reasoning engine with a divergent-convergent strategy.
In the divergent phase, we generated five independent candidate
proposals, each using chain-of-thought
prompting. The following step repeated five times: Firstly, we elicited relative knowledge from LLM by itself with the prompt design shown in the Appendix: Figure~\ref{fig:p1}. Secondly, LLM generated a design based on the knowledge following the prompt shown in Appendix: Figure~\ref{fig:p2}. In the convergent phase, a meta-synthesis prompt filtered biomechanically implausible configurations and consolidated
robust design principles into a unified initial layout ($L_0$). The detailed prompt design is shown in Appendix: Figure~\ref{fig:p3}.




\subsubsection{Rapid Validation}
\label{sec:validation}

The Rapid Validation pipeline translates each candidate sensor layout into a deployable prototype and produces a standardized performance evaluation within approximately 15 minutes per layout. As shown in Figure~\ref{fig:architecture}, this pipeline is shared identically across all experimental conditions, ensuring that any observed performance differences can be attributed to the layout design rather than to external confounding factors. The pipeline comprises five sequential stages: prototype deployment, data acquisition, ground-truth extraction, model training, and performance reporting. Below we describe each stage in detail.

\begin{figure}[htbp]
    \centering
    \includegraphics[width=0.65\linewidth]{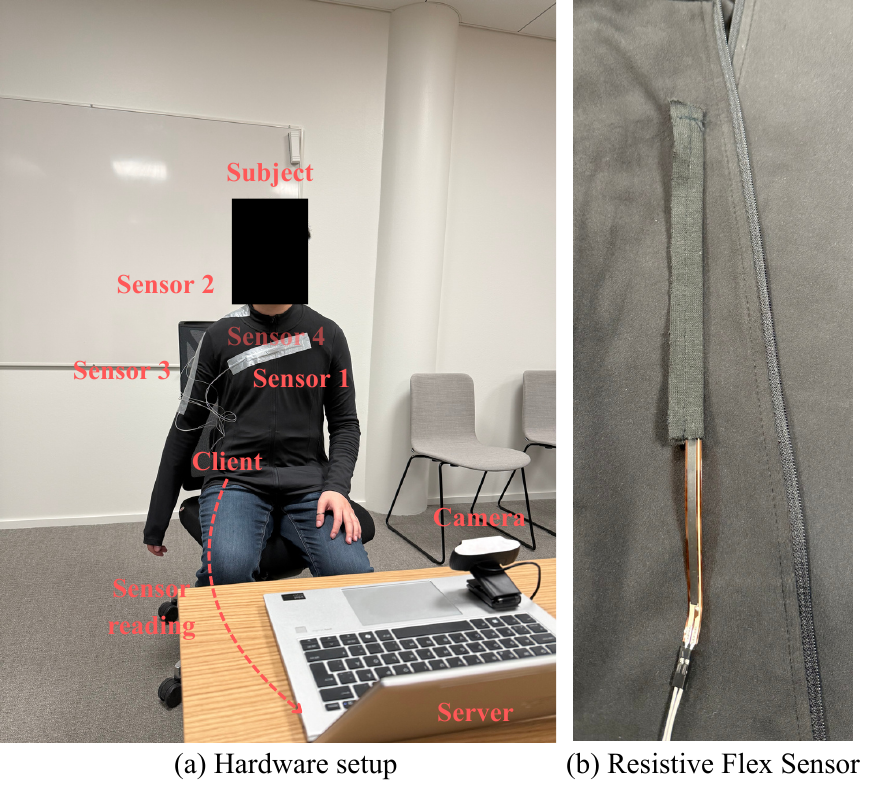}
    \caption{Data collection setup. (a)~Hardware overview: four resistive flex sensors are attached to a compression shirt worn by the subject; a client microcontroller reads the sensor signals and transmits them to a server laptop, while an RGB webcam captures reference video. (b)~Close-up of a SpectraFlex\texttrademark{} flex sensor with a custom-fitted fabric container.}
    \label{fig:hardware_setup}
\end{figure}

As shown in Fig.~\ref{fig:hardware_setup}, the physical prototype consists of a tight-fitting compression shirt that conforms closely to the body surface, ensuring consistent sensor--skin coupling across postures. Four resistive SpectraFlex\texttrademark{} Flex Sensors\footnote{\url{https://www.spectrasymbol.com/resistive-flex-sensors/spectraflex-flex-sensors}} (95\,mm active length, 10\,k$\Omega$ nominal resistance) serve as the sensing elements. Resistive flex sensors were selected for their high signal stability and low drift over repeated bending cycles~\cite{saggio2016resistive}, which is critical for an iterative prototyping study. A custom-fitted fabric container was prepared for each sensor (Fig.~\ref{fig:hardware_setup}(b)) to prevent direct contact between the sensor and the adhesive tape, avoiding residue buildup on the sensing surface and enabling rapid removal and reattachment across layout iterations. The containers were secured to the garment surface using elastic antistatic tape, pressing each sensor against the fabric to capture bending dynamics during motion. Given a layout specification, the four sensor containers can be attached to the designated positions on the garment in approximately five minutes.

\begin{figure}[htbp]
    \centering
    \includegraphics[width=1\linewidth]{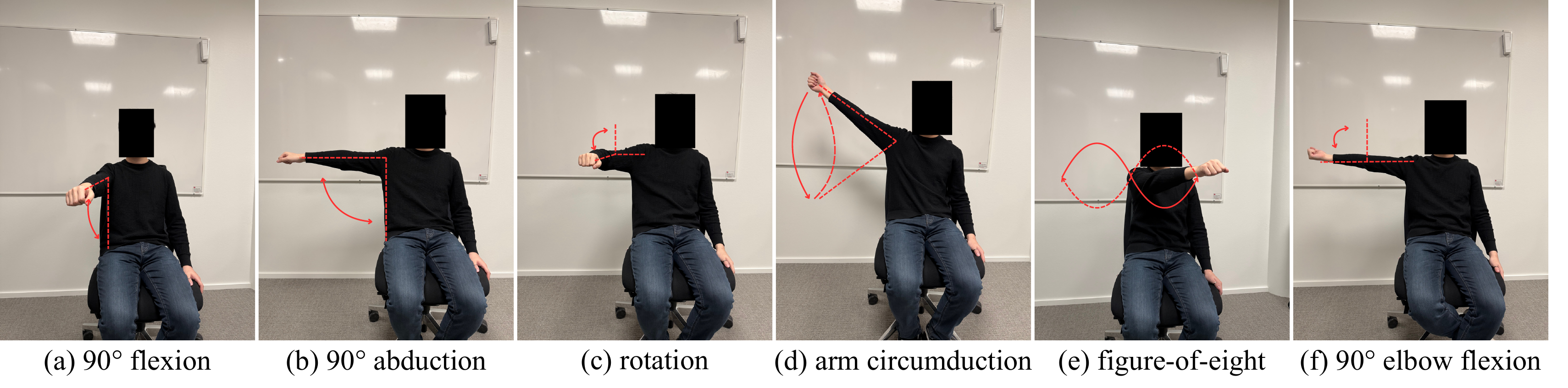}
    \caption{Illustration of the six standardized motions used for performance evaluation.}
    \label{fig:all_motion}
\end{figure}

\textbf{Data acquisition}
Two data streams were recorded simultaneously during each evaluation: sensor readings and reference video. An Arduino-based acquisition board samples the four flex sensors at 40\,Hz, while a single RGB webcam captures reference video at 60\,fps (Fig.~\ref{fig:hardware_setup}(a)). An online world clock provides a shared epoch-time reference to synchronize the two recording systems. Note that all raw data containing personally identifiable information---including video recordings of the motion participant and photographs of the garment during fitting---were destroyed after the relevant features (joint angles, sensor coordinates) had been extracted and anonymized. Only the derived, de-identified data are retained for analysis.

To ensure a fair comparison across layouts and conditions, the same data collection process was used in each iteration. This began with a neutral-pose calibration phase lasting 15 seconds. Subsequently, a training set ($\sim$3 minutes) was gathered, consisting of six standardized motions adapted from the shoulder movement protocol defined by Qu et al.~\cite{qu2025laytex}---90\textdegree{} flexion, 90\textdegree{} abduction, internal--external rotation, arm circumduction, figure-of-eight, and 90\textdegree{} elbow flexion (Fig.~\ref{fig:all_motion})---, with each motion repeated five times. A separate test set ($\sim$45\,s) used the same six motions performed once in reversed order for generalization assessment. The training set was used to fit the prediction model; the test set was utilized for generating reported performance metrics.

\textbf{Ground-truth extraction from video:}
Raw video frames were processed through SMPLest-X~\cite{yin2025smplest}, a foundation model for expressive human pose and shape estimation based on SMPL-X parametric fitting, to extract per-frame shoulder joint angles for three clinical degrees of freedom: flexion/extension, abduction/adduction, and internal/external rotation. The resulting pose sequences were further refined using SmoothNet~\cite{zeng2022smoothnet}, a temporal-only network that mitigates frame-to-frame jitter while preserving motion detail, yielding temporally smooth ground-truth angle trajectories stored as per-frame arrays of shape $(N, 3)$.

Because the sensor and video streams operate at different sampling rates (40\,Hz vs.\ 60\,fps), a temporal alignment procedure is required before model training. First, both streams were cropped to their overlapping interval. The per-frame joint angle labels were then aligned to the cropped video frames using frame indices. Next, the joint angle time series was linearly interpolated from the video-frame timestamps to the sensor sampling timestamps, effectively downsampling the labels to 40\,Hz. Then, the sensor readings were normalized to zero mean and unit variance using a standard scaler fitted on the training set only; the same scaler parameters were applied to the test set to prevent information leakage. Finally, a sliding window with a length of 40 samples (approximately one second) with a stride of one was applied to construct input--output pairs: each input window comprises 40 consecutive readings across all four sensor channels, and the corresponding output is the three-dimensional joint angle vector at the last timestep of the window.

\textbf{Machine learning models for joint angle estimate}
A hybrid 1D-CNN + LSTM architecture was designed~\cite{chen2022analyzing, lopez2025upper} to ensure the robustness of model training. The spatial feature extraction stage consists of two stacked one-dimensional convolutional layers: the first applies 32 filters of kernel size~3 with zero-padding of~1 to the four input sensor channels, followed by batch normalization, ReLU activation, and dropout ($p{=}0.1$); the second expands to 64 filters with the same kernel and padding configuration. The resulting feature maps are then passed to a single-layer LSTM with 64 hidden units for temporal modeling. Only the final hidden state is forwarded to a regression head comprising a fully connected layer ($64 \to 32$ units) with ReLU and dropout ($p{=}0.1$), followed by a linear output layer ($32 \to 3$) that produces the predicted flexion, abduction, and rotation angles in degrees. The model was trained from scratch for every layout configuration using the Adam optimizer with a learning rate of $10^{-3}$, mean squared error loss, a batch size of 32, and gradient clipping at a maximum norm of 1.0. Training runs for 50 epochs, and the checkpoint with the lowest training loss was retained for evaluation. The complete training and evaluation cycle completed within approximately five minutes, enabling rapid turnaround during the iterative design workshop. Finally, a performance dashboard was generated by applying trained model to the test set, which will be introduced in Section~\ref{sec:feedback_integration}.

\subsubsection{Feedback \& Refinement}
\label{sec:feedback_integration}

\begin{figure}
    \centering
    \includegraphics[width=1\linewidth]{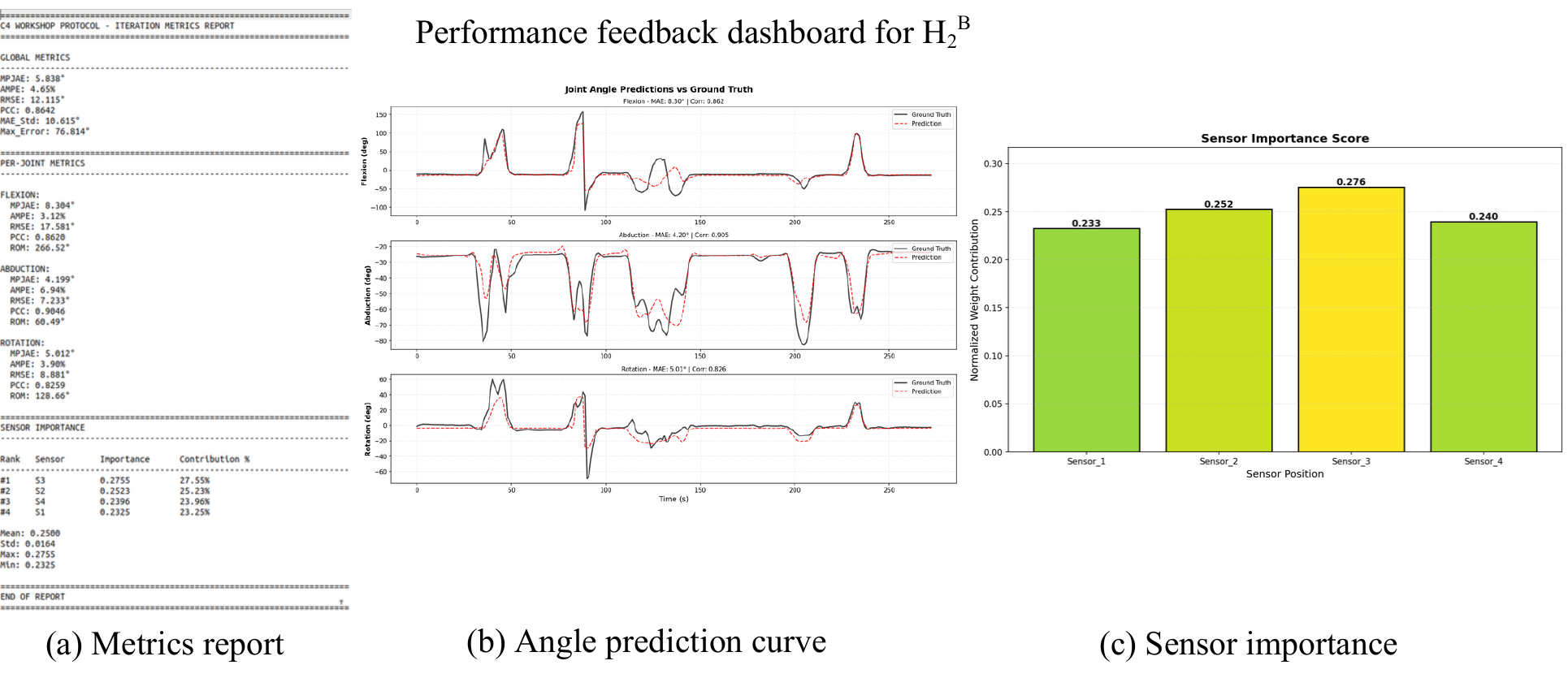}
    \caption{Screenshots of the Quantitative Feedback dashboard. (a) a metrics report lists global and per-joint prediction errors (MPJAE, AMPE, RMSE, PCC) and sensor importance rankings, (b) joint angle prediction curves versus ground truth for flexion, abduction, and rotation, respectively and (c) a bar chart of normalized sensor importance scores across four sensors.}
    \label{fig:example dashboard}
\end{figure}

At the conclusion of each Rapid Validation cycle in Experiment 2, the performance evaluation results were presented to the designers via a Quantitative Feedback dashboard (see Figure~\ref{fig:example dashboard}). The dashboard presents global metrics, per-joint error decomposition, and sensor contribution rankings based on feature importance scores. The definition for each metric will be presented in in Section~\ref{sec:metrics}. In the H-Series and LH-Series, the feedback was interpreted and translated into design refinements using different approaches. In the H-Series, the human designer follows an independent diagnosis-to-revision pathway. They inspected both the quantitative evaluation results and the physical prototype, independently diagnosed performance
issues---drawing on their own domain expertise to connect observed errors to physical causes---and directly repositioned sensors on the garment.




In the LH-Series, feedback integration and design Refinement
proceed through a three-stage process that combines human physical
observations with LLM-driven analytical reasoning
(Figure~\ref{fig:architecture}, right).

\begin{figure}
    \centering
    \includegraphics[width=1\linewidth]{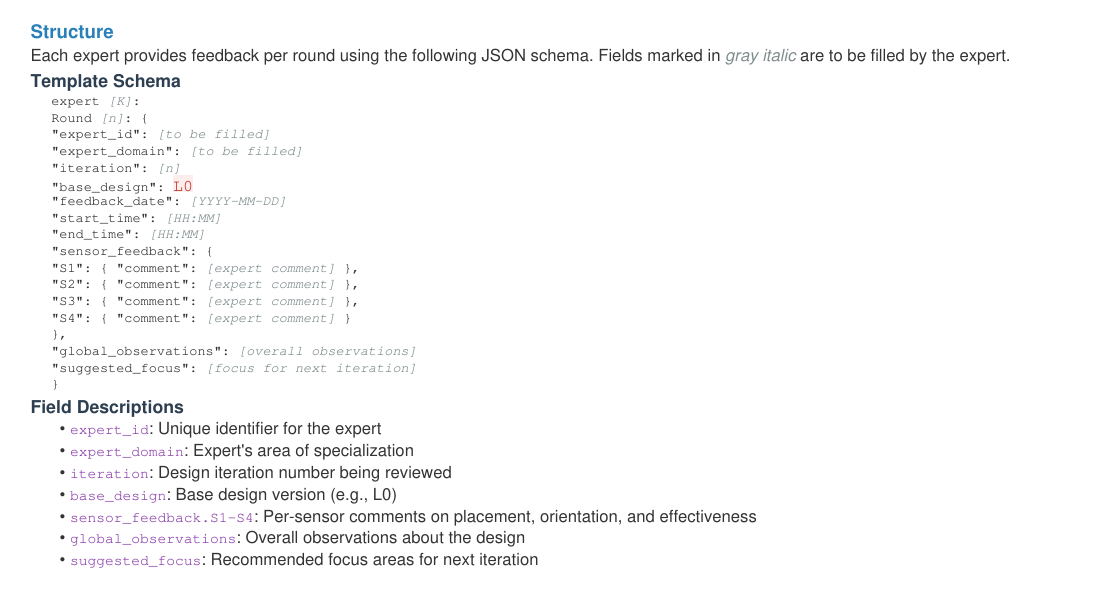}
    \caption{Template for Structured Human Feedback. Note that only feedback information is used for LLM prompting to avoid misleading.}
    \label{fig:template_feedback}
\end{figure}

\textit{Stage~1: Structured Human Feedback.}
Rather than directly revising the layout, the human designer provided structured feedback following the template shown in Figure~\ref{fig:template_feedback}. These per-sensor and global observations capture information that is inaccessible to the LLM from quantitative data alone: physical phenomena observed during motion, sensor--body coupling quality, positional adjustment suggestions based on hands-on inspection, and priority judgments indicating which sensors or issues the human designer considers most critical. 

\textit{Stage~2: Diagnostic Integration.}
The structured human designer observations from Stage~1 are combined with the
quantitative performance metrics from the dashboard into an integrated
diagnostic prompt. This prompt is submitted to the LLM, which performs
a diagnosis step: analyzing the current design's failure modes by
cross-referencing the quantitative indicators (e.g., low sensor
importance, high per-joint error) with the human designer's physical
observations (e.g., poor coupling, misalignment with motion axes).
The LLM explicitly identifies root causes before proposing any
changes, reasoning across biomechanical, material, and signal
processing considerations to connect symptoms to underlying design
issues. The detailed prompt is shown in Appendix: Figure~\ref{fig:p4}.

\textit{Stage~3: Synthesis and Refinement.}
Building on the diagnosis, the LLM generates a refined layout
specification in which each proposed modification is explicitly traced
to a finding from the diagnosis step. The output is a JSON
specification containing updated sensor coordinates (relative to
anatomical landmarks), orientations, and a structured design rationale
that links each change to the identified root cause. It creates an auditable
record of design evolution across iterations. The detailed prompt is shown in Appendix: Figure~\ref{fig:p5}.

The refined layout is then evaluated following the Rapid
Validation pipeline. This three-stage structure enforces a separation between observation (human), diagnosis (LLM with human input), and synthesis(LLM with traceability), ensuring that human embodied knowledge and LLM cross-domain reasoning are integrated systematically.

During the user study, besides quantitative performance metrics, we recorded complete LLM interaction logs (prompt-response chains), human designer's structured feedback, design evolution trajectories (sensor coordinate changes across iterations), and photographs of the garment at each iteration.


\subsubsection{LLM Configuration}
\label{sec:llm_config}

All LLM interactions used
\texttt{gpt-4.1-2025-04-14} accessed through the Azure OpenAI API. For text generation temperature setting, in Experiment 1, two different temperature is used for N-to-1 generation. The first divergent phase uses a higher temperature ($T{=}0.7$) to encourage
diversity across the five independent candidate proposals; the
convergent synthesis phase uses a lower temperature ($T{=}0.2$) to
favor deterministic consolidation; In Experiment 2, we used an intermediate temperature ($T{=}0.4$) to balance creative and consistent generation.




\subsubsection{Quantitative Performance Metrics}
\label{sec:metrics}

We employed four complementary metrics to assess the quality of the layout design.
We denote the total number of time samples as $N$, the number of degrees
of freedom as $J$, the ground-truth joint angle at time $t$ for joint $j$ as
$y_{t,j}$, and the predicted joint angle as $\hat{y}_{t,j}$.

\paragraph{Mean Per-Joint Angular Error (MPJAE)}
The primary metric, representing the average absolute deviation in degrees:
\begin{equation}
  \text{MPJAE} = \frac{1}{J} \sum_{j=1}^{J}
    \left( \frac{1}{N} \sum_{t=1}^{N} | \hat{y}_{t,j} - y_{t,j} | \right)
\end{equation}

\paragraph{Root Mean Square Error (RMSE)}
Penalizes large deviations more heavily, highlighting occasional spikes:
\begin{equation}
  \text{RMSE}_j = \sqrt{ \frac{1}{N} \sum_{t=1}^{N}
    ( \hat{y}_{t,j} - y_{t,j} )^2 }
\end{equation}

\paragraph{Pearson Correlation Coefficient (PCC)}
Evaluates how well the predicted signal tracks the shape and temporal
dynamics of the ground truth, independent of offset or scaling:
\begin{equation}
  \text{PCC}_j = \frac{ \sum_{t=1}^{N}
    (y_{t,j} - \bar{y}_j)(\hat{y}_{t,j} - \bar{\hat{y}}_j) }
  { \sqrt{\sum_{t=1}^{N} (y_{t,j} - \bar{y}_j)^2}\;
    \sqrt{\sum_{t=1}^{N} (\hat{y}_{t,j} - \bar{\hat{y}}_j)^2} }
\end{equation}

\paragraph{Average Mean Percentage Error (AMPE)}
A scale-independent measure normalized by the range of motion, enabling
comparison across joints with different movement ranges:
\begin{equation}
  \text{AMPE}_j = \frac{1}{N} \sum_{t=1}^{N}
    \left( \frac{ | \hat{y}_{t,j} - y_{t,j} | }
    { \max(y_j) - \min(y_j) } \right) \times 100\%
\end{equation}

\paragraph{Sensor Importance Score (SIS)}
Quantifies each sensor's contribution to the prediction based on the learned weights of the first convolutional layer in the prediction model. Let $\mathbf{W} \in \mathbb{R}^{F \times K \times M}$ denote the weight tensor of the first Conv1d layer, where $F$ is the number of output filters, $K$ is the number of input channels (one per sensor), and $M$ is the kernel size. The importance of sensor $i$ is defined as:
\begin{equation}
  \text{SIS}_i = \frac{ \sum_{f=1}^{F} \sum_{m=1}^{M} |W_{f,i,m}| }
  { \sum_{k=1}^{K} \sum_{f=1}^{F} \sum_{m=1}^{M} |W_{f,k,m}| }
\end{equation}
The resulting scores sum to one across all sensors. A higher score indicates that the model relies more heavily on the corresponding sensor’s signal for prediction. Ideally, equal SIS values for each sensor reflect a balanced design for joint-angle prediction.

Overall, we define the design quality as the combined measure of the final prediction performance—primarily represented by MPJAE—supplemented by the additional metrics discussed above.

\subsection{Participant Recruitment}
\label{sec:participants}

Test subject for sensing performance evaluation: we employed a single-subject design (male, age 28, athletic build, no
shoulder pathologies). This choice was motivated by two considerations:
(1)~holding biomechanical variation constant isolates the effect of layout
design methodology, enabling direct comparison across conditions; and
(2)~a single participant enables more iteration cycles within practical
time constraints. The participant performed standardized motions only
and did not participate in design decisions.

Human designers: we recruited three designers with distinct domain backgrounds to
explore how expertise influences both H-Series and LH-Series, Recruitment criteria included: (1) prior experience of designing and implementing e-textile-based motion sensing systems; (2)~willingness to participate in both
H-Series and LH-Series. 

\begin{table}[h]
  \caption{Human Designer Profiles}
  \label{tab:experts}
  \begin{tabular}{llll}
    \toprule
    Human Designer & Background & Experience  \\
    \midrule
    A & Kinesiology Research & 8 years (Senior)  \\
    B & E-Textile Engineering & 4 years (Intermediate)  \\
    C & Electrical Engineering & 2 years (Junior)  \\
    \bottomrule
  \end{tabular}
\end{table}

The three human designers have diverse knowledge backgrounds and varying levels of experience. They all have prior experience using LLMs; however, they have not used LLMs specifically for sensor layout design generation. This provides a suitable foundation for understanding general LLM usage practices while ensuring that their feedback is not biased toward this particular task.



\section{Results}

\subsection{Experiment 1: Initial Design Comparison}

\begin{figure*}[t]
  \centering
  \includegraphics[width=\textwidth]{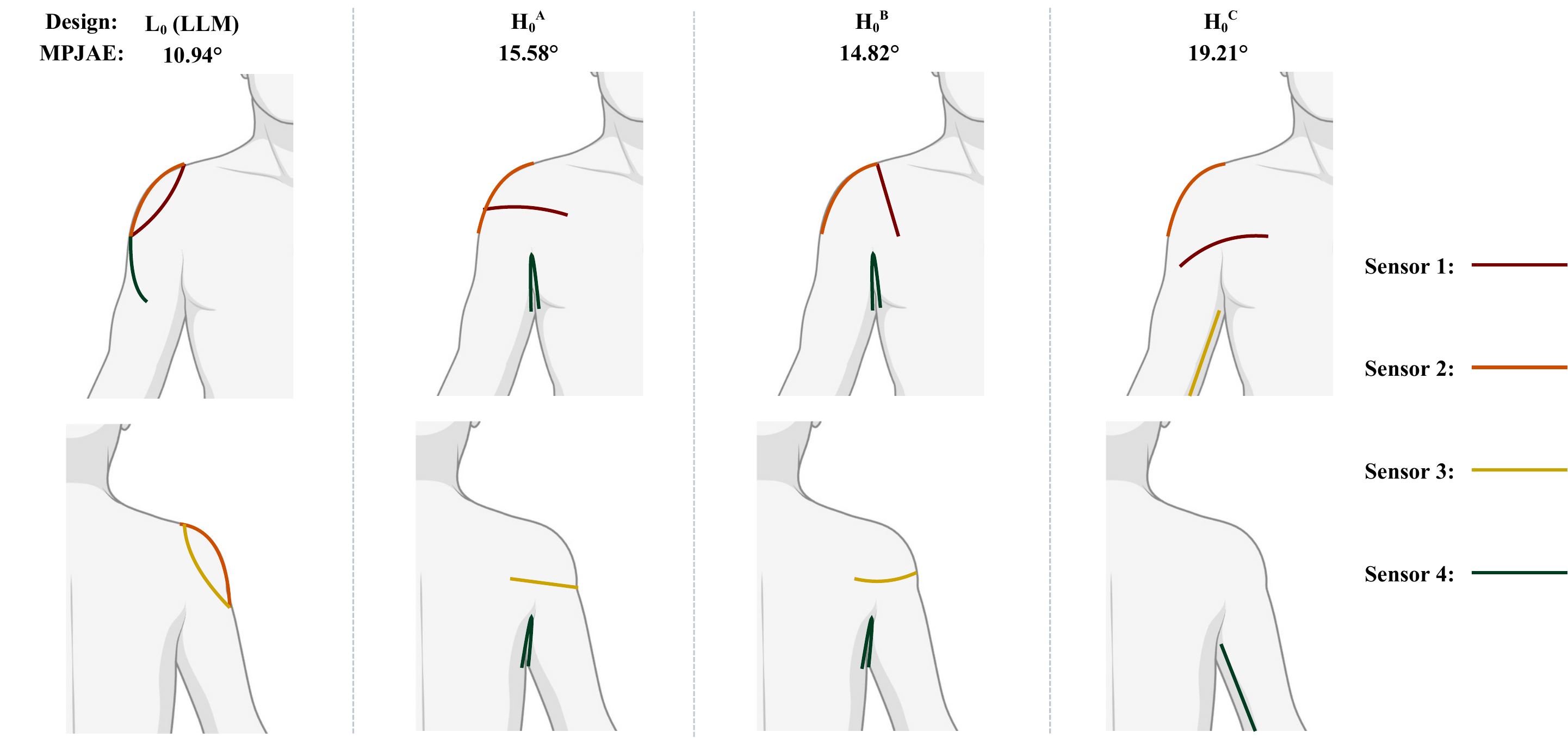}
  \caption{Initial sensor layouts proposed by LLM and human designers, shown on a standardized shoulder model. Each layout comprises 4 flex sensors (colored rectangles).}
  \Description{Four 3D shoulder models showing different sensor placement strategies, with colored rectangles representing flex sensors at different positions and orientations.}
  \label{fig:initial_layouts}
\end{figure*}

To address \textbf{RQ1}, we compared the quality of the four initial designs including the LLM-generated layout ($L_0$) and the three human designer-generated layouts ($H_0^A$, $H_0^B$, $H_0^C$), as illustrated in Figure~\ref{fig:initial_layouts}. The detailed LLM generated outputs using divergent-convergent strategy can be found in Appendix~\ref{app:iteration_llm}. Following the results in Table~\ref{tab:exp1_detailed}, the LLM-generated layout achieved the lowest MPJAE of 10.94°, compared to
15.58° (Designer~A), 14.82° (Designer~B), and 19.21° (Designer~C). $L_0$
also achieved the highest PCC and the lowest AMPE and RMSE. 

\begin{table}[h]
  \caption{Initial Design Performance Metrics. Bold values indicate best performance per metric.}
  \label{tab:exp1_detailed}
  \begin{tabular}{lcccccc}
    \toprule
    Design & MPJAE (°) & PCC & AMPE (\%) & RMSE (°) \\
    \midrule
    $L_0$ (LLM) & \textbf{10.94}  & \textbf{0.794} & \textbf{8.75} & \textbf{17.03} \\
    $H_0^A$ (Senior) & 15.58 & 0.566 & 9.28 & 27.31 \\
    $H_0^B$ (Intermediate) & 14.82 & 0.390 & 9.66 & 25.42 \\
    $H_0^C$ (Junior) & 19.21 & 0.340 & 13.58 & 28.02 \\
    \bottomrule
  \end{tabular}
\end{table}


We also analysed the per-joint error decomposition as shown in Table~\ref{tab:perjoint}. $L_0$ achieved the lowest error for flexion and rotation. For abduction, Designer A's design $H_0^A$ achieved the best performance.

\begin{table}[h]
  \caption{Per-Joint MPJAE in degrees. Bold indicates the best per degree of freedom.}
  \label{tab:perjoint}
  \begin{tabular}{lcccc}
    \toprule
    Degree of Freedom & $L_0$ & $H_0^A$ & $H_0^B$ & $H_0^C$ \\
    \midrule
    Flexion  & \textbf{18.01} & 30.21 & 25.36 & 30.98 \\
    Abduction  & 9.62 & \textbf{7.76} & 9.91 & 13.77 \\
    Rotation  & \textbf{5.20} & 8.78 & 9.20 & 12.88 \\
    \bottomrule
  \end{tabular}
\end{table}

\subsection{Experiment 2: Iterative Design Refinement}

To address \textbf{RQ2}, we compared iteration trajectories across H-Series and LH-Series for all three designers over two iteration cycles. The detailed sensor layout design (Appendix~ \ref{app:iteration_results}), performance dashboard (Appendix~\ref{app:dashboard}), structured human designer feedback logs (Appendix~\ref{sec:appendix-lh-human}) and LLM outputs (Appendix~\ref{sec:appendix-lh-llm}) for different iteration conditions can be found in the Appendix.

As listed in Table~\ref{tab:trajectories}, in the H-Series, Designer~A (Senior) achieved consistent performance improvement over two iterations, reaching the second-lowest MPJAE overall. Designer~B (Intermediate) improved the design quality steadily across both iterations, achieving the lowest MPJAE in the H-Series and LH-Series. Designer~C (Junior) improved substantially in iteration~1 but regressed in iteration~2.

In the LH-Series, designers modified sensor positions based on the evaluation results of $L_0$ in the first iteration. Designer~A modified all four sensor positions, resulting in an increase of MPJAE from 10.94° to 12.75°. Designer~A chose to discontinue the LH-Series after Iteration 1. Designer~B achieved a 28.4\% improvement in Iteration 1 but regressed in Iteration 2. Designer~C was the only participant who achieved consistent improvement over the two iterations. Compared with the best design Designer C achieved in the H-Series, Designer C achieved better design with the assistance of LLM.

\begin{figure}[h]
  \centering
  \includegraphics[width=\linewidth]{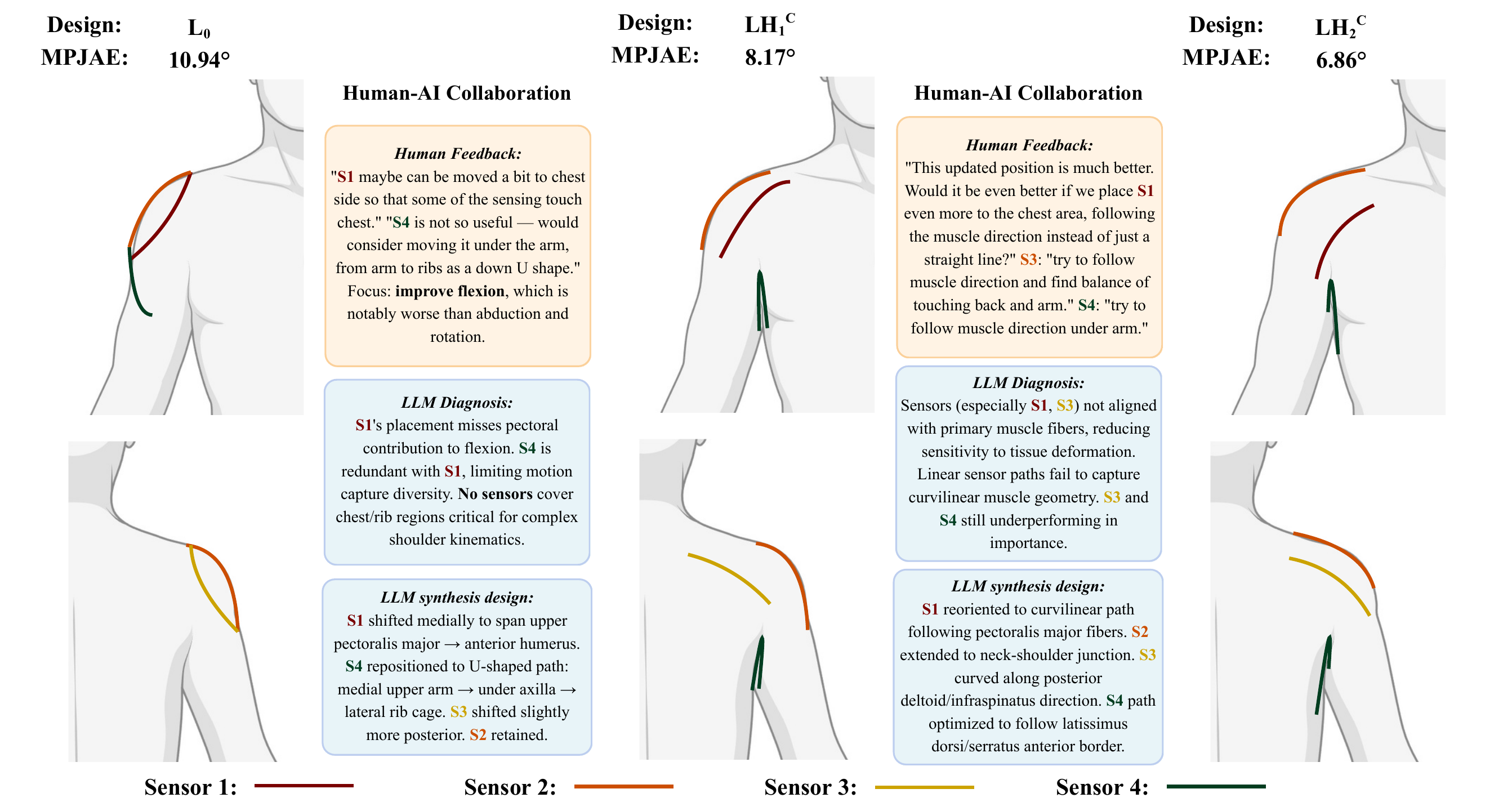}
  \caption{Designer C's LH-Series layout evolution. Annotations show corresponding feedback excerpts.}
  \Description{Three shoulder models showing the progression of sensor positions across iterations, with arrows indicating movement directions and text annotations showing designer feedback.}
  \label{fig:iteration_evolution}
\end{figure}

Figure~\ref{fig:iteration_evolution} illustrates Designer~C's layout design evolution in the LH-Series. Designer~C suggested moving sensors ``a bit to chest side'' in Iteration 1, and placing sensors to ``follow the muscle direction'' in Iteration 2. Each iteration involved incremental adjustments to one or two sensors rather than comprehensive redesign, and the LLM translated these directional observations into concrete coordinate updates. 



\begin{table}[h]
  \caption{MPJAE in each iteration.}
  \label{tab:trajectories}
  \begin{tabular}{lcccc}
    \toprule
    Condition & Initial & Iter 1 & Iter 2 &  Pattern \\
    \midrule
    \multicolumn{5}{l}{\textit{H-Series (Independent Human Iteration)}} \\
    Designer A & 15.58° & 6.49° & \textbf{6.33°} &  $\searrow\searrow$ \\
    Designer B & 14.82° & 7.80° & \textbf{5.84°} &  $\searrow\searrow$ \\
    Designer C & 19.21° & 10.08° & 13.03° &  $\searrow\nearrow$ \\
    \midrule
    \multicolumn{5}{l}{\textit{LH-Series (LLM-Human Collaboration)}} \\
    Designer A & 10.94° & 12.75° & ---$^\dagger$ &  $\nearrow$ \\
    Designer B & 10.94° & 7.83° & 10.18° &  $\searrow\nearrow$ \\
    Designer C & 10.94° & 8.17° & \textbf{6.86°} &  $\searrow\searrow$ \\
    \bottomrule
  \end{tabular}
  
  \footnotesize{$\searrow$ = improvement; $\nearrow$ = regression.
  $^\dagger$Designer A terminated after iteration 1.}
\end{table}



\subsubsection{Human Designers' Structured Feedback}

To address \textbf{RQ3}, we examined the feedback records from LH-Series iterations. Table~\ref{tab:feedbacktypes_detailed} presents representative excerpts from each designer, the observation column is concluded from the corresponding feedback excerpt. We further observe the following
  patterns across the three cases:

  \paragraph{Feedback framing differed across designers.} Designer A framed feedback in terms of anatomical principles, prescribing specific positions based on muscle anatomy. For instance, Designer~A identified each sensor
   position by its corresponding muscle (e.g., ``m.deltoideus anterior part,'' ``m.deltoideus middle part'') and consistently recommended that sensors ``should touch or pass the GH joint.'' Designer~A also noted that
  certain placements ``will not detect IR/ER rotations of the shoulder''. Designer B referenced quantitative metrics and sensor physical constraints. In Iteration~1, Designer~B observed that ``the sensors bend only in one
  direction'' and highlighted that sensors are ``placed on clothes'' rather than directly on the body, requesting the system to ``compensate for that.'' In Iteration~2, Designer~B noted that ``S2 importance is quite low''
  and that ``rotation detection got worse.'' Designer C described observations from physical interaction with the prototype, using tentative language (``maybe,'' ``I am wondering''). Designer~C grounded suggestions in
  personal testing experience (e.g., ``this suggestion is especially based on my own testing results'') and referred to perceived sensor behavior during movement, such as suggesting a sensor be moved ``a bit to chest
  side so that some of the sensing are touch chest.''
  
  \paragraph{Scope of proposed changes varied.} Designer A proposed modifications to all four sensors with detailed per-sensor anatomical reasoning. Designer B
  provided only global feedback without targeting specific sensors in either of the two iterations, offering directives such as ``move it so that the sensors better move with the joint'' and ``distributing them a bit
  wider'' around the shoulder seam. Designer C proposed two to three incremental adjustments per iteration with global observations, building on the previous iteration's outcome. In Iteration~2, Designer~C acknowledged that ``the flexion has improved and is much better now'' and noted that ``the importance of sensors are more equal and balanced'', while referencing current metric values (e.g., ``MPJAE is around 11.7, 6.7 and 6.1 for
  flexion, abduction and rotation'') to guide further refinement.
  
  \paragraph{Observed relationship between feedback style and outcome.} In this case study, the human designer who proposed fewer changes per round and described physical observations (Designer~C) achieved the best LH-Series
  outcome. Designer~C completed two iterations, progressively refining sensor positions while retaining effective placements from the previous iteration. The designer who proposed the most changes grounded in domain theory
  (Designer~A) completed only one iteration and experienced performance degradation. Designer~B also completed two iterations with global-level suggestions and achieved intermediate results.

\begin{table*}[t]
  \caption{Representative Designer Feedback Excerpts from the LH-Series}
  \label{tab:feedbacktypes_detailed}
  \begin{tabular}{p{0.08\linewidth}p{0.04\linewidth}p{0.60\linewidth}p{0.20\linewidth}}
    \toprule
    Designer & Iter. & Feedback Excerpt & Observation \\
    \midrule
    A & 1 & ``This sensor position mimics the m.deltoideus anterior part placement/location in the body. The flex sensor should touch or pass the GH joint.'' & References specific anatomy to prescribe sensor positions \\
    \cmidrule{2-4}
    & 1 & ``If the sensor is placed like in the description, there will be curves (waves) that the sensor will make, which will not align with the human body surface during flexion and abduction movement.'' & Predicts physical behavior from anatomical reasoning \\
    \midrule
    B & 1 & ``The detection results have a lot of error. Move it so that the sensors better move with the joint. Note that the sensors bend only in one direction, so orient the sensors accordingly.'' & References error metrics and sensor physics \\
    \cmidrule{2-4}
    & 2 & ``S2 importance is quite low... the orientation of the sensor is not aligned well enough to the direction of bending of the joints. Currently the three sensors are placed quite close to one another, maybe distributing them a bit wider would capture different movements.'' & References computed importance; suggests spatial redistribution \\
    \midrule
    C & 1 & ``Maybe can be moved a bit to chest side so that some of the sensing touch chest... This suggestion is especially based on my own testing results.'' & Reports observation from physical testing \\
    \cmidrule{2-4}
    & 2 & ``This updated position is much better. I am wondering would it be even better if we place the sensor even more to the chest area? Maybe we could place the sensor to be more following the muscle direction instead of just a straight line.'' & Builds on previous round; uses exploratory language \\
    \bottomrule
  \end{tabular}
\end{table*}

\section{Discussion}

\subsection{LLM as a Cross-Domain Knowledge Integrator (RQ1)}

RQ1 asked whether LLMs can generate sensor layouts that match or exceed the quality of initial designs produced by human designers. Our results indicate that, the answer is yes: $L_0$ outperformed all three human designs across every evaluation metric.

A closer look at the designs helps explain why. Each designer approached placement through the lens of their own discipline. Designer~A, drawing on kinesiology training, concentrated sensors on anatomically meaningful muscle groups---a strategy that yielded the best abduction prediction among all initial designs (7.76°) but left anterior--posterior coverage gaps that penalized flexion and rotation. Designer~B distributed sensors along garment structural features, reflecting an e-textile engineering perspective, while Designer~C clustered sensors around the most visibly mobile region of the shoulder. Each design thus bore the imprint of its creator's disciplinary focus, capturing some motion axes well but underserving others.

$L_0$, by contrast, spread sensors across both the anterior and posterior shoulder---a broader spatial strategy that none of the three designers adopted independently. This pattern is consistent with the hypothesis that LLMs can act as cross-domain knowledge integrators, simultaneously drawing on anatomy, biomechanics, and textile--skin coupling in ways that are difficult for a single practitioner~\cite{xu2024penetrative, makatura2024can}. The per-joint error decomposition reinforces this interpretation: $L_0$ achieved the lowest errors for flexion and rotation---the two axes that benefit most from spatially distributed sensing---while Designer~A's anatomically focused layout retained an advantage for abduction, a motion well captured by localized deltoid placement alone.

The N-to-1 divergent--convergent prompting strategy also contributed to $L_0$'s robustness. Our synthesis analysis (Appendix~\ref{app:iteration_llm}) revealed that while all five independent LLM proposals converged on lateral and posterior deltoid coverage, they diverged most on the fourth sensor's path and on flexion sensor orientation. The meta-synthesis step filtered implausible configurations and consolidated the most robust principles into a balanced layout.

\subsection{Iterative Refinement: When Does Collaboration Help? (RQ2)}
\label{sec:discuss_2}

RQ2 asked how human--AI collaborative iteration compares to purely human-driven iteration. Our data reveal that the answer depends critically on the designer's experience level---a pattern we did not anticipate when designing the study.

In the H-Series, the two more experienced designers (A and B) steadily improved their layouts across both cycles, with Designer~B ultimately reaching the lowest MPJAE of any condition. Designer~C, the least experienced, improved initially but regressed in the second cycle. The LH-Series told a strikingly different story: Designer~C was the only participant to achieve consistent improvement throughout, while Designer~A experienced degradation after a single round and chose to discontinue.

This inverted expertise--outcome relationship resonates with recent findings on AI-assisted work in other domains. Brynjolfsson et al.~\cite{brynjolfsson2025generative} observed that generative AI productivity gains among customer support agents accrued primarily to novice workers, arguing that AI effectively disseminates the tacit knowledge of high performers. A similar dynamic appears at play here: the LLM's cross-domain reasoning compensated for Designer~C's limited biomechanical background, while Designer~A's well-established domain models left less room for complementary AI contributions---and may have even introduced friction when the LLM's suggestions conflicted with their existing mental models. From a practical standpoint, this skill-leveling effect is notable: through collaboration, Designer~C (2~years of experience) achieved performance comparable to what Designer~B (4~years) reached through fully independent effort.

It is worth noting that the fixed condition ordering (H-Series before LH-Series) makes this finding complex. Any experience accumulated during the H-Series could have carried over to benefit subsequent LH-Series performance, meaning the Designer C’s strong performance on the LH-Series may also stem from tacit knowledge gained during the H‑series experiments. Another consideration regarding the experimental design is the starting point for the LH-Series. Instead of using the standard LLM-generated initial design ($L_0$), a potential alternative is to use a human designer's initial design ($H_0$). While $L_0$ provides a fair baseline for comparison, using $H_0$ might offer better insights into how LLMs assist designers in cross-domain knowledge integration by addressing missing layout concerns.

\begin{figure}[h]
  \centering
  \includegraphics[width=\linewidth]{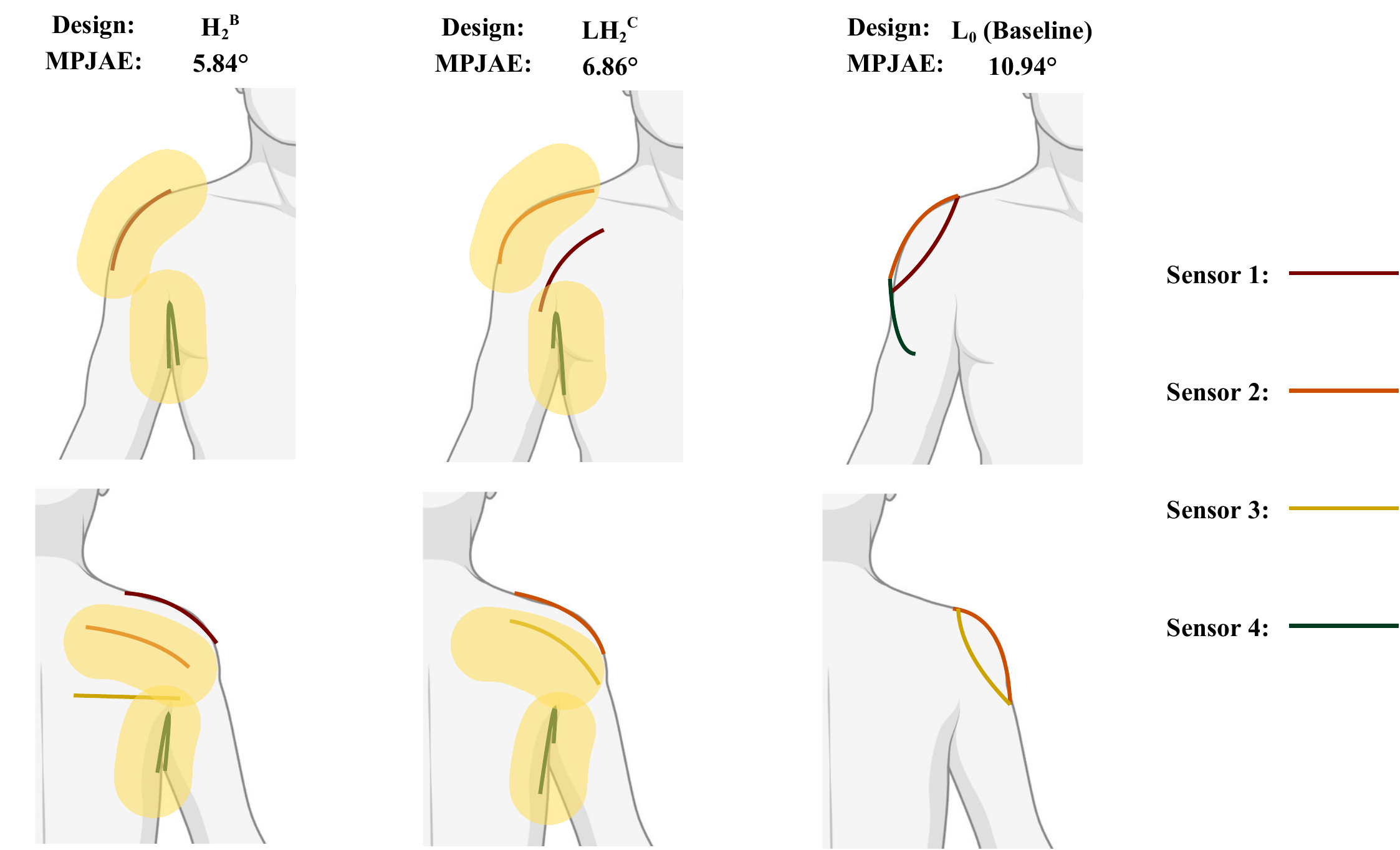}
  \caption{Best-performing layouts. (a) $H_2^B$ and (b) $LH_2^C$ show convergent placement across the acromion, posterior scapula, and axilla, outperforming the (c) $L_0$ baseline. Highlighted zones indicate these common sensing regions.}
  \Description{Three shoulder models comparing the best layouts from different experimental conditions, illustrating the convergence of sensors around the acromion, back, and underarm.}
  \label{fig:best_layouts}
\end{figure}

An intriguing piece of convergent evidence further supports this interpretation. The two best-performing layouts---$H_2^B$ (5.84°, human-only) and $LH_2^C$ (6.86°, human--AI)---arrived at remarkably similar tri-zonal configurations (Figure~\ref{fig:best_layouts}): a superior sensor crossing the acromion, posterior coverage spanning the infraspinatus and posterior deltoid, and an inferior sensor anchored near the axilla. That two independent paths---one guided entirely by an experienced human, the other by a novice collaborating with an LLM---converged on the same spatial solution suggests both a relatively constrained optimum for this task and, more importantly, that LLM-assisted collaboration can guide less experienced designers toward that optimum.

\subsection{What Shapes Collaboration Effectiveness? (RQ3)}

RQ3 aimed to identify factors that influence the effectiveness of human--AI collaboration. Our qualitative analysis of the LH-Series feedback records points to three interrelated factors: the granularity and abstraction level of human feedback, and the way domain background shapes both.

\paragraph{Granularity: incremental adjustments outperform sweeping redesigns.}
The contrast between Designer~A and Designer~C is instructive. Designer~A proposed modifications to all four sensors simultaneously, each backed by detailed anatomical reasoning, yet the resulting layout performed worse. Designer~C adjusted only one or two sensors per round, preserving what had worked and building on it incrementally. This pattern is consistent with classical iterative experimental design principles: small changes make it possible to attribute performance shifts to specific decisions, while sweeping modifications conflate effects and raise the risk of regression~\cite{box2005statistics}. In the human--AI context, this principle has an additional dimension---targeted feedback constrains each LLM reasoning step to a manageable subproblem, whereas coordinating simultaneous changes across all sensors may stretch the model's capacity for maintaining internal consistency.

Designer~C's feedback trajectory illustrates this progressive strategy. In Iteration~1, after hands-on inspection of the prototype, Designer~C suggested shifting S1 (Sensor 1) toward the chest and repositioning S4 into a U-shaped path under the arm---targeted changes that produced a substantial flexion improvement. In Iteration~2, Designer~C explicitly acknowledged the prior gains (``the flexion has improved and is much better now'') and turned attention to finer adjustments, asking sensors to ``follow the muscle direction.'' Each round thus built on the last rather than starting from scratch.

\paragraph{Abstraction: observations outperform prescriptions.}
How designers framed their feedback proved equally consequential. Designer~A communicated in prescriptive anatomical terms---specifying muscle names and insisting that sensors ``should touch or pass the GH joint''---effectively dictating exact placements and leaving the LLM to serve as a coordinate translator. Designer~C described what they observed and wondered about (``maybe can be moved a bit to chest side''; ``I am wondering would it be even better if we place the sensor even more to the chest area?''), giving the LLM space to integrate these observations with the quantitative metrics and its own biomechanical knowledge.

This distinction echoes a broader pattern in human--AI co-creative systems: overly directive instructions tend to suppress complementary AI reasoning, while observation-level prompts enable synergistic outcomes in which each party contributes its distinct strengths~\cite{rezwana2023designing, gmeiner2023exploring}. Designer~B's experience occupies a middle ground---global observations (``sensors bend only in one direction''; ``distributing them a bit wider'') without per-sensor specificity---and produced intermediate results, reinforcing the idea that collaboration effectiveness varies along a continuum from directive to observational feedback.

\section{Limitations}

Our study has several limitations. First, the single-subject design ensures controlled comparison across layout conditions but limits the generalizability of absolute performance values to other body types and anatomical variations. Second, with only three human designer participants, our findings capture preliminary patterns in feedback diversity and collaboration dynamics but do not support statistical inference; a larger designer panel is needed to validate the observed relationships between domain background and collaboration effectiveness. Third, although markerless motion capture has been widely adopted as a practical reference system and offers acceptable measurement error in prior work~\cite{yin2025smplest}, its inherent uncertainty means that the lowest reported errors may partially reflect reference noise. Fourth, considering the rapid development and diversity of LLM technologies, a broader range of models needs to be evaluated.

To address these limitations, our next step is to conduct a large-scale multi-subject study with diverse body types and a broader human designer panel, enabling statistical validation of the design patterns and collaboration dynamics observed in the current case study. Additionally, we plan to integrate simulation-based methods into the iterative design pipeline to examine whether combining LLM-driven reasoning with simulation feedback can improve generalizability and reduce reliance on costly physical prototyping and testing. In the future, we will explore more diverse and advanced LLMs to investigate the potential effects of human–AI collaboration on sensor layout design.

\section{Conclusion}
This paper investigated LLMs as cross-domain knowledge integrators for e-textile sensor layout design through a case study on shoulder joint angle prediction involving three human designers with different technical backgrounds. In an initial design comparison, the LLM-generated layout outperformed all three human designers initial designs, demonstrating the potential of LLMs to synthesize cross-domain knowledge. In iterative refinement, comparing human-only iteration with human--AI collaboration revealed an intriguing relationship: the least experienced designer achieved improvement through collaboration, reaching performance comparable to the best human-only result, while the most experienced designer experienced performance degradation. Qualitative analysis identified two critical factors governing collaboration effectiveness: feedback granularity, where incremental adjustments outperformed sweeping redesigns by enabling focused LLM reasoning, and feedback abstraction, where observation-oriented input yielded better outcomes than prescriptive directives by preserving space for complementary AI reasoning. These findings suggest that LLMs can effectively democratize access to cross-domain design expertise in physical prototyping tasks, provided that human feedback is structured to complement rather than constrain AI reasoning. Future work should validate these patterns with larger human designer panels, diverse body types, and integration of simulation-based virtual prototyping and evaluation to further reduce reliance on costly physical prototyping and tests.



\bibliographystyle{ACM-Reference-Format}
\bibliography{sample-base}

\clearpage
\appendix

\section{Prompt Template}
\label{app:prompt}
\begin{figure}[htbp]
    \centering
    \includegraphics[width=1\linewidth]{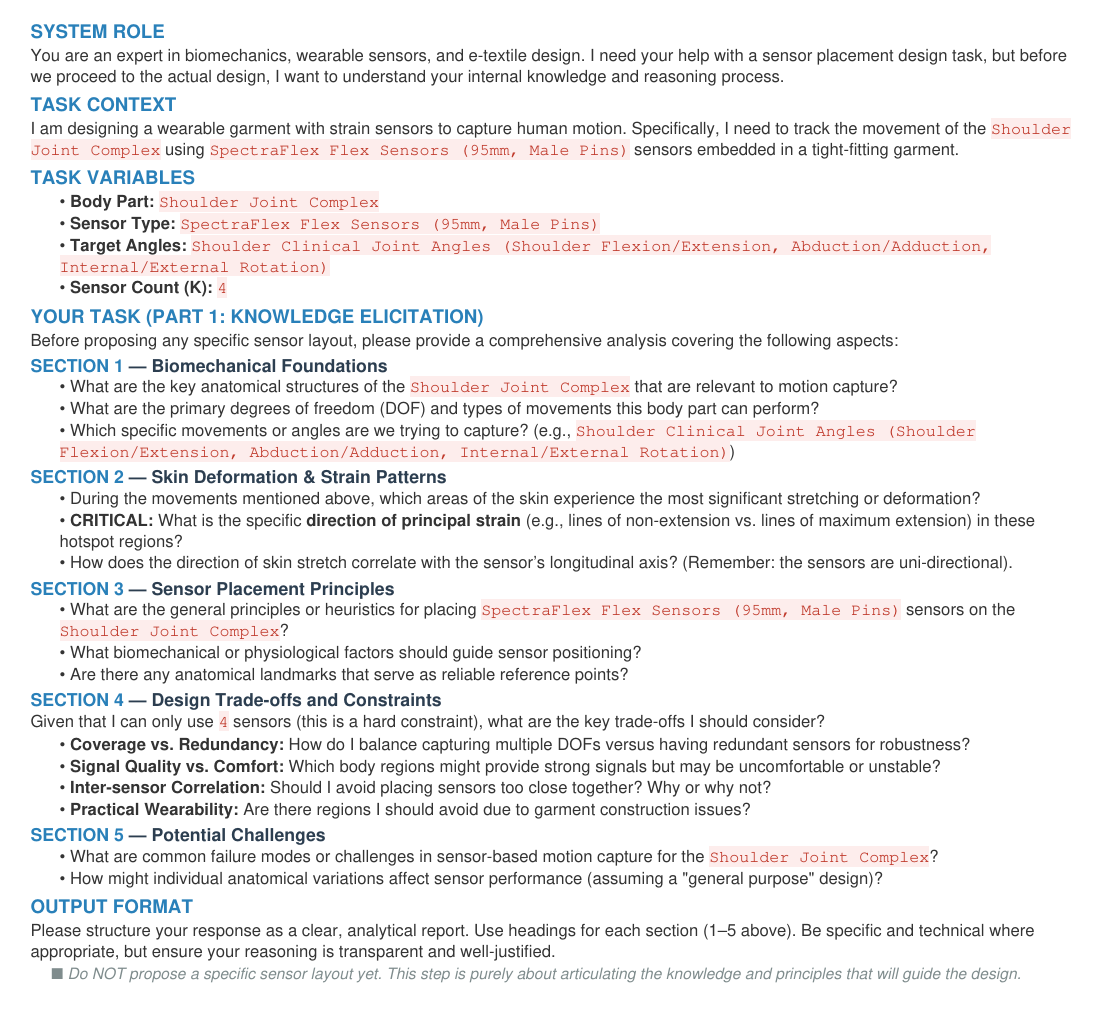}
    \caption{Prompt for divergent phase (step 1). This prompt extracts the LLM's internal knowledge prior to any design proposal}
    \label{fig:p1}
\end{figure}

\begin{figure}[htbp]
    \centering
    \includegraphics[width=1\linewidth]{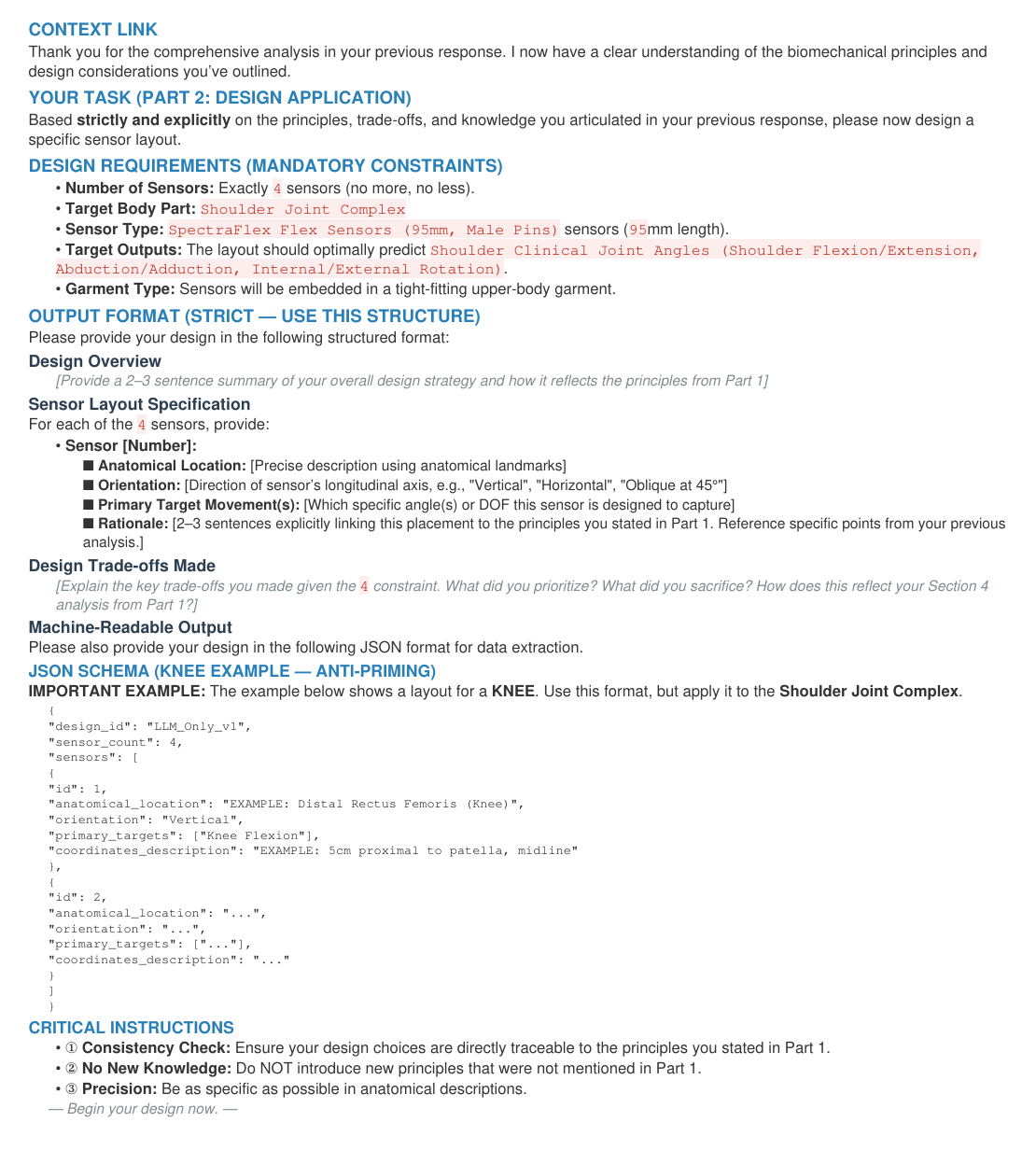}
    \caption{Prompt for divergent phase (step 1). Conditioned on the knowledge elicited by P1, this prompt requests a structured sensor layout. The JSON example uses a knee scenario to prevent misleading.}
    \label{fig:p2}
\end{figure}

\begin{figure}
    \centering
    \includegraphics[width=1\linewidth]{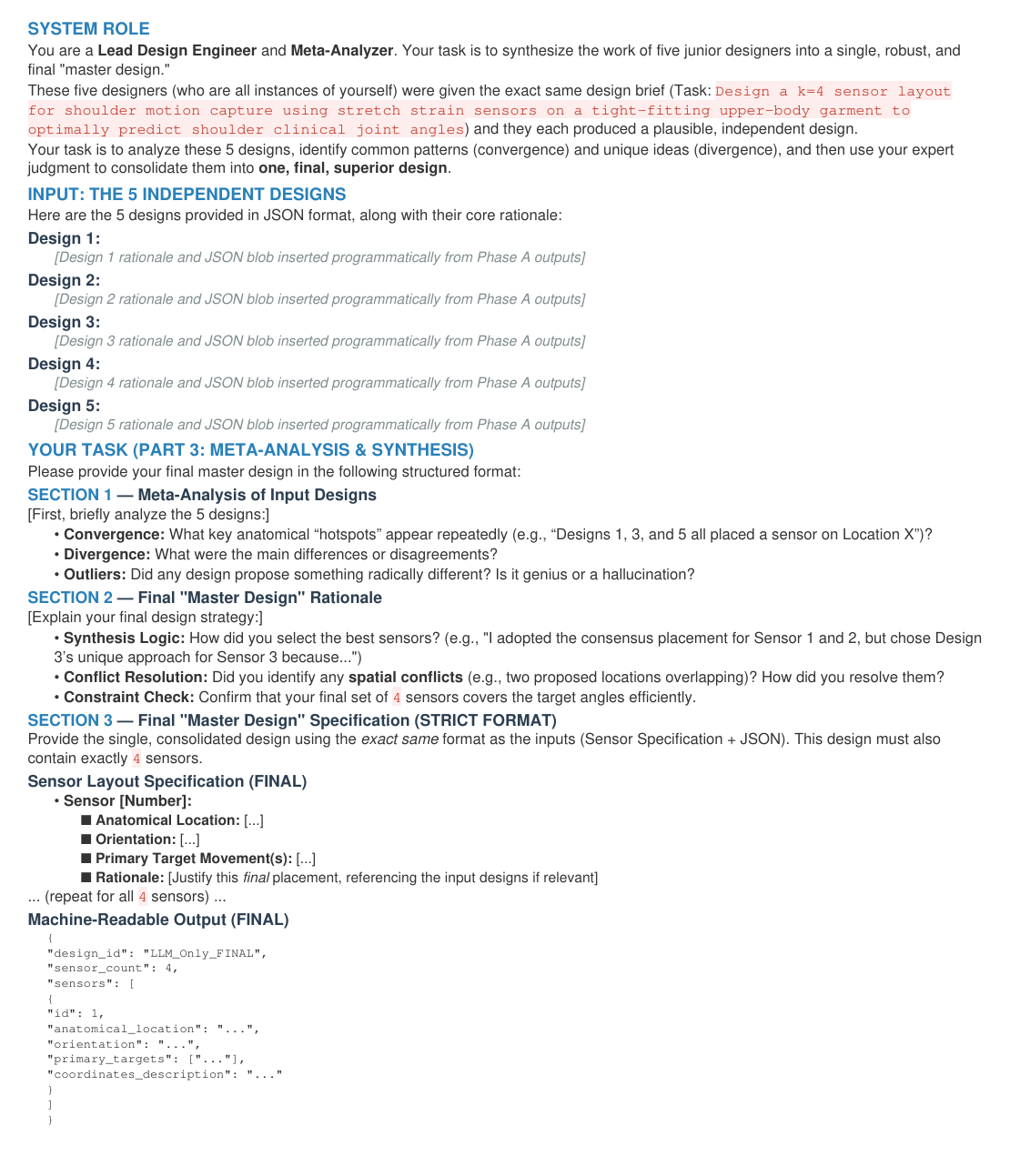}
    \caption{Prompt for convergent phase. This prompt synthesizes N = 5 independent designs from Phase A into a single consolidated master design.}
    \label{fig:p3}
\end{figure}

\begin{figure}
    \centering
    \includegraphics[width=1\linewidth]{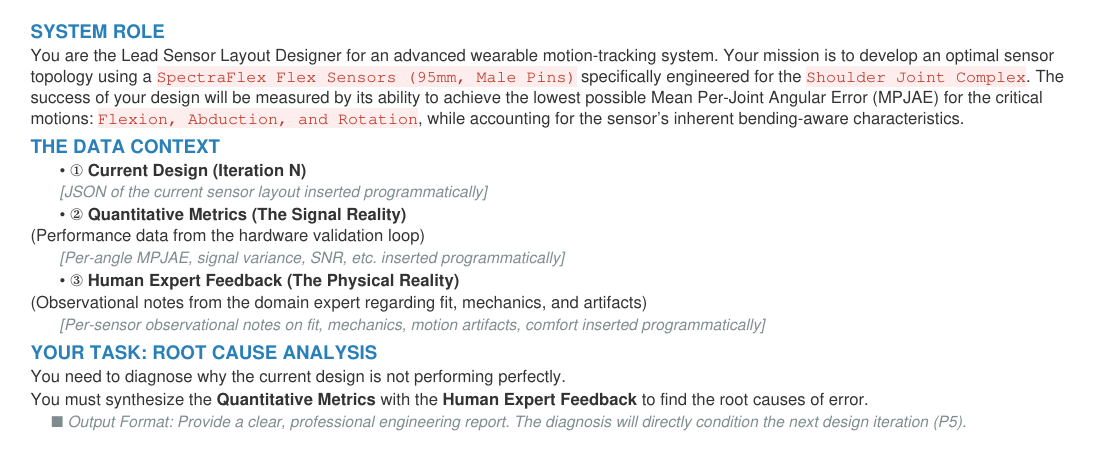}
    \caption{Prompt for Diagnostic Integration. The LLM receives the current design, quantitative performance metrics, and qualitative human designer feedback, then diagnoses failure modes.}
    \label{fig:p4}
\end{figure}

\begin{figure}
    \centering
    \includegraphics[width=1\linewidth]{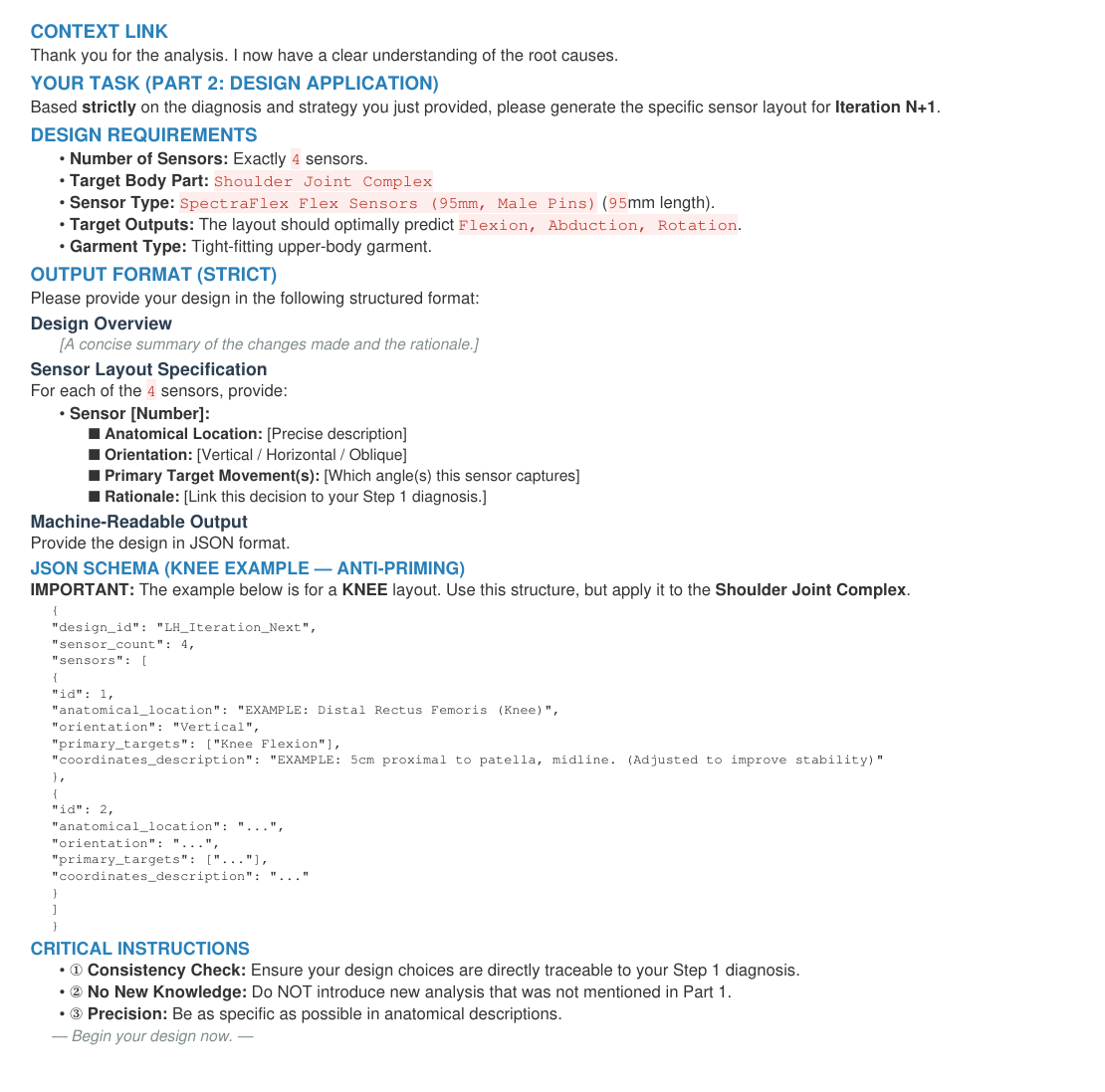}
    \caption{Prompt for synthesis and refinement. Conditioned on the diagnosis,
this prompt generates the next-iteration sensor layout.
}
    \label{fig:p5}
\end{figure}

\clearpage

\section{LLM Initial Design Outputs}
\label{app:iteration_llm}

   
\begin{figure}[H]                                                                           
\centering                                                
\includegraphics[width=\textwidth]{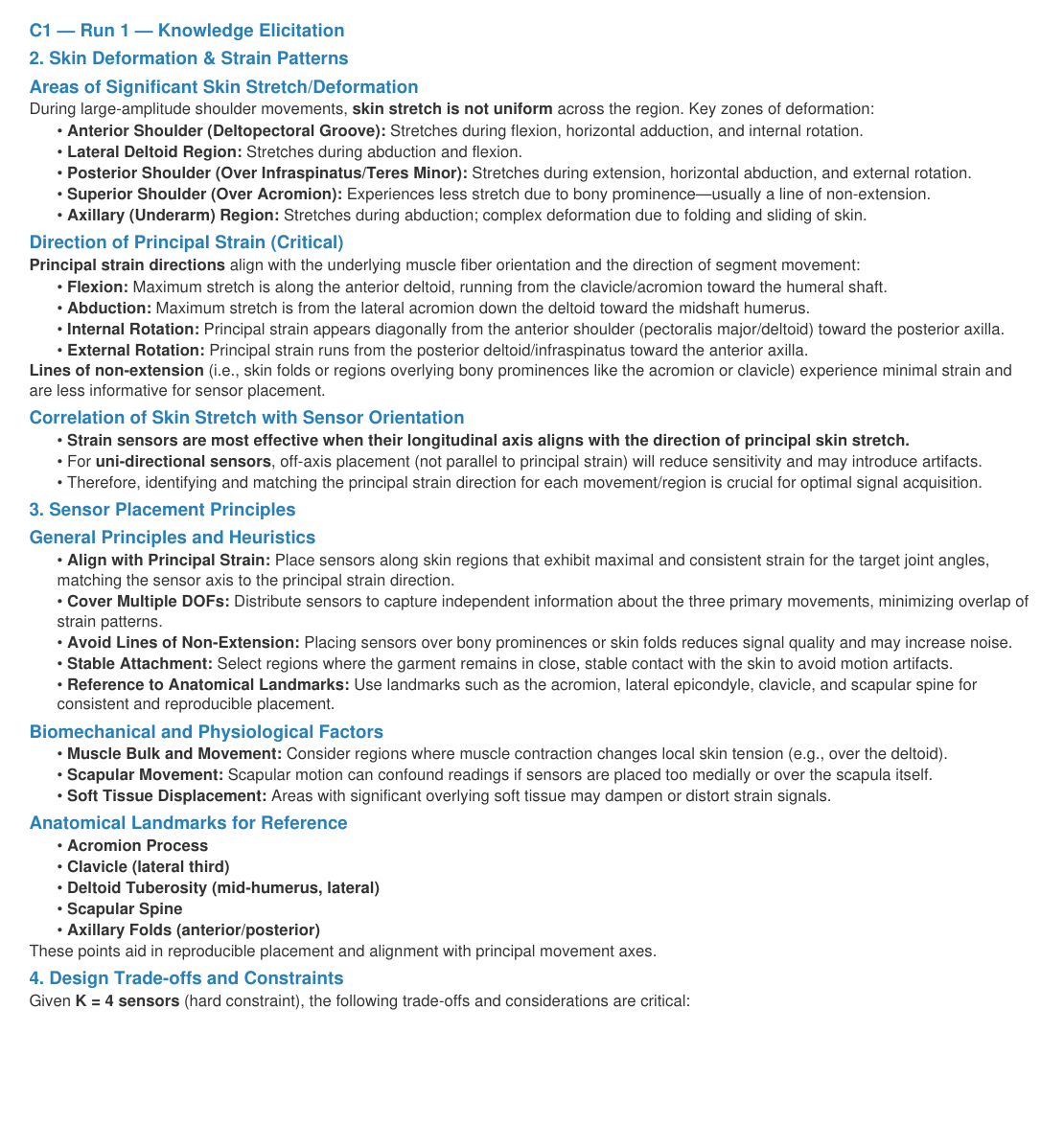}
\caption{C1 Run 1 — Knowledge Elicitation (Part 1).}
\label{fig:c1-r1-t1-p1}
\end{figure}
\begin{figure}[H]
\centering
\includegraphics[width=\textwidth]{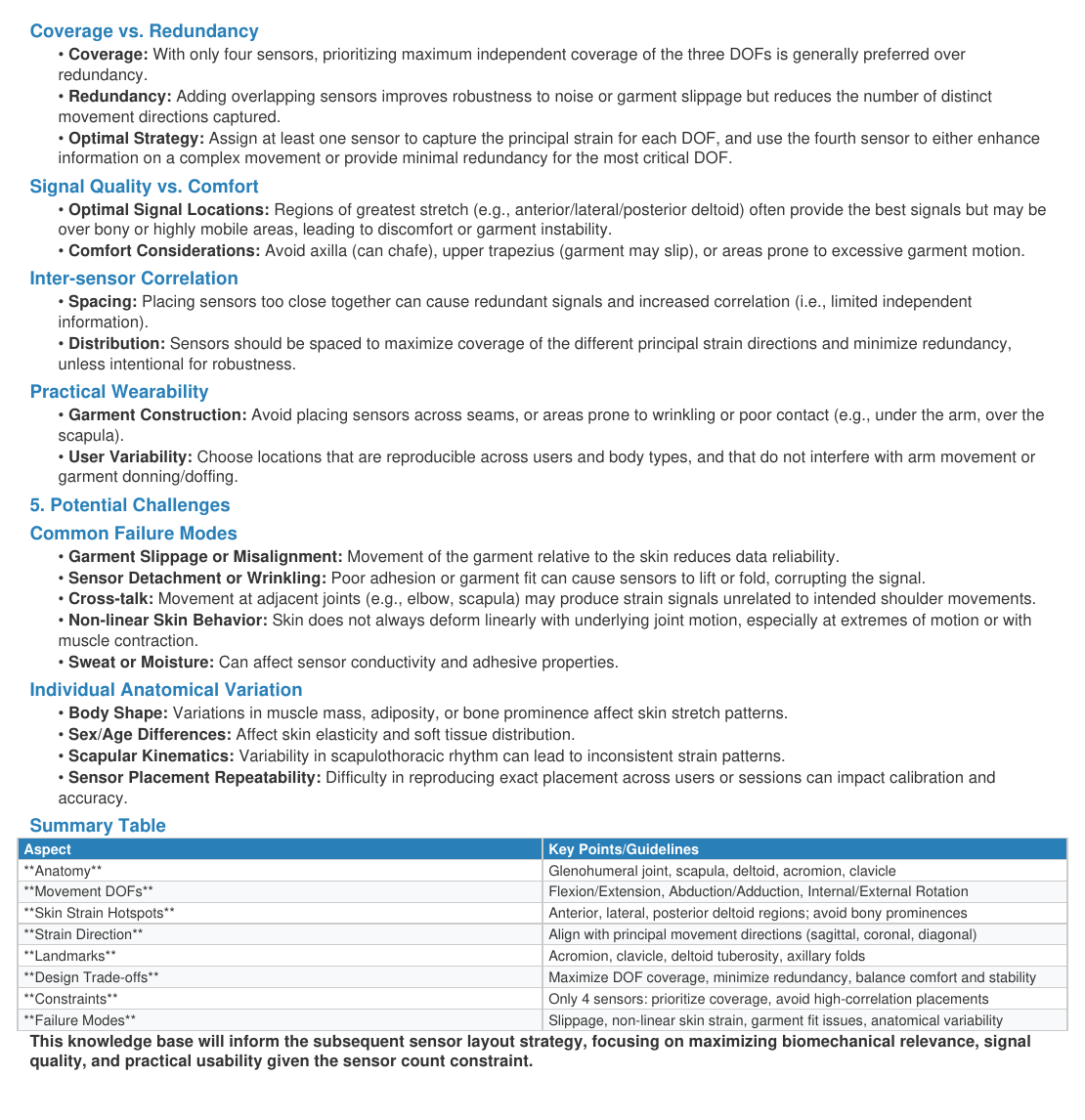}
\caption{C1 Run 1 — Knowledge Elicitation (Part 2).}
\label{fig:c1-r1-t1-p2}
\end{figure}
\begin{figure}[H]
\centering
\includegraphics[width=\textwidth]{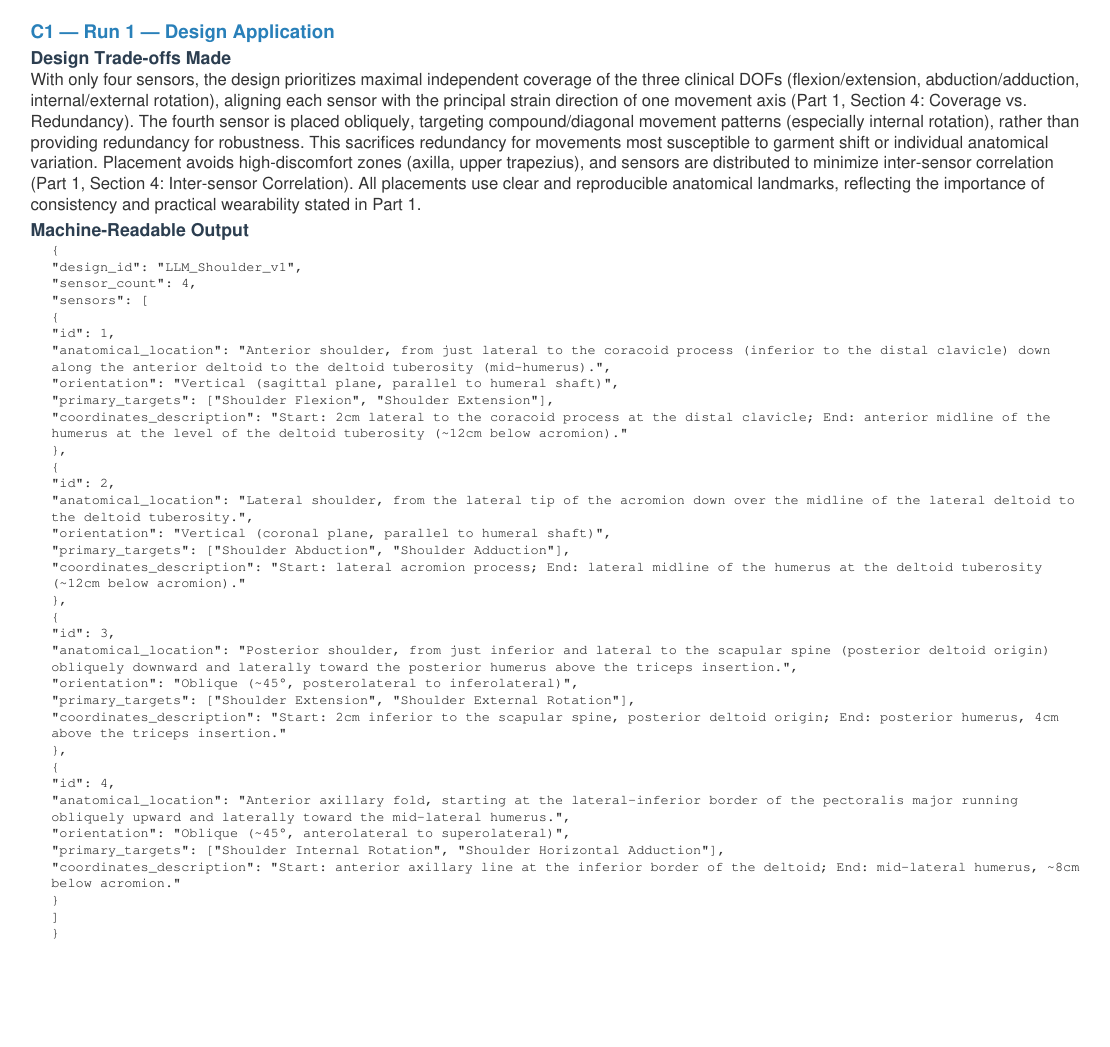}
\caption{C1 Run 1 — Design Application.}
\label{fig:c1-r1-t2}
\end{figure}
\begin{figure}[H]
\centering
\includegraphics[width=\textwidth]{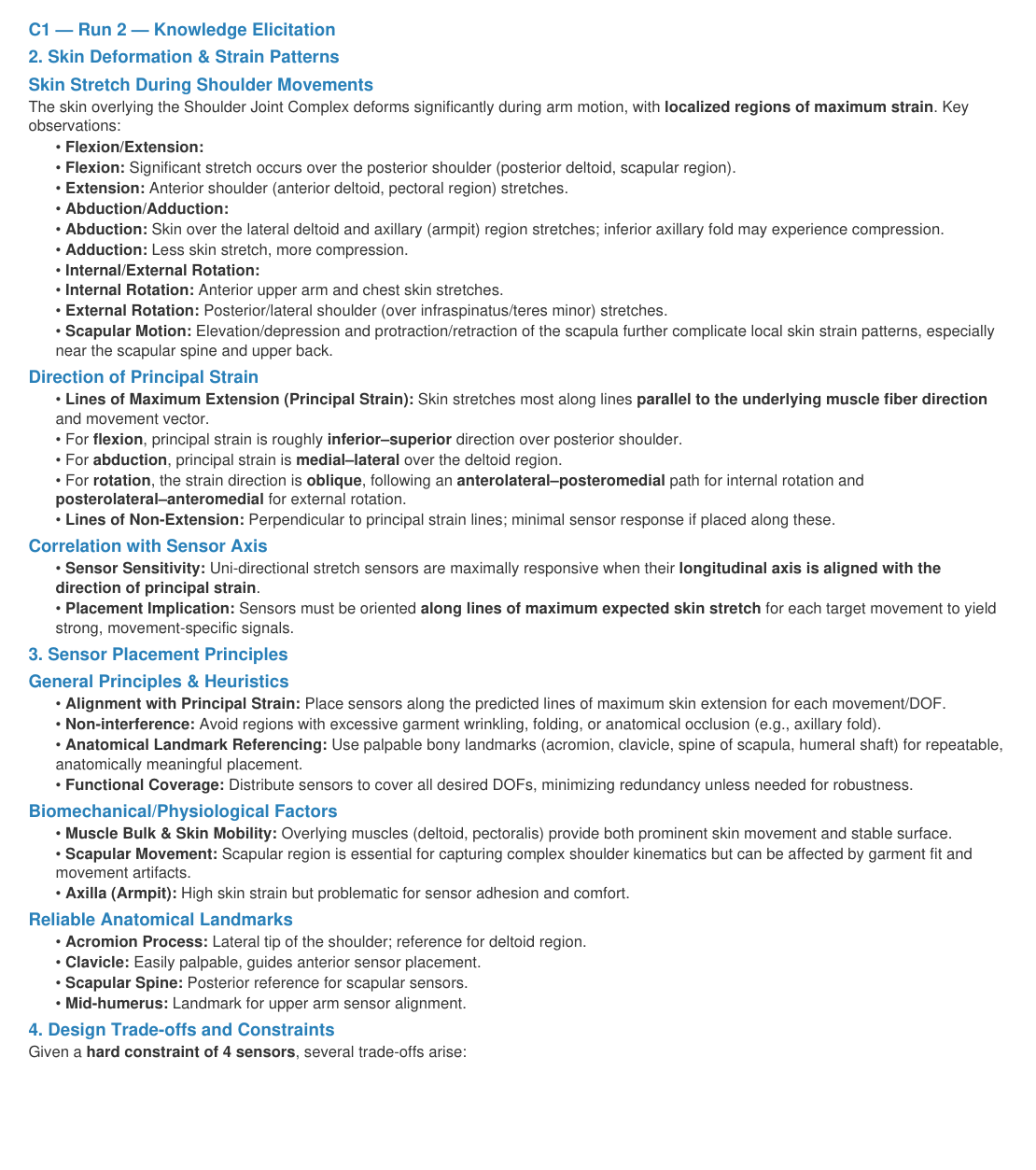}
\caption{C1 Run 2 — Knowledge Elicitation (Part 1).}
\label{fig:c1-r2-t1-p1}
\end{figure}

\begin{figure}[H]
\centering
\includegraphics[width=\textwidth]{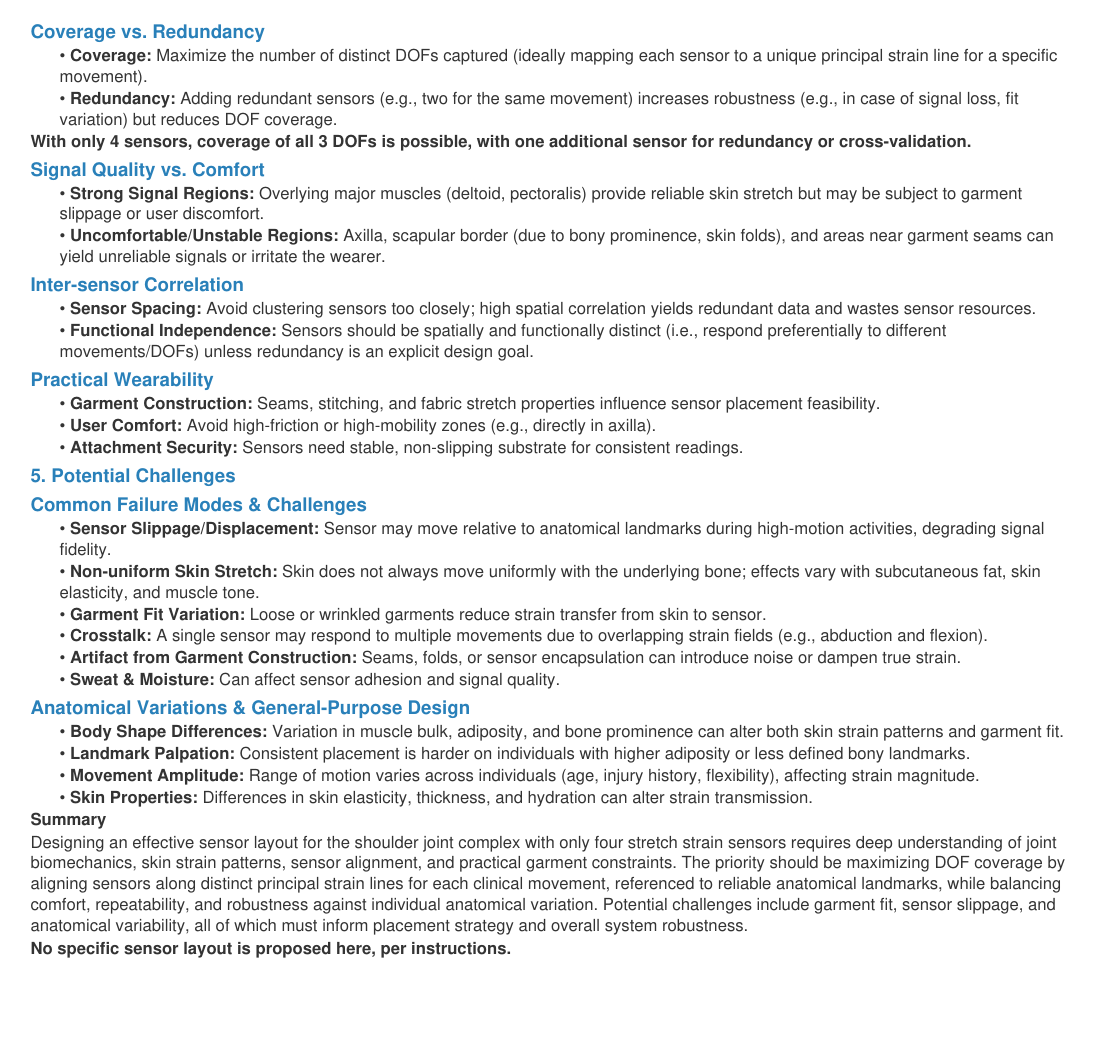}
\caption{C1 Run 2 — Knowledge Elicitation (Part 2).}
\label{fig:c1-r2-t1-p2}
\end{figure}


\begin{figure}[H]
\centering
\includegraphics[width=\textwidth]{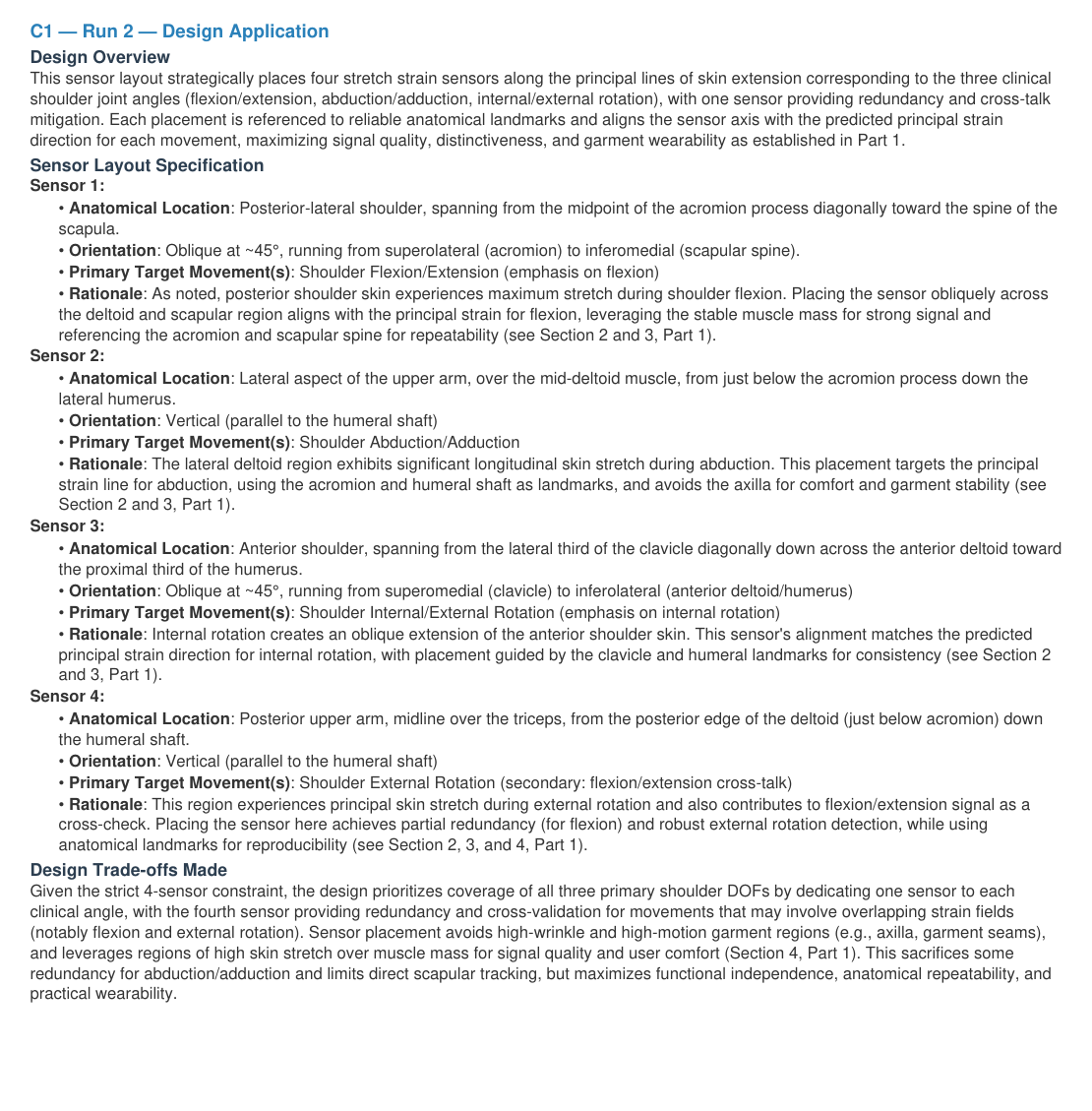}
\caption{C1 Run 2 — Design Application (Part 1).}
\label{fig:c1-r2-t2-p1}
\end{figure}

\begin{figure}[H]
\centering
\includegraphics[width=\textwidth]{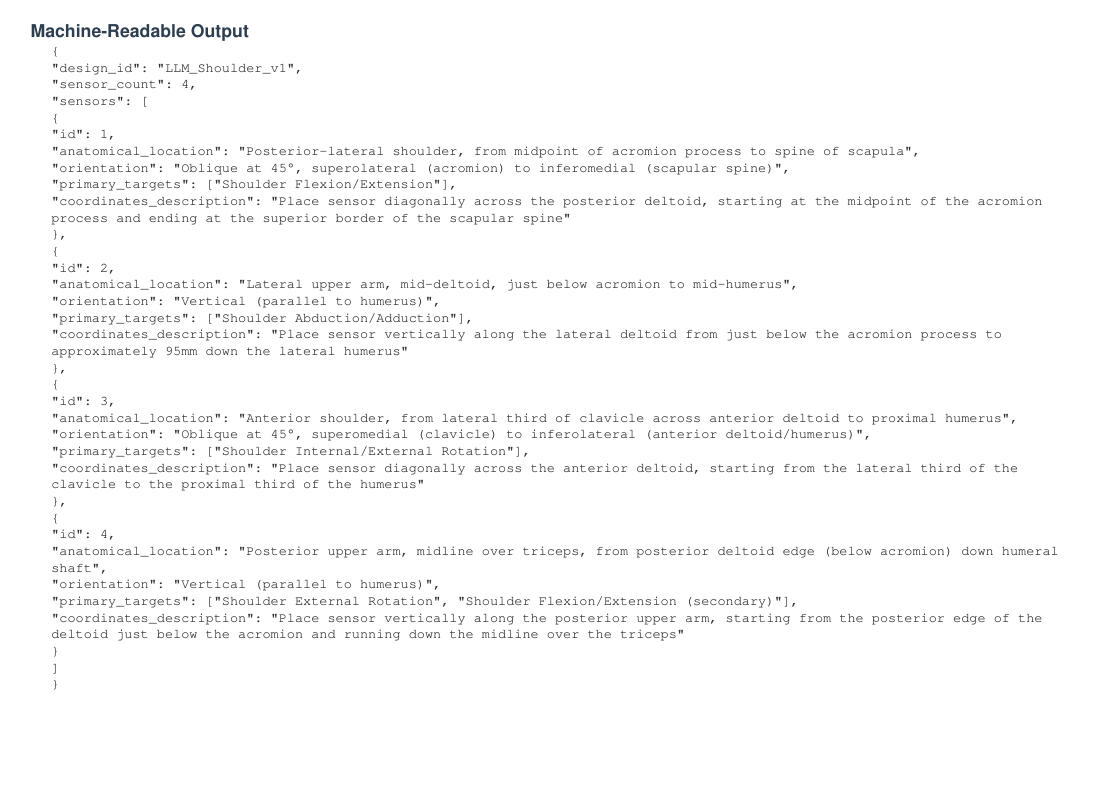}
\caption{C1 Run 2 — Design Application (Part 2).}
\label{fig:c1-r2-t2-p2}
\end{figure}


\begin{figure}[H]
\centering
\includegraphics[width=\textwidth]{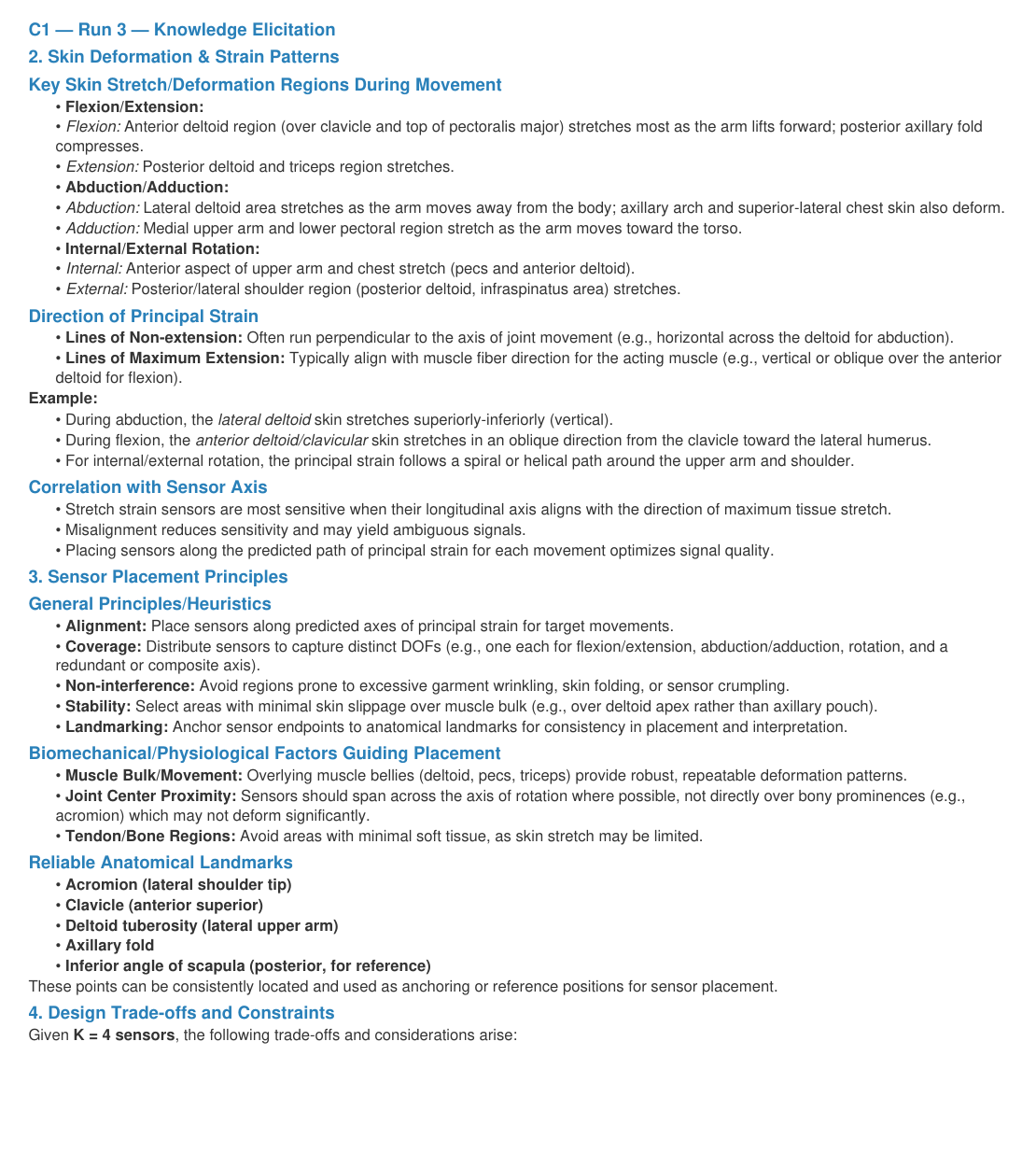}
\caption{C1 Run 3 — Knowledge Elicitation (Part 1).}
\label{fig:c1-r3-t1-p1}
\end{figure}

\begin{figure}[H]
\centering
\includegraphics[width=\textwidth]{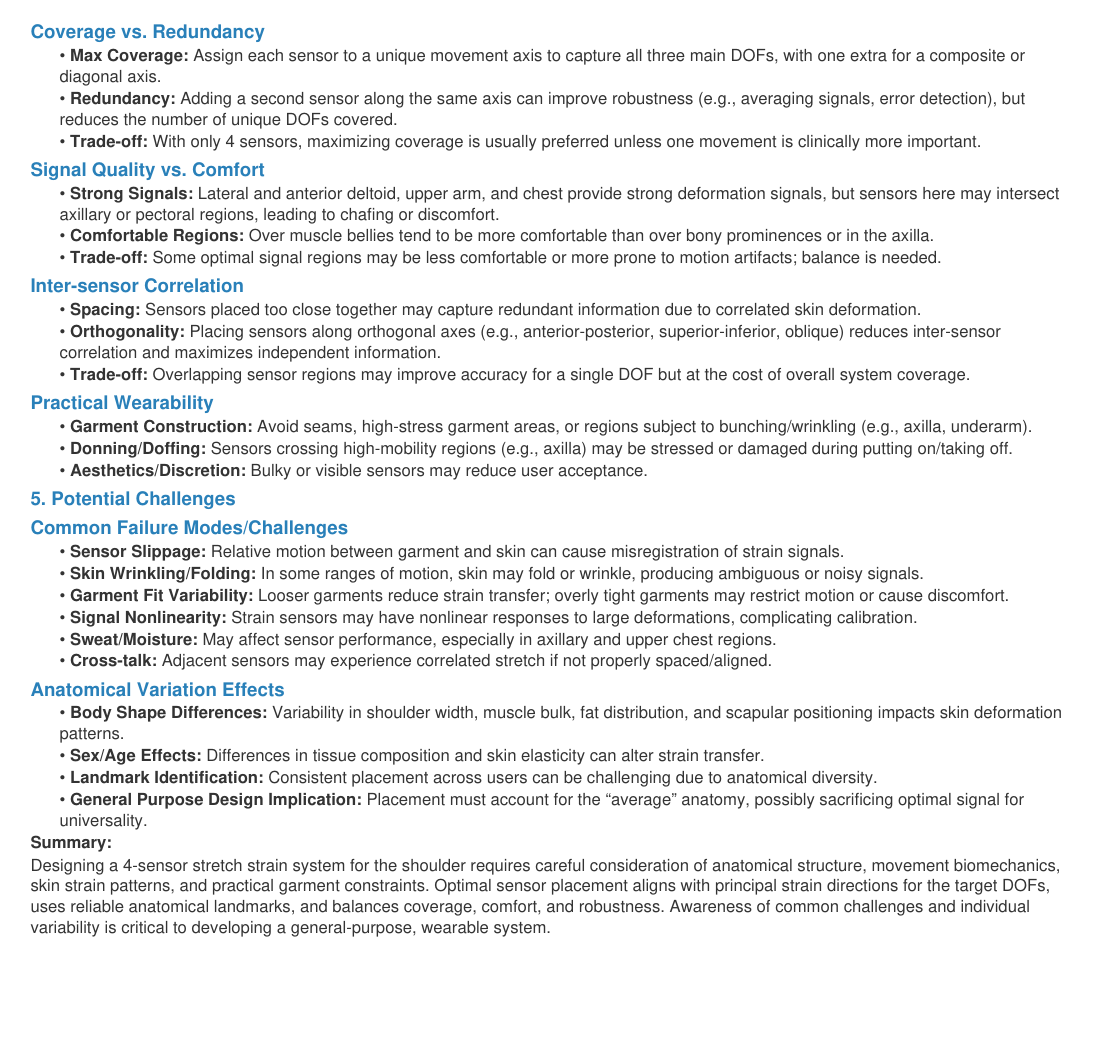}
\caption{C1 Run 3 — Knowledge Elicitation (Part 2).}
\label{fig:c1-r3-t1-p2}
\end{figure}


\begin{figure}[H]
\centering
\includegraphics[width=\textwidth]{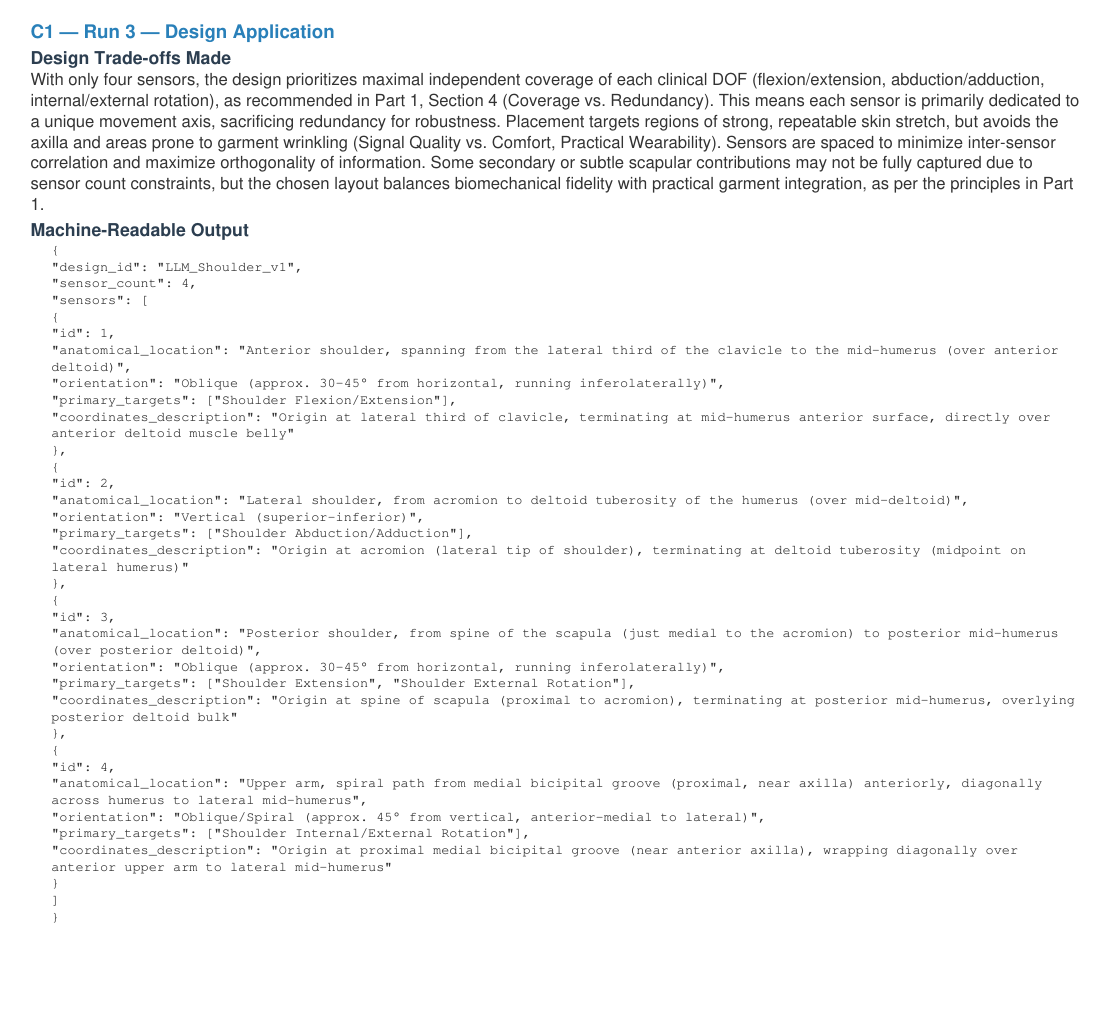}
\caption{C1 Run 3 — Design Application.}
\label{fig:c1-r3-t2}
\end{figure}


\begin{figure}[H]
\centering
\includegraphics[width=\textwidth]{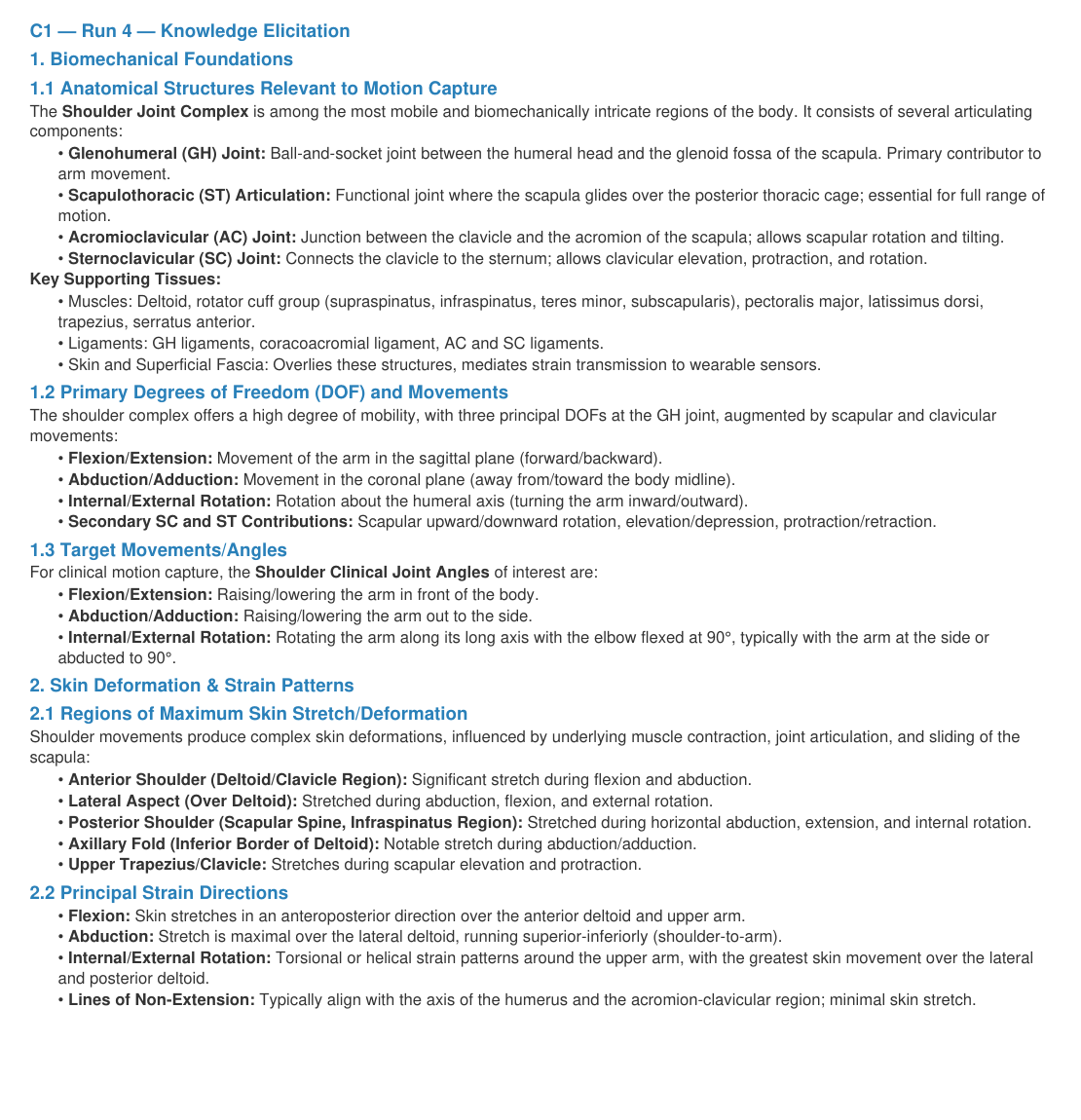}
\caption{C1 Run 4 — Knowledge Elicitation (Part 1).}
\label{fig:c1-r4-t1-p1}
\end{figure}

\begin{figure}[H]
\centering
\includegraphics[width=\textwidth]{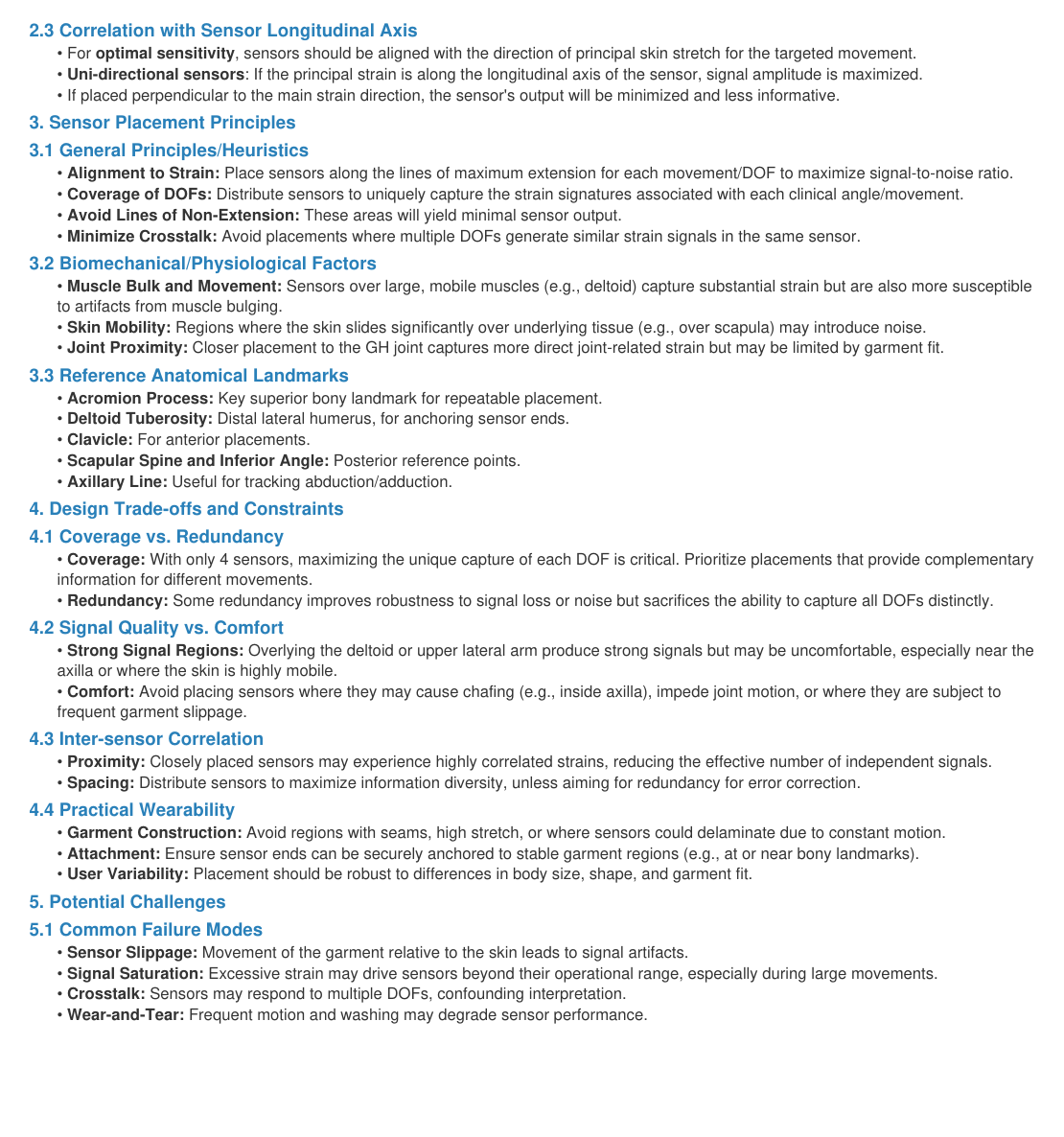}
\caption{C1 Run 4 — Knowledge Elicitation (Part 2).}
\label{fig:c1-r4-t1-p2}
\end{figure}

\begin{figure}[H]
\centering
\includegraphics[width=\textwidth]{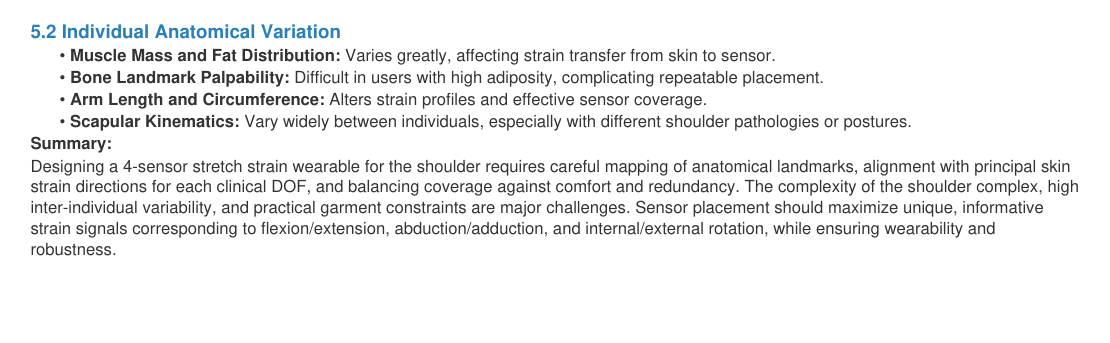}
\caption{C1 Run 4 — Knowledge Elicitation (Part 3).}
\label{fig:c1-r4-t1-p3}
\end{figure}


\begin{figure}[H]
\centering
\includegraphics[width=\textwidth]{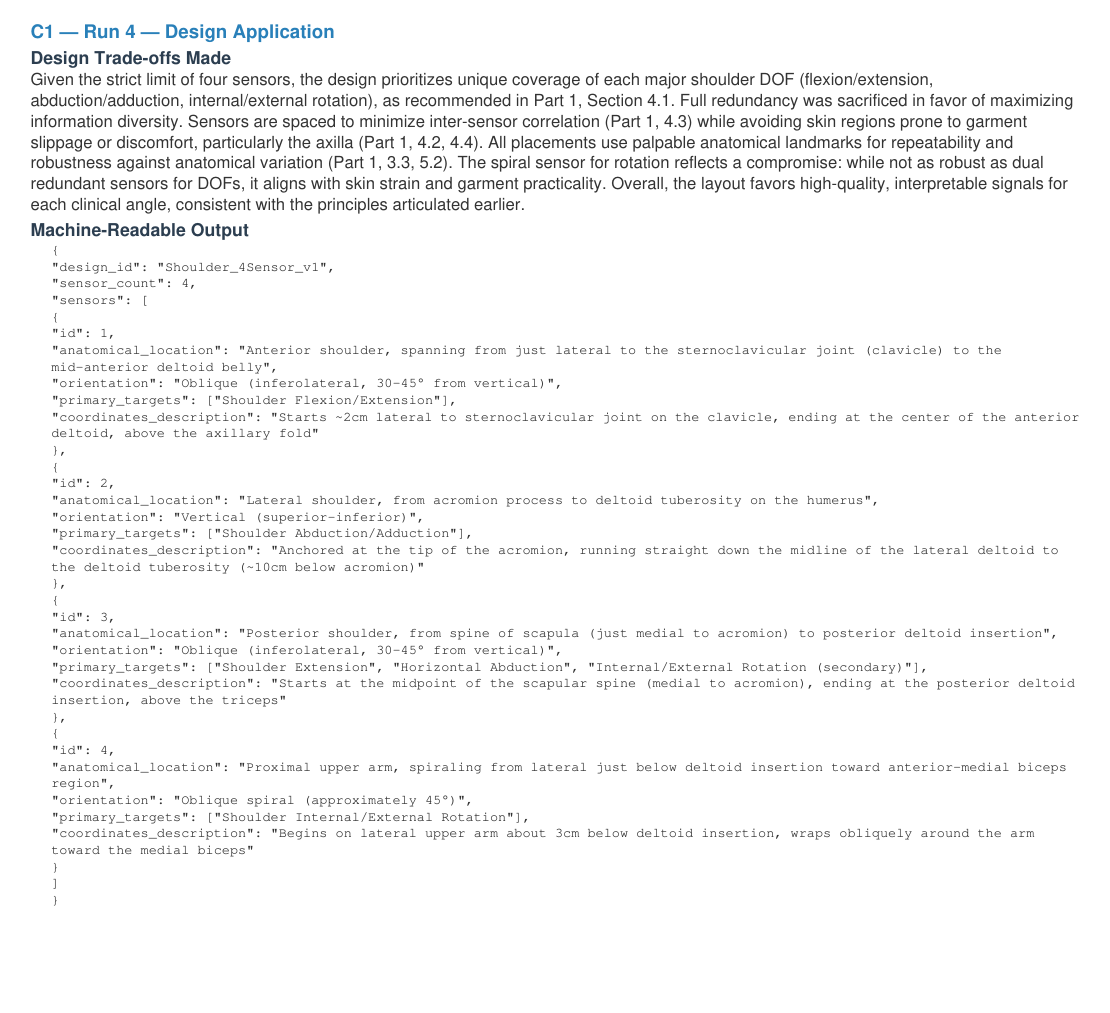}
\caption{C1 Run 4 — Design Application.}
\label{fig:c1-r4-t2}
\end{figure}


\begin{figure}[H]
\centering
\includegraphics[width=\textwidth]{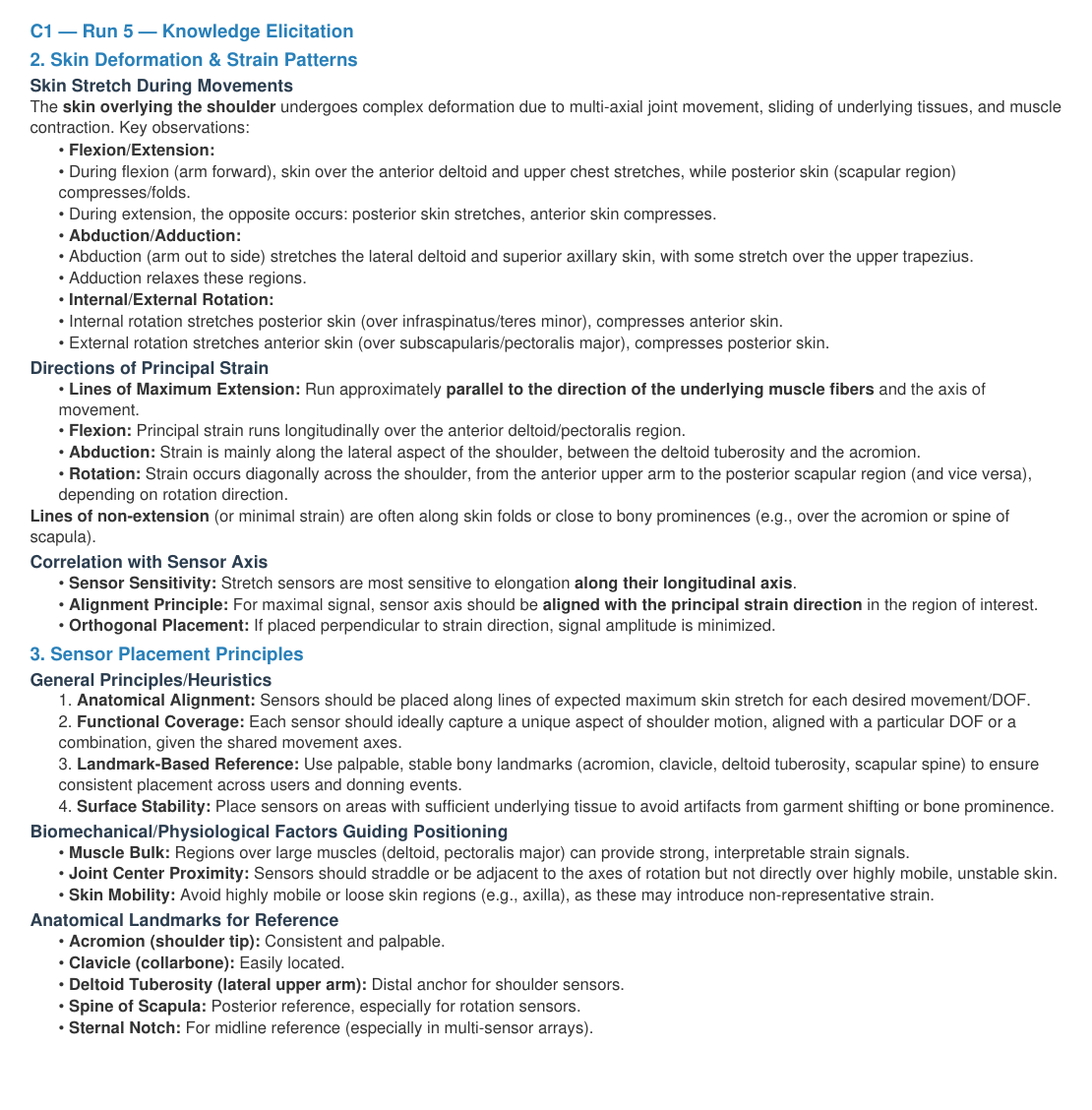}
\caption{C1 Run 5 — Knowledge Elicitation (Part 1).}
\label{fig:c1-r5-t1-p1}
\end{figure}

\begin{figure}[H]
\centering
\includegraphics[width=\textwidth]{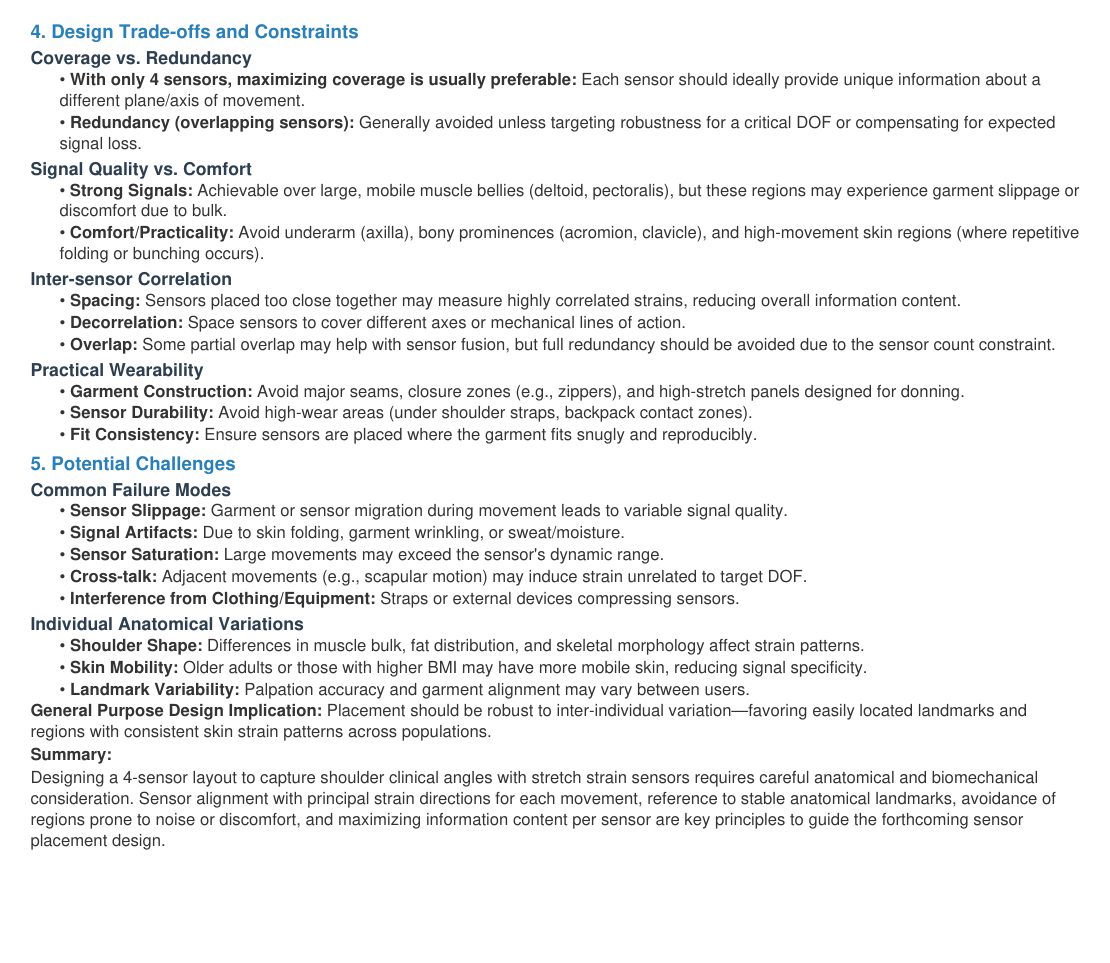}
\caption{C1 Run 5 — Knowledge Elicitation (Part 2).}
\label{fig:c1-r5-t1-p2}
\end{figure}


\begin{figure}[H]
\centering
\includegraphics[width=\textwidth]{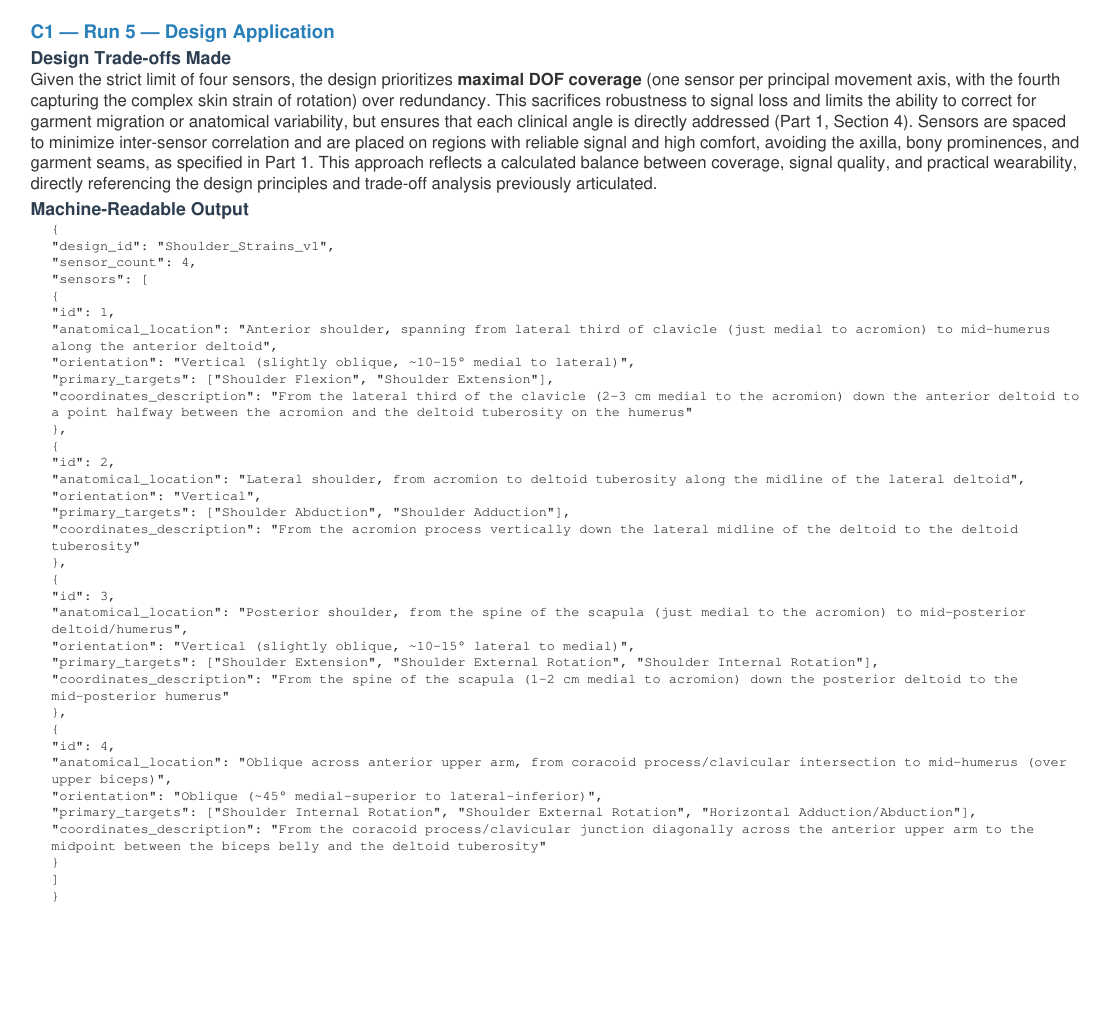}
\caption{C1 Run 5 — Design Application.}
\label{fig:c1-r5-t2}
\end{figure}


\begin{figure}[H]
\centering
\includegraphics[width=\textwidth]{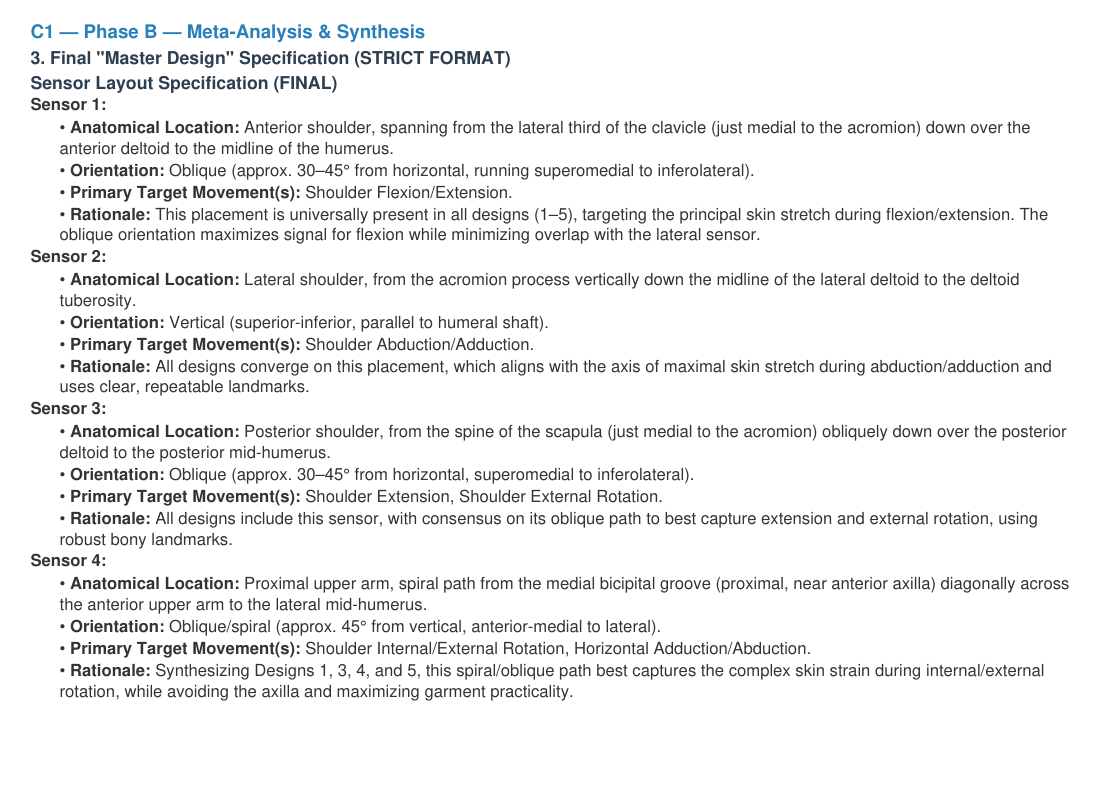}
\caption{C1 Phase B — Meta-Analysis \& Synthesis (Part 1).}
\label{fig:c1-synth-p1}
\end{figure}

\begin{figure}[H]
\centering
\includegraphics[width=\textwidth]{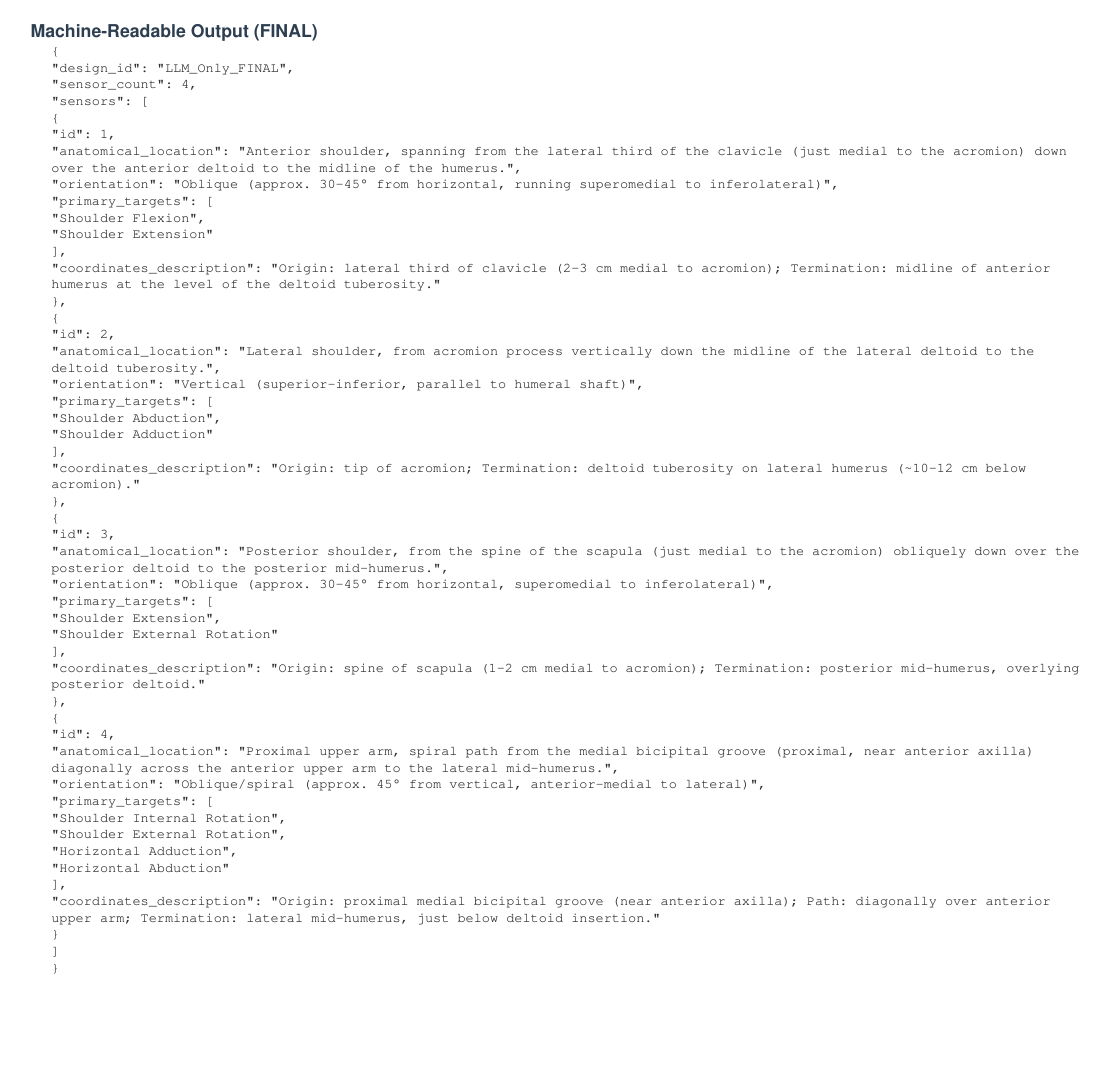}
\caption{C1 Phase B — Meta-Analysis \& Synthesis (Part 2).}
\label{fig:c1-synth-p2}
\end{figure}

\section{Iteration Layouts}
\label{app:iteration_results}

\begin{figure*}[htbp]
    \centering
    \includegraphics[width=\textwidth]{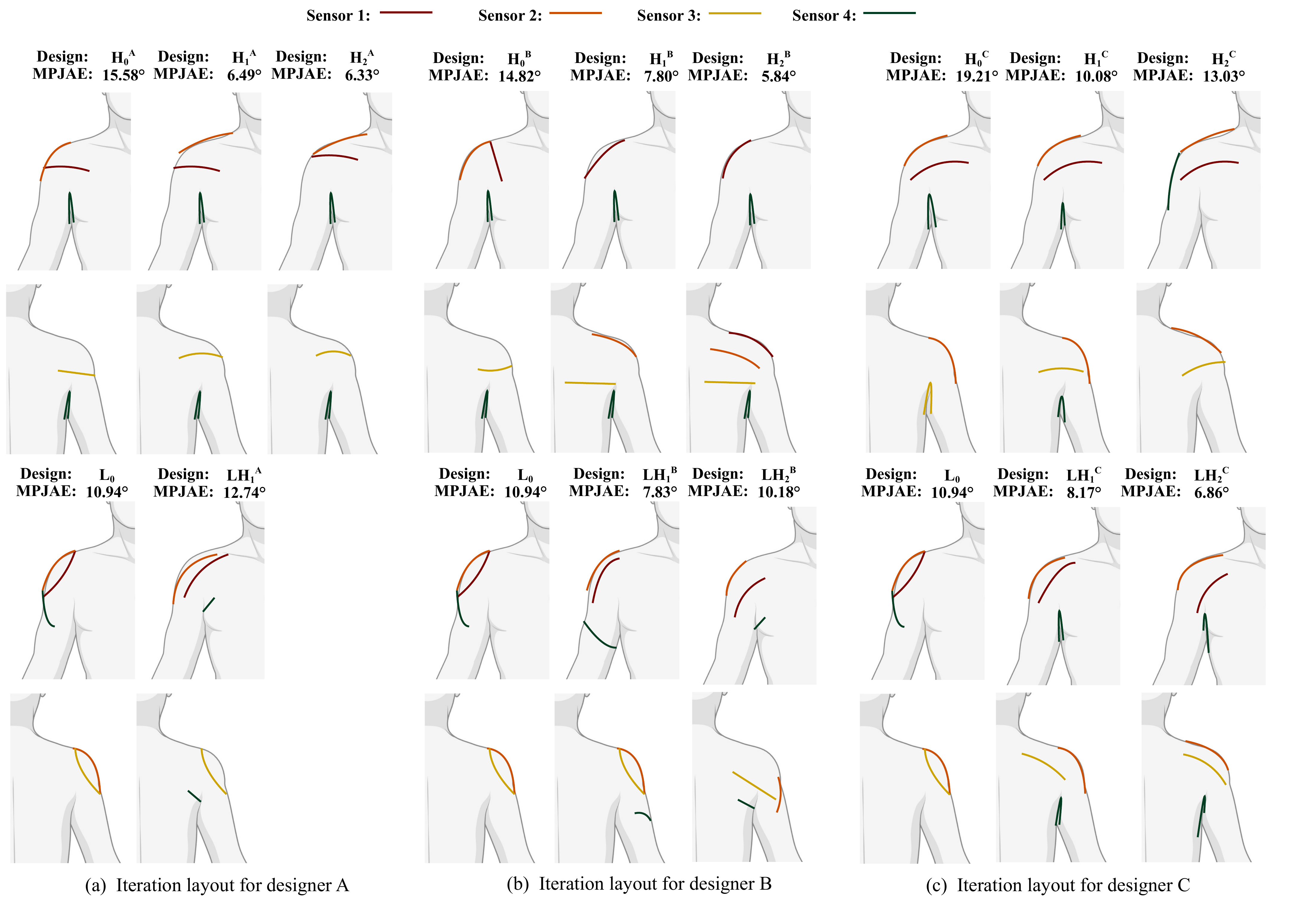}
    \caption{Complete sensor layout evolution across all experimental
    conditions. Each sub-figure corresponds to one designer:
    (a)~Designer~A (kinesiology), (b)~Designer~B (e-textile engineering),
    (c)~Designer~C (electrical engineering). Within each sub-figure, the
    top row shows the H-Series (human-only iteration from $H_0$) and
    the bottom row shows the LH-Series (human--AI collaboration from
    $L_0$), each depicted in anterior and posterior views. Sensors are
    color-coded: Sensor~1 (red), Sensor~2 (orange), Sensor~3 (yellow),
    Sensor~4 (teal). MPJAE values are annotated for each layout.}
    \label{fig:all_design}
\end{figure*}

Figure~\ref{fig:all_design} presents the complete set of sensor layouts
evaluated across both experiments, providing a visual reference for the
iteration trajectories summarized in Table~\ref{tab:trajectories}.

Several spatial trends are visible. In the H-Series, Designers~A and~B
progressively shifted sensors toward the posterior shoulder and scapular
regions across iterations, converging on similar coverage despite
different starting configurations. Designer~C's H-Series regression in
iteration~2 ($H_1^C$: 10.08°~$\rightarrow$~$H_2^C$: 13.03°)
corresponds to a visible repositioning that reduced posterior coverage.
In the LH-Series, Designer~C's monotonic improvement is accompanied by
gradual anterior-to-posterior migration and alignment of sensors along
muscle fiber directions, while Designer~A's single-round modification
resulted in a substantially different spatial distribution that
increased error. Designer~B's LH-Series shows moderate spatial
adjustments in iteration~1 followed by a redistribution in iteration~2
that did not sustain the initial improvement.

\section{Performance Feedback Dashboard}
\label{app:dashboard}
\begin{figure}[htbp]
    \centering
    \includegraphics[width=1\linewidth]{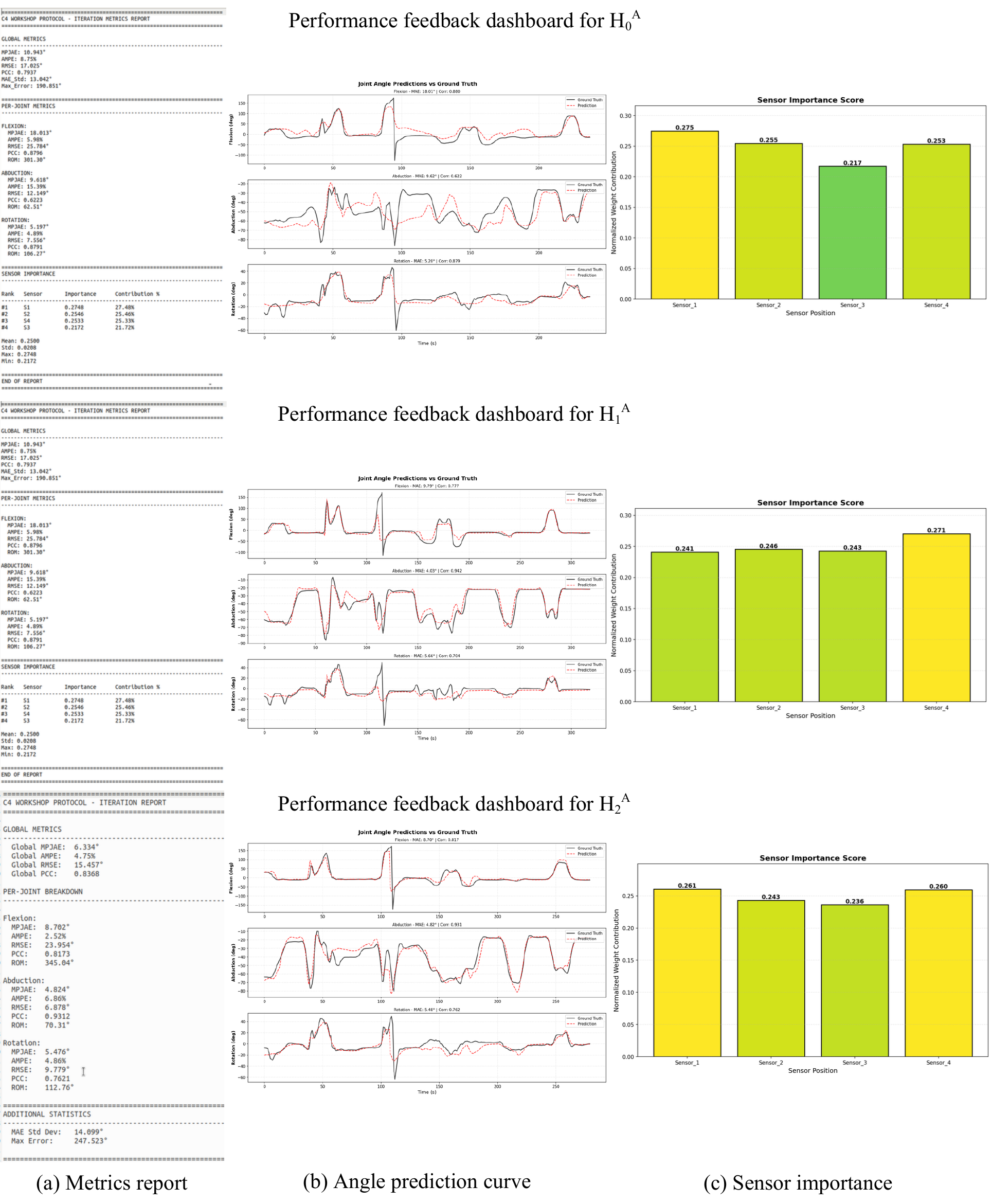}
    \caption{Performance feedback dashboard for Designer~A, H-Series.}
    \label{fig:ha}
\end{figure}

\begin{figure}[htbp]
    \centering
    \includegraphics[width=1\linewidth]{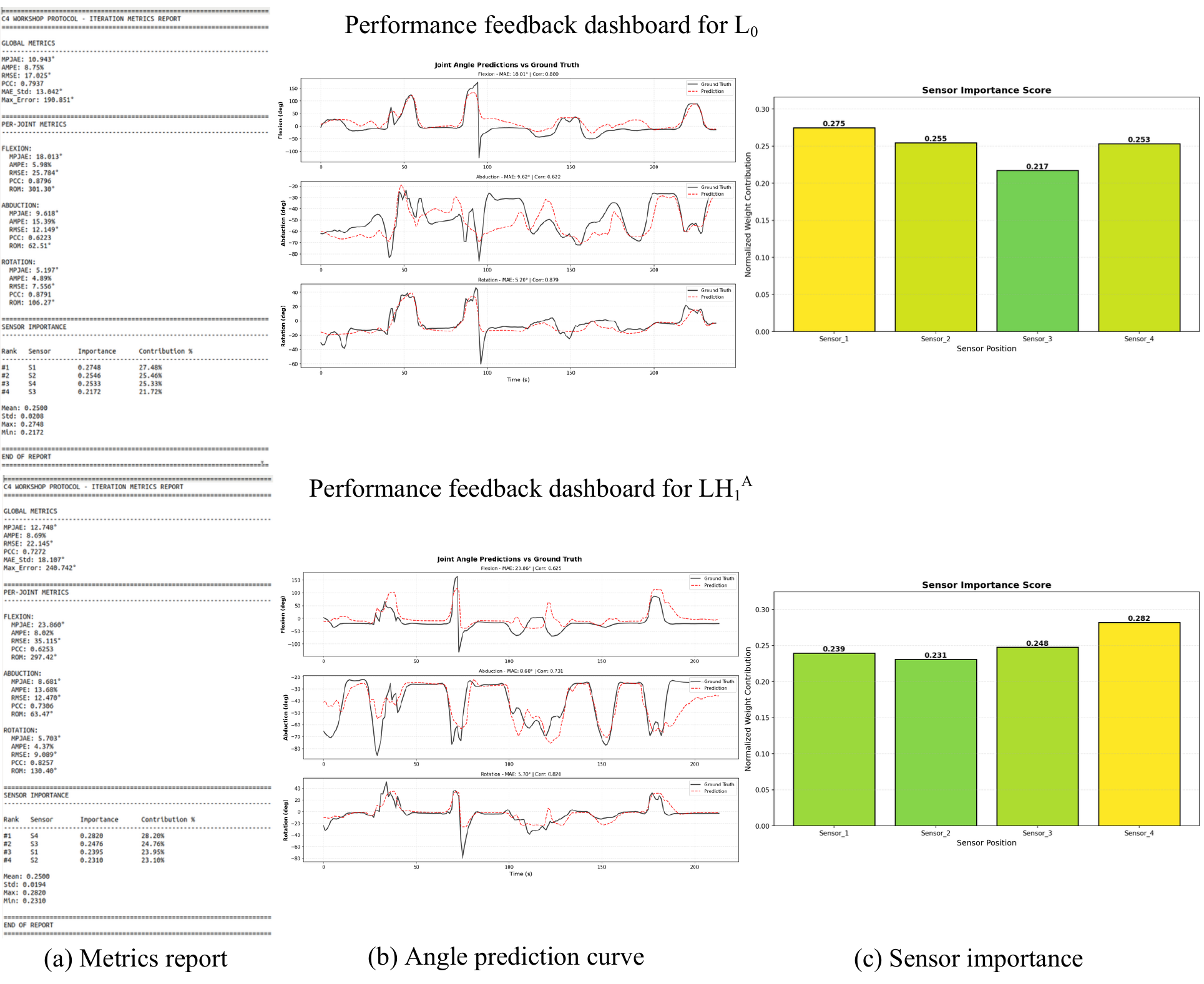}
    \caption{Performance feedback dashboard for Designer~A, LH-Series.}
    \label{fig:lha}
\end{figure}

\begin{figure}[htbp]
    \centering
    \includegraphics[width=1\linewidth]{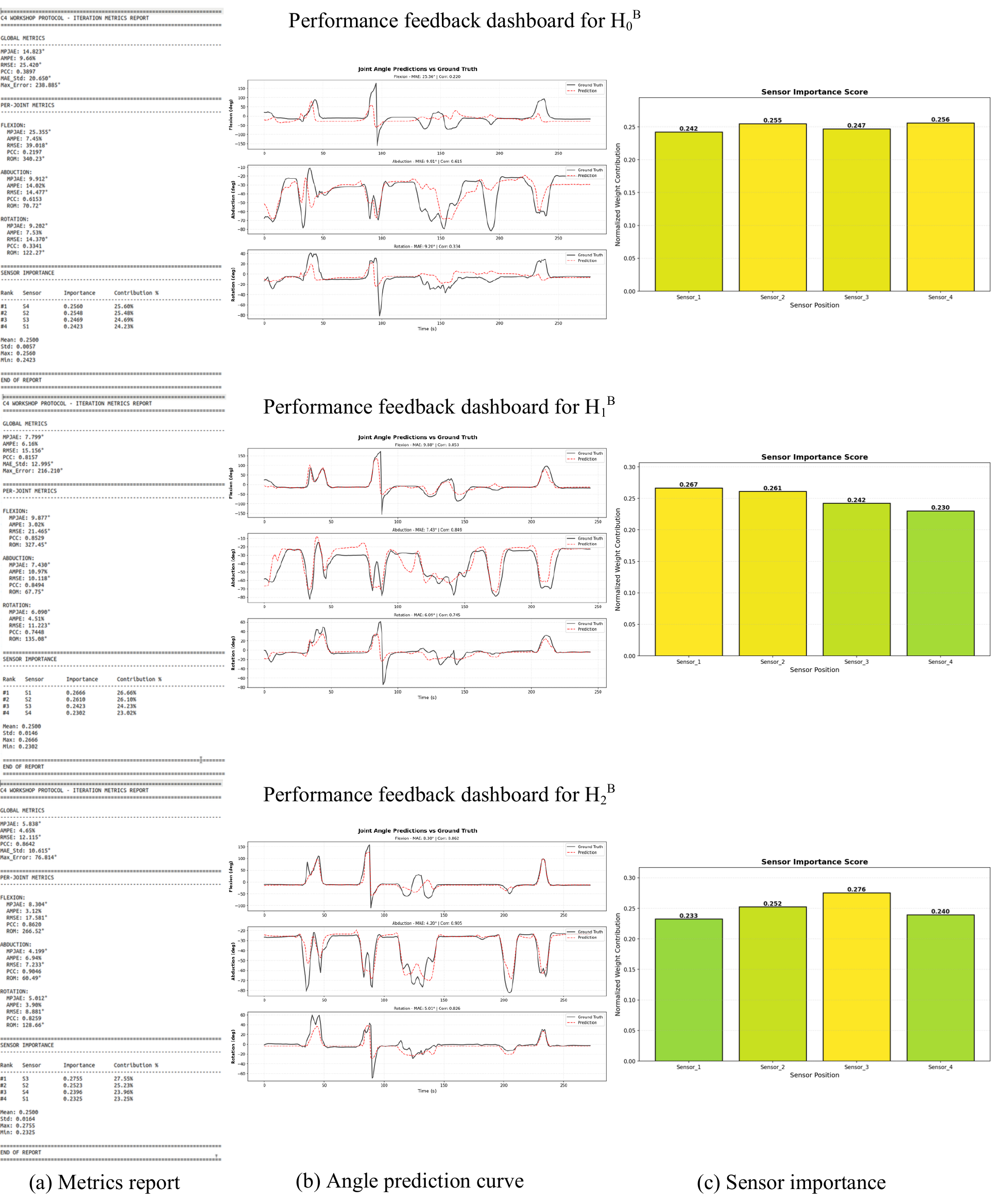}
    \caption{Performance feedback dashboard for Expert~B, H-Series.}
    \label{fig:hb}
\end{figure}

\begin{figure}[htbp]
    \centering
    \includegraphics[width=1\linewidth]{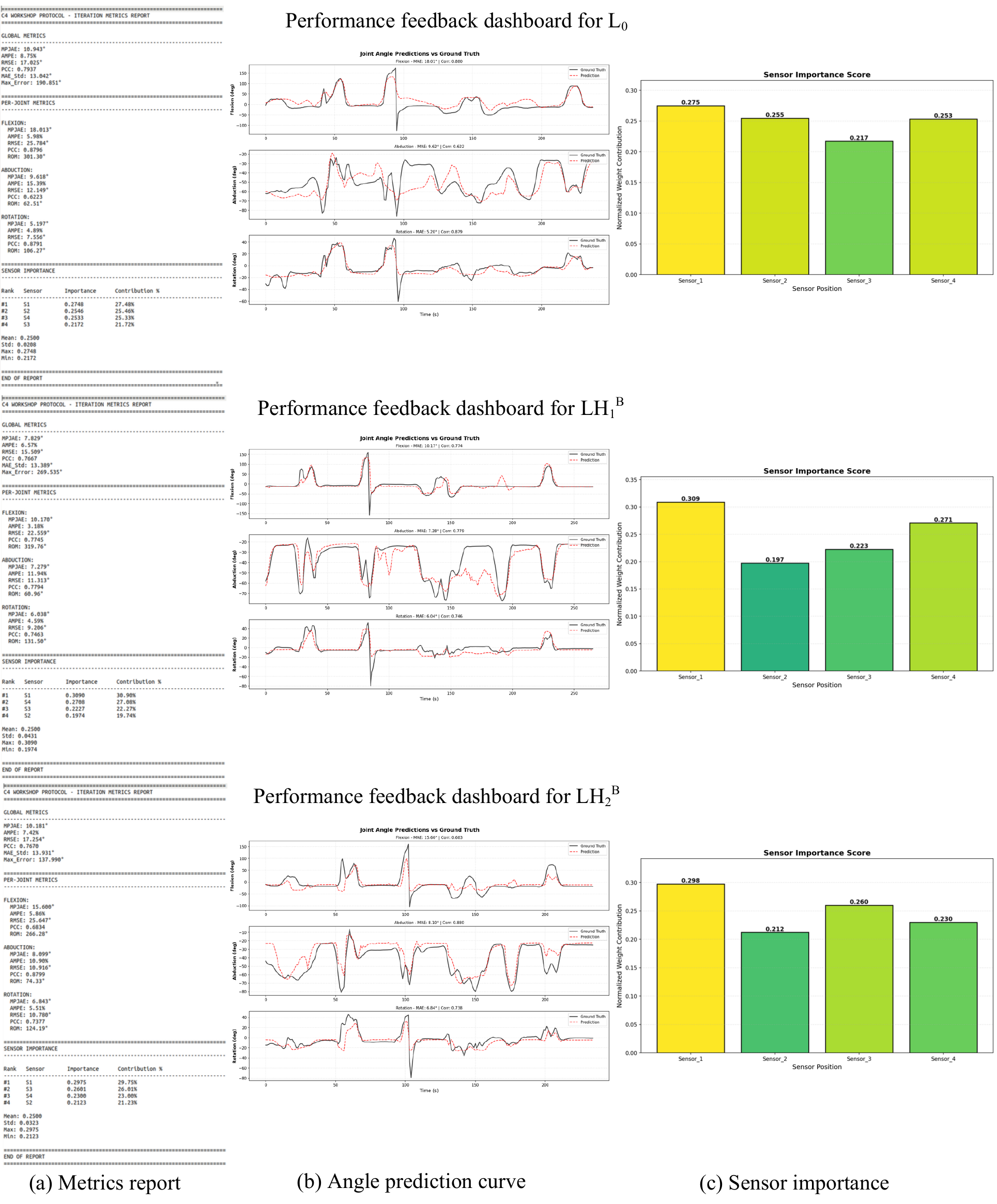}
    \caption{Performance feedback dashboard for Designer~B, LH-Series.}
    \label{fig:lhb}
\end{figure}

\begin{figure}[htbp]
    \centering
    \includegraphics[width=1\linewidth]{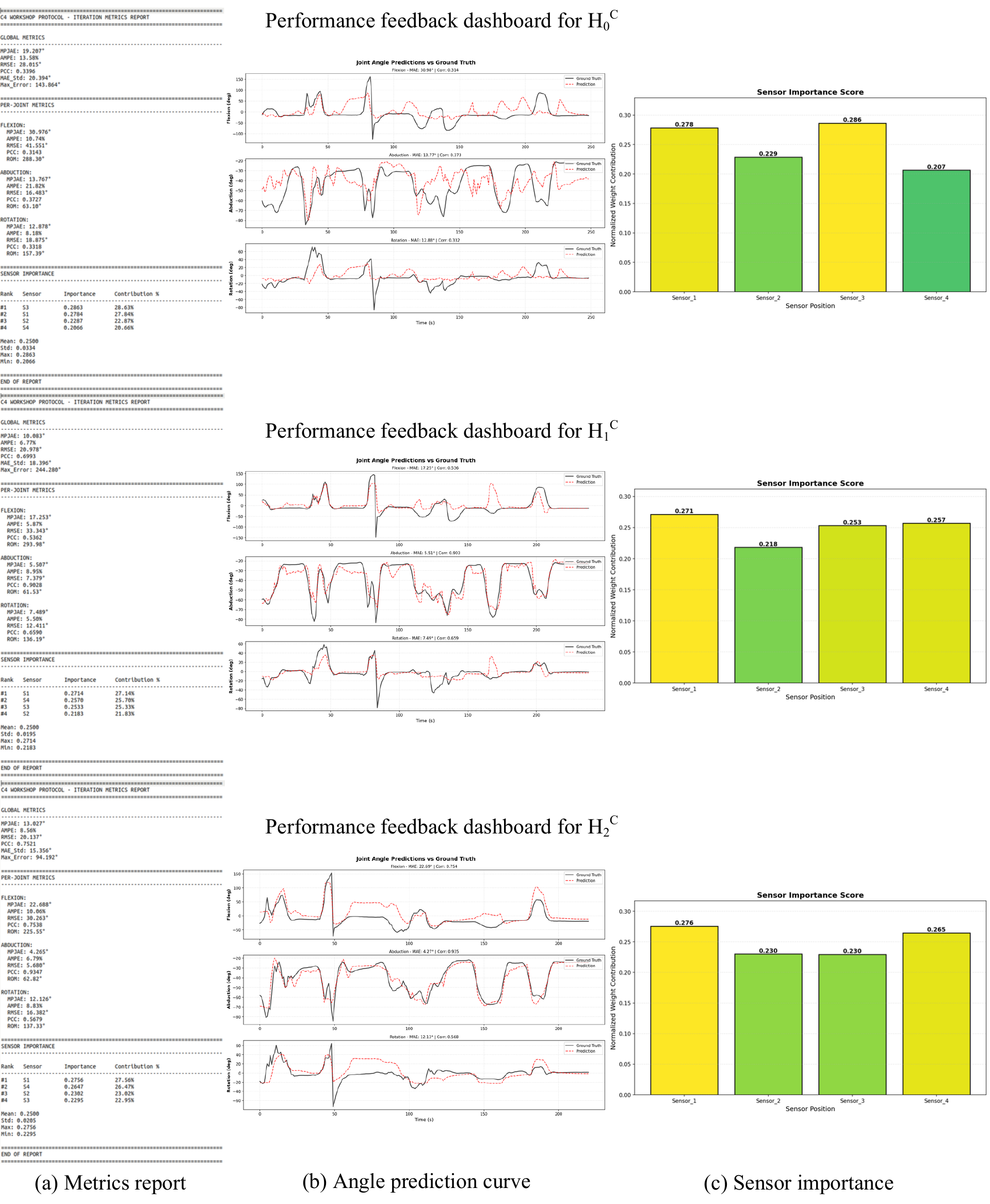}
    \caption{Performance feedback dashboard for Designer~C, H-Series.}
    \label{fig:hc}
\end{figure}

\begin{figure}[htbp]
    \centering
    \includegraphics[width=1\linewidth]{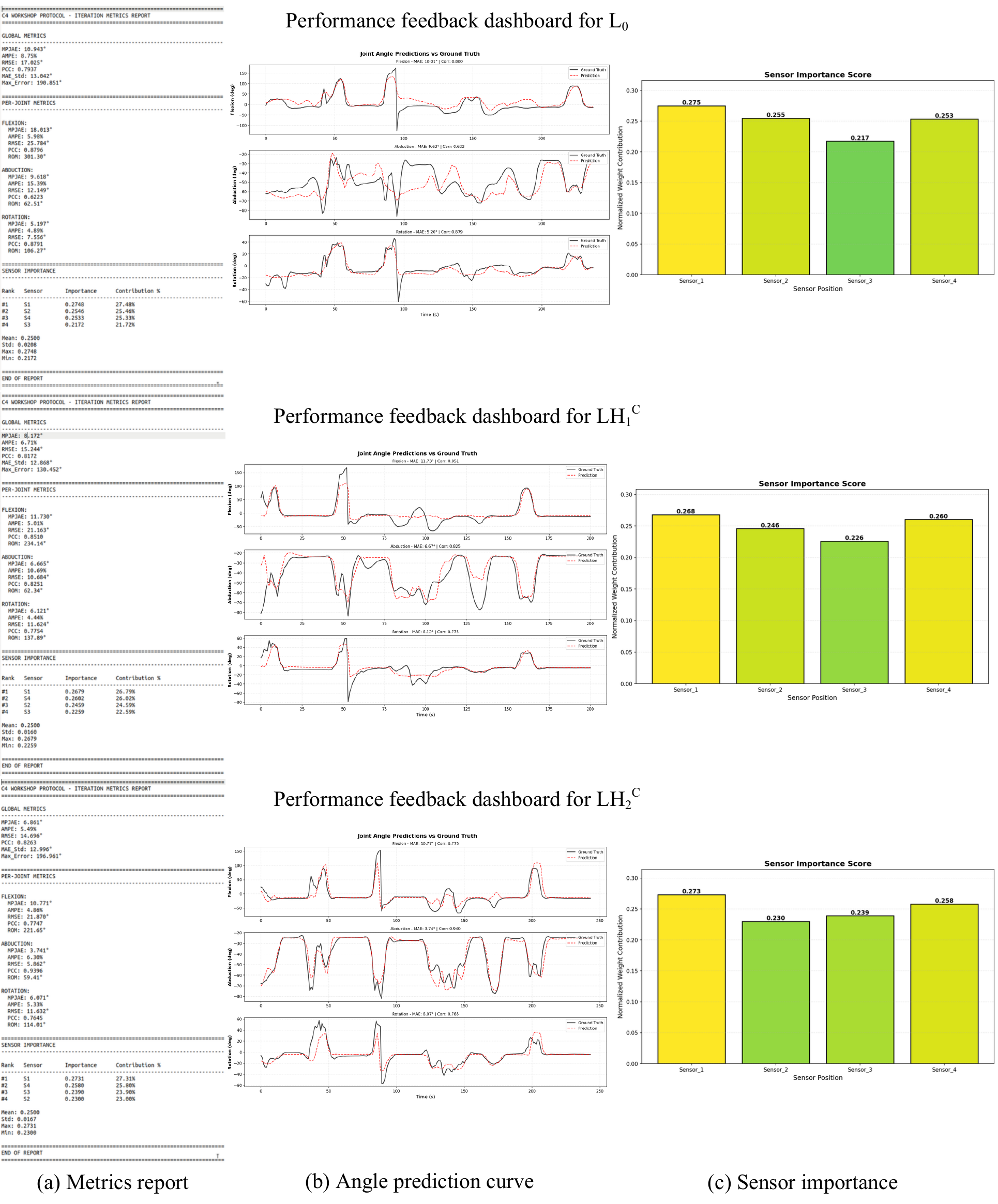}
    \caption{Performance feedback dashboard for Designer~C, LH-Series.}
    \label{fig:lhc}
\end{figure}

Figures~\ref{fig:ha}--\ref{fig:lhc} present the performance feedback dashboards provided to each designer after every iteration cycle, showing global metrics (MPJAE, AMPE, PCC, RMSE), per-joint error decomposition, angle prediction curves, and sensor contribution rankings.

\clearpage

\section{LH-Series Human Designer Structured Feedbacks}
\label{sec:appendix-lh-human}


\begin{figure}[H]
\centering
\includegraphics[width=\textwidth]{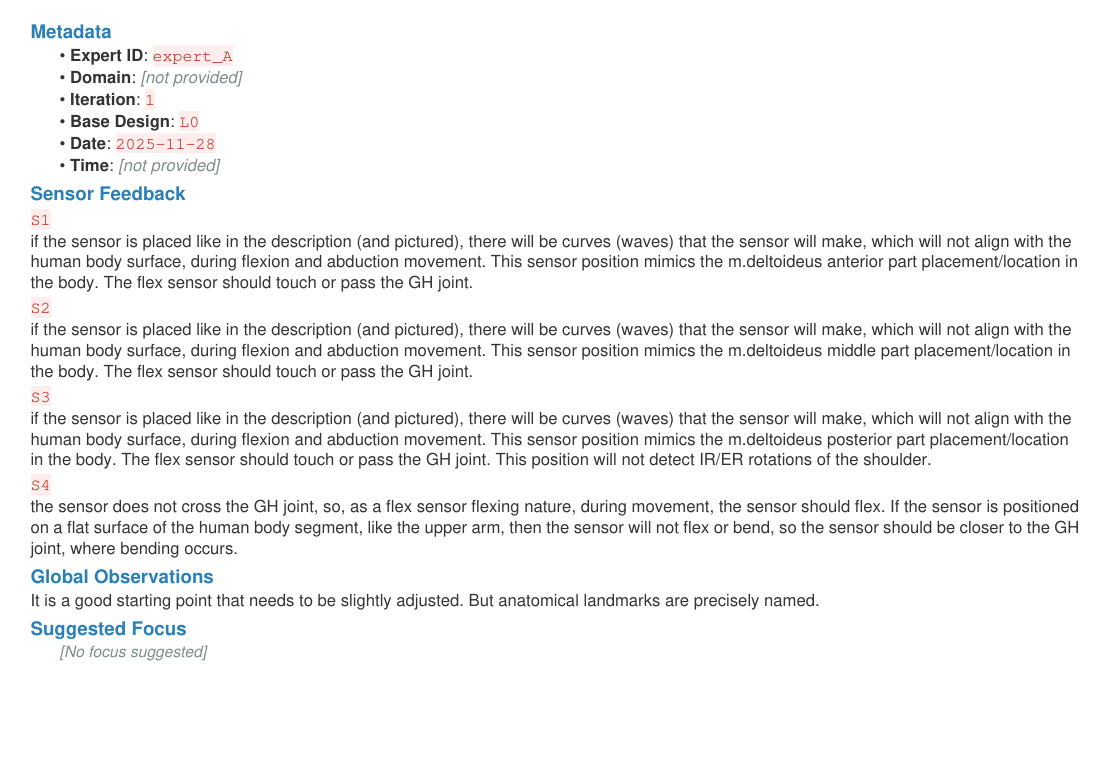}
\caption{Designer A -- Round 1 structured feedback.}
\label{fig:lh-human-a-r1}
\end{figure}


\begin{figure}[H]
\centering
\includegraphics[width=\textwidth]{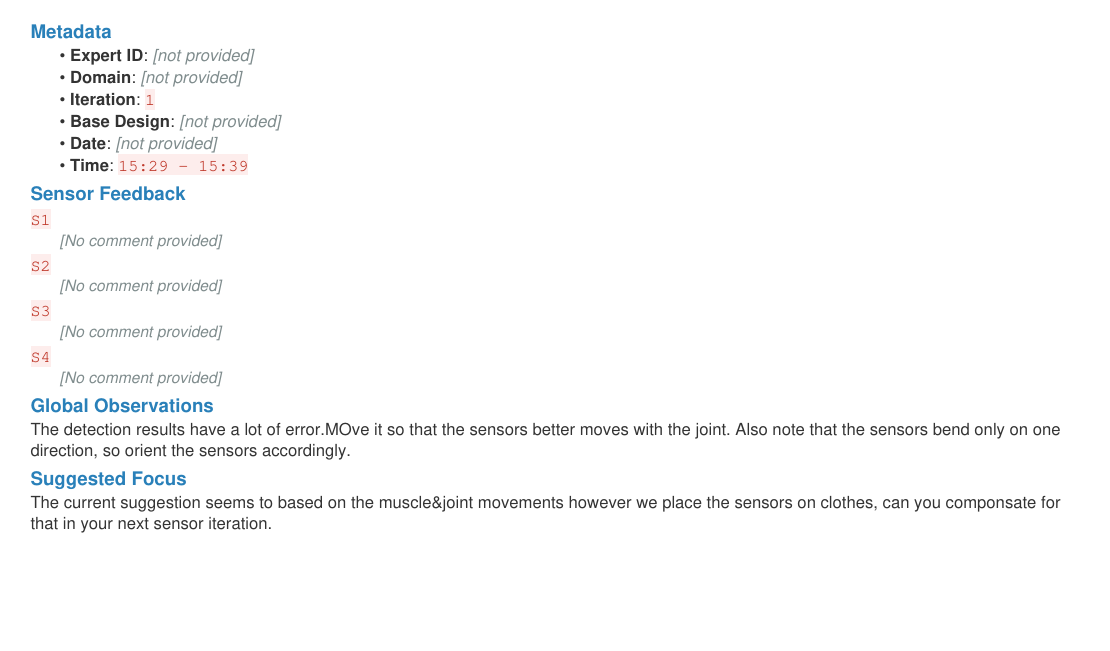}
\caption{Designer B -- Round 1 structured feedback.}
\label{fig:lh-human-b-r1}
\end{figure}

\begin{figure}[H]
\centering
\includegraphics[width=\textwidth]{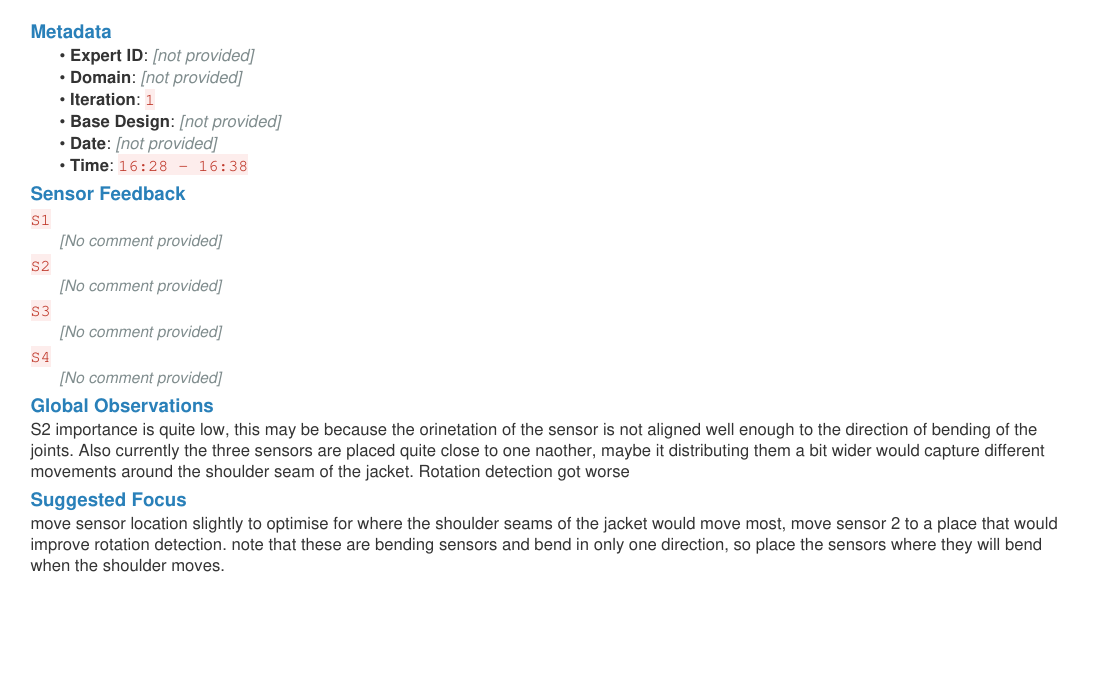}
\caption{Designer B -- Round 2 structured feedback.}
\label{fig:lh-human-b-r2}
\end{figure}


\begin{figure}[H]
\centering
\includegraphics[width=\textwidth]{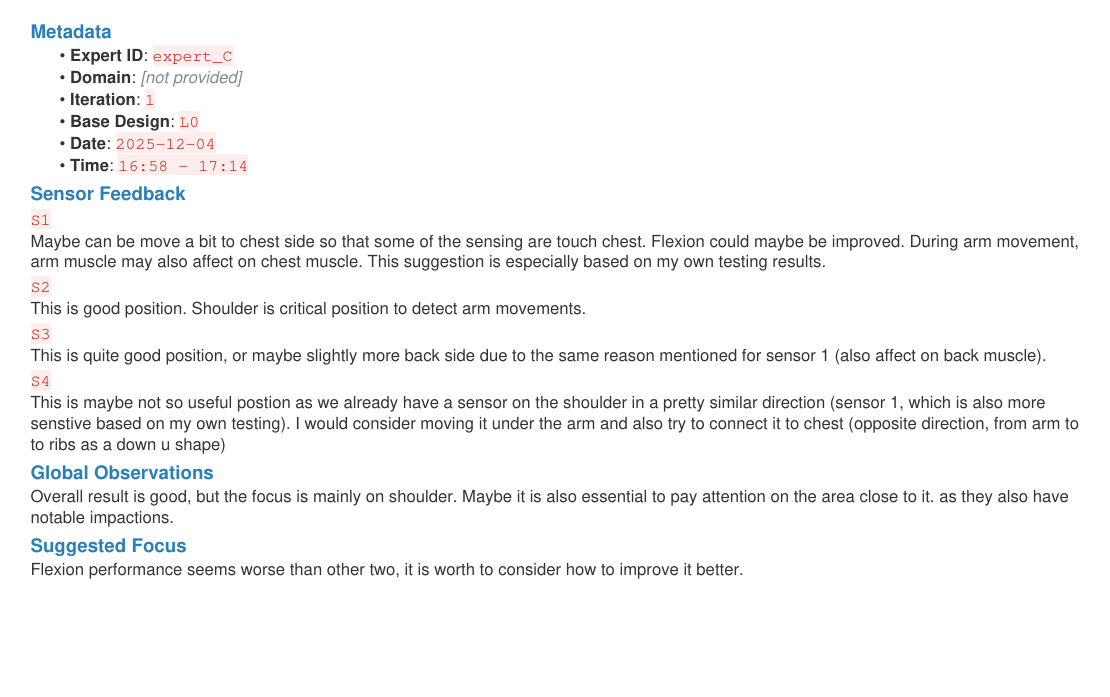}
\caption{Designer C -- Round 1 structured feedback.}
\label{fig:lh-human-c-r1}
\end{figure}

\begin{figure}[H]
\centering
\includegraphics[width=\textwidth]{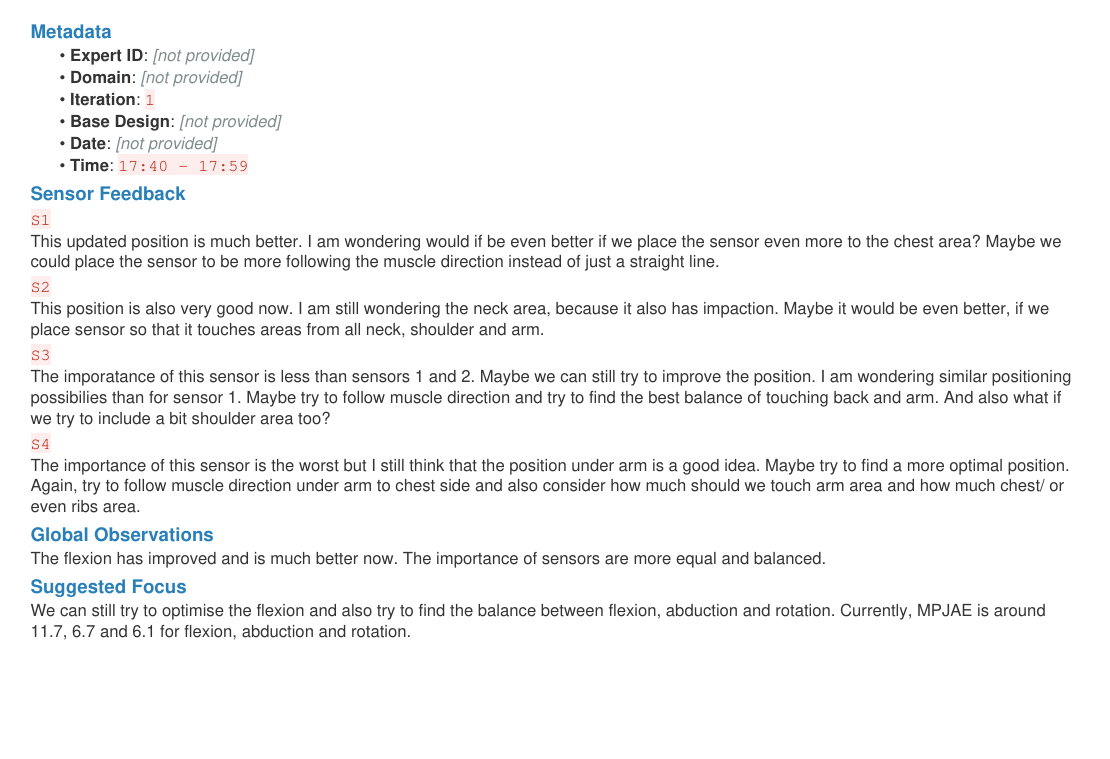}
\caption{Designer C -- Round 2 structured feedback.}
\label{fig:lh-human-c-r2}
\end{figure}

\section{LH-Series LLM Outputs}
\label{sec:appendix-lh-llm}


\begin{figure}[H]
\centering
\includegraphics[width=\textwidth]{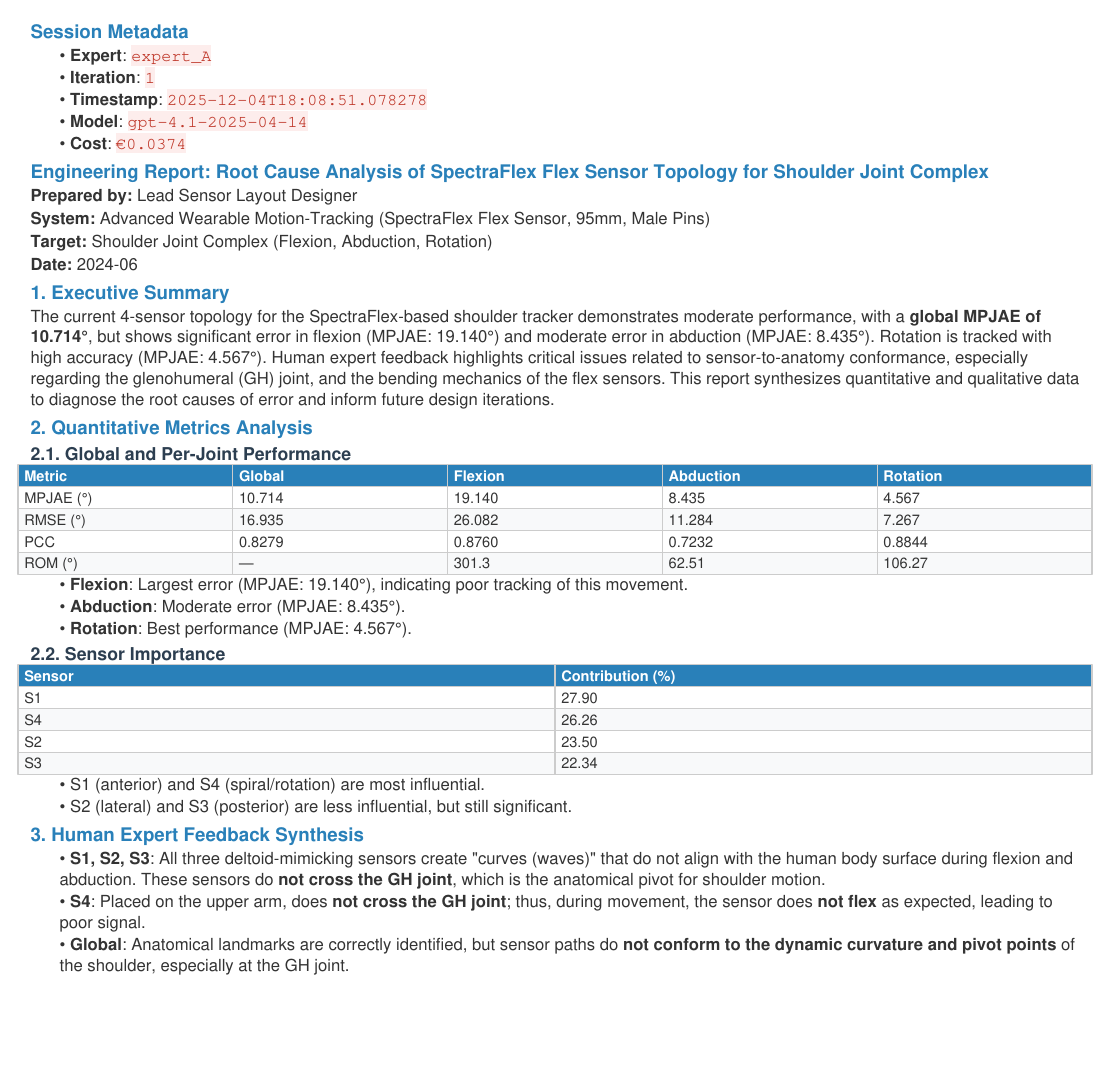}
\caption{LLM output for Designer A, Iteration 1 (Part 1).}
\label{fig:lh-llm-a-i1-p1}
\end{figure}

\begin{figure}[H]
\centering
\includegraphics[width=\textwidth]{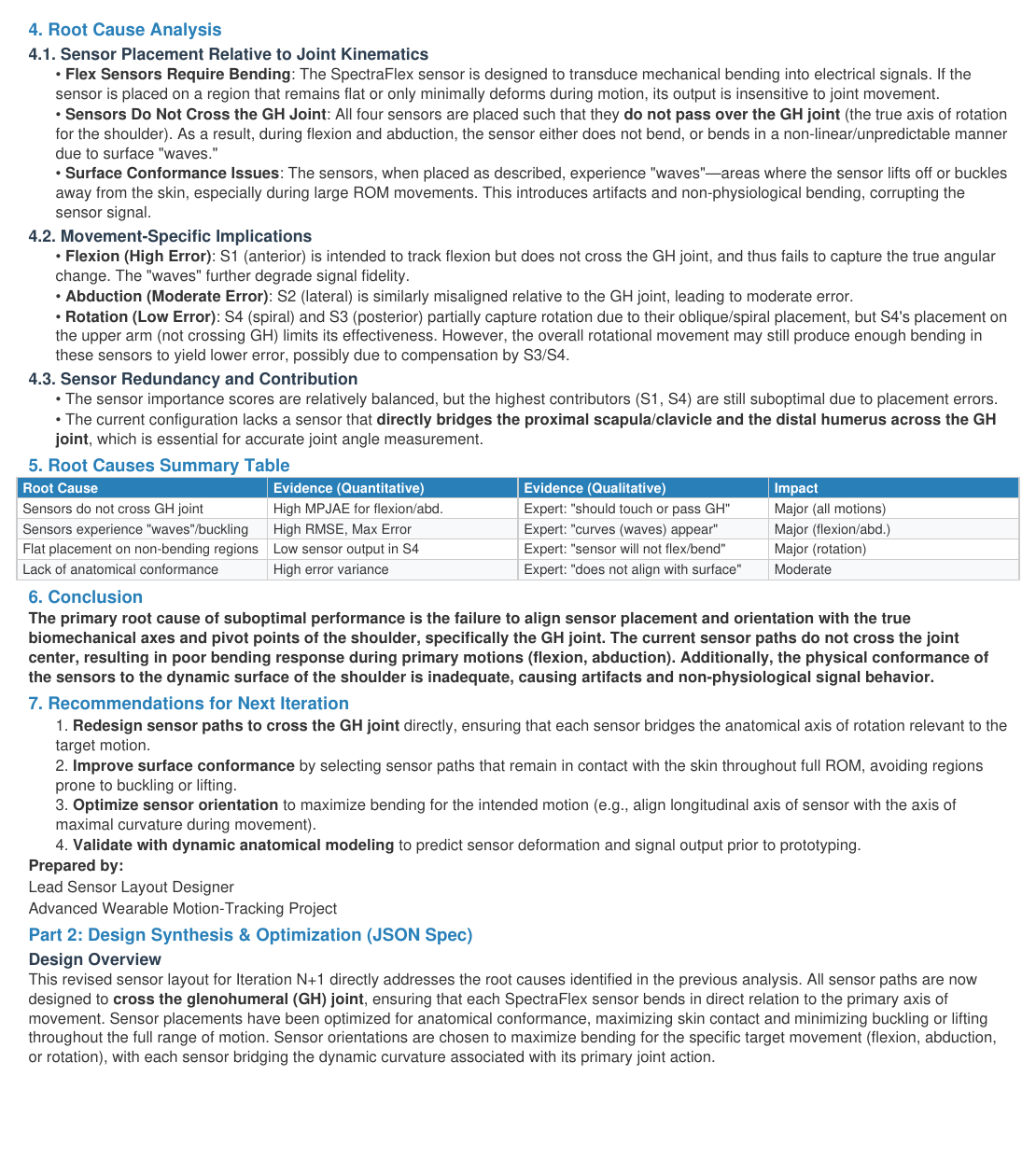}
\caption{LLM output for Designer A, Iteration 1 (Part 2).}
\label{fig:lh-llm-a-i1-p2}
\end{figure}

\begin{figure}[H]
\centering
\includegraphics[width=\textwidth]{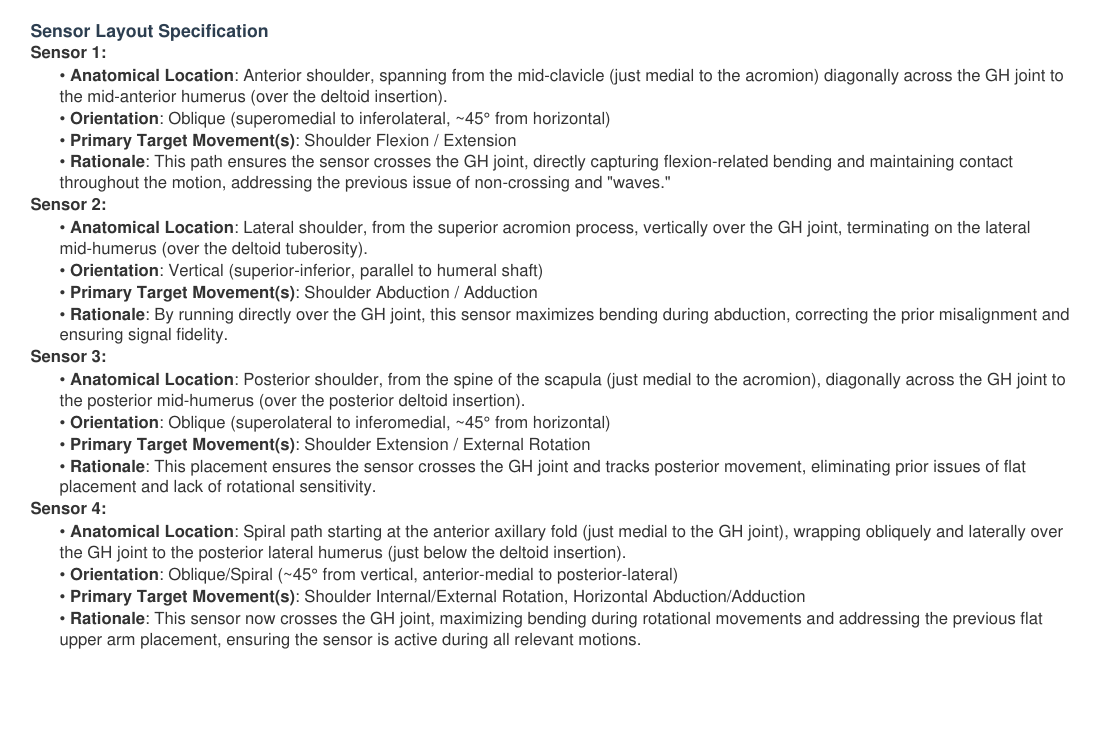}
\caption{LLM output for Designer A, Iteration 1 (Part 3).}
\label{fig:lh-llm-a-i1-p3}
\end{figure}

\begin{figure}[H]
\centering
\includegraphics[width=\textwidth]{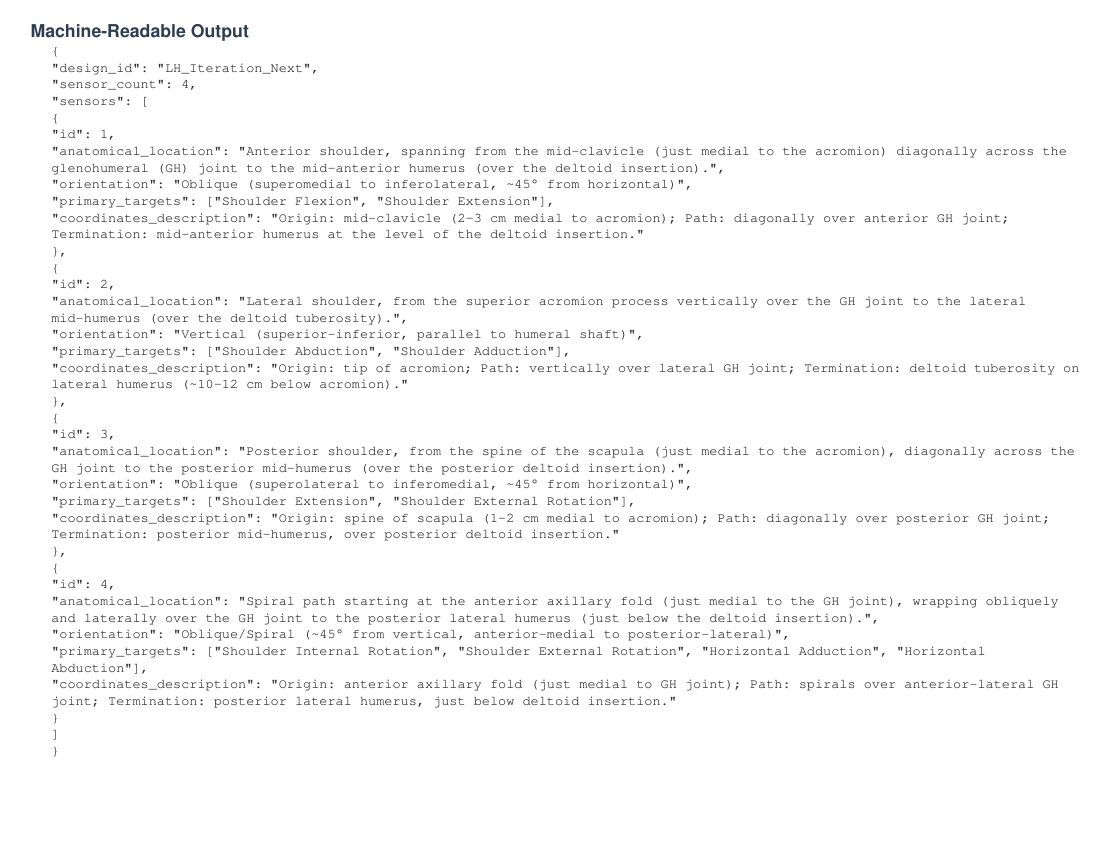}
\caption{LLM output for Designer A, Iteration 1 (Part 4).}
\label{fig:lh-llm-a-i1-p4}
\end{figure}


\begin{figure}[H]
\centering
\includegraphics[width=\textwidth]{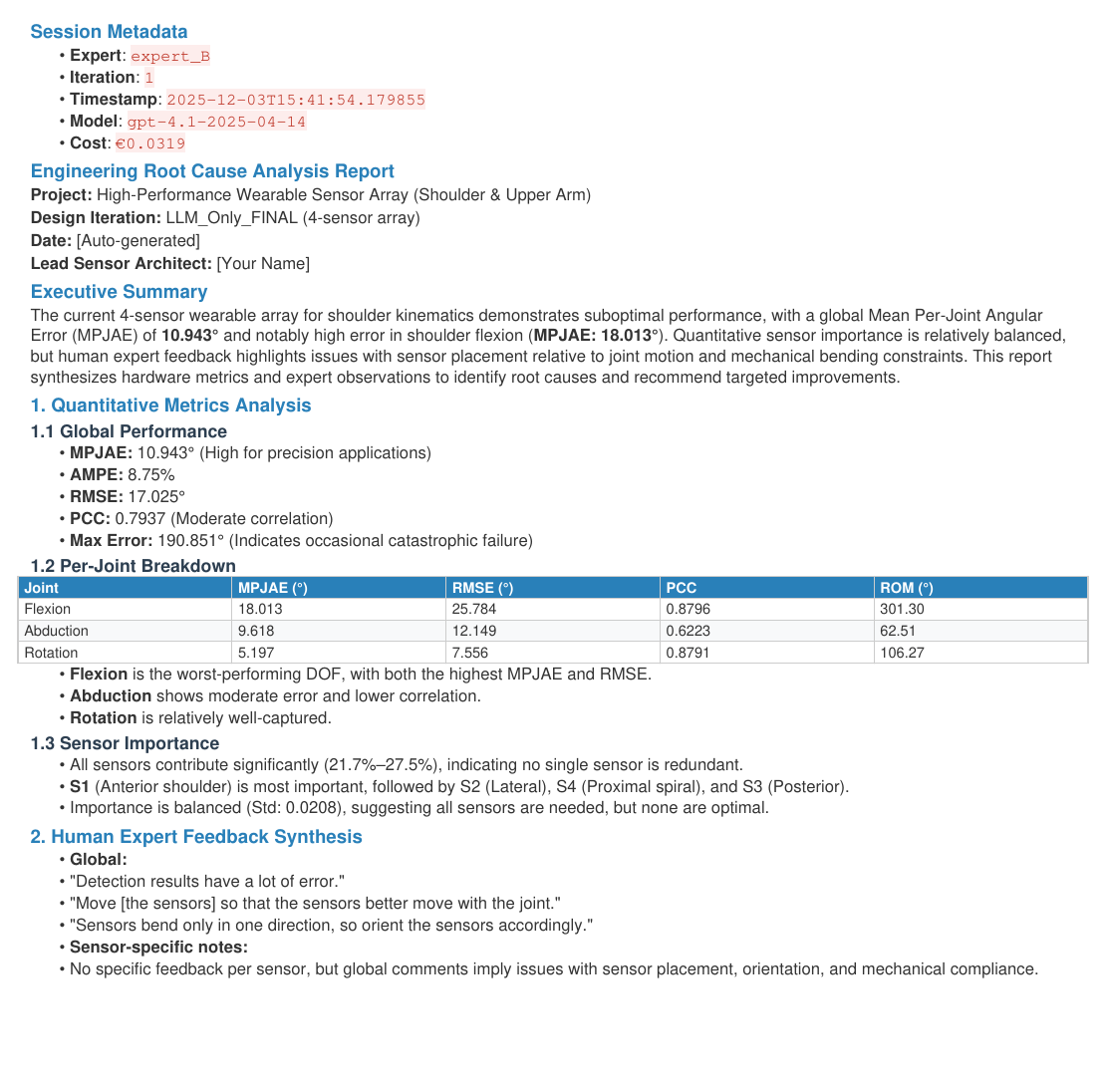}
\caption{LLM output for Expert B, Iteration 1 (Part 1).}
\label{fig:lh-llm-b-i1-p1}
\end{figure}

\begin{figure}[H]
\centering
\includegraphics[width=\textwidth]{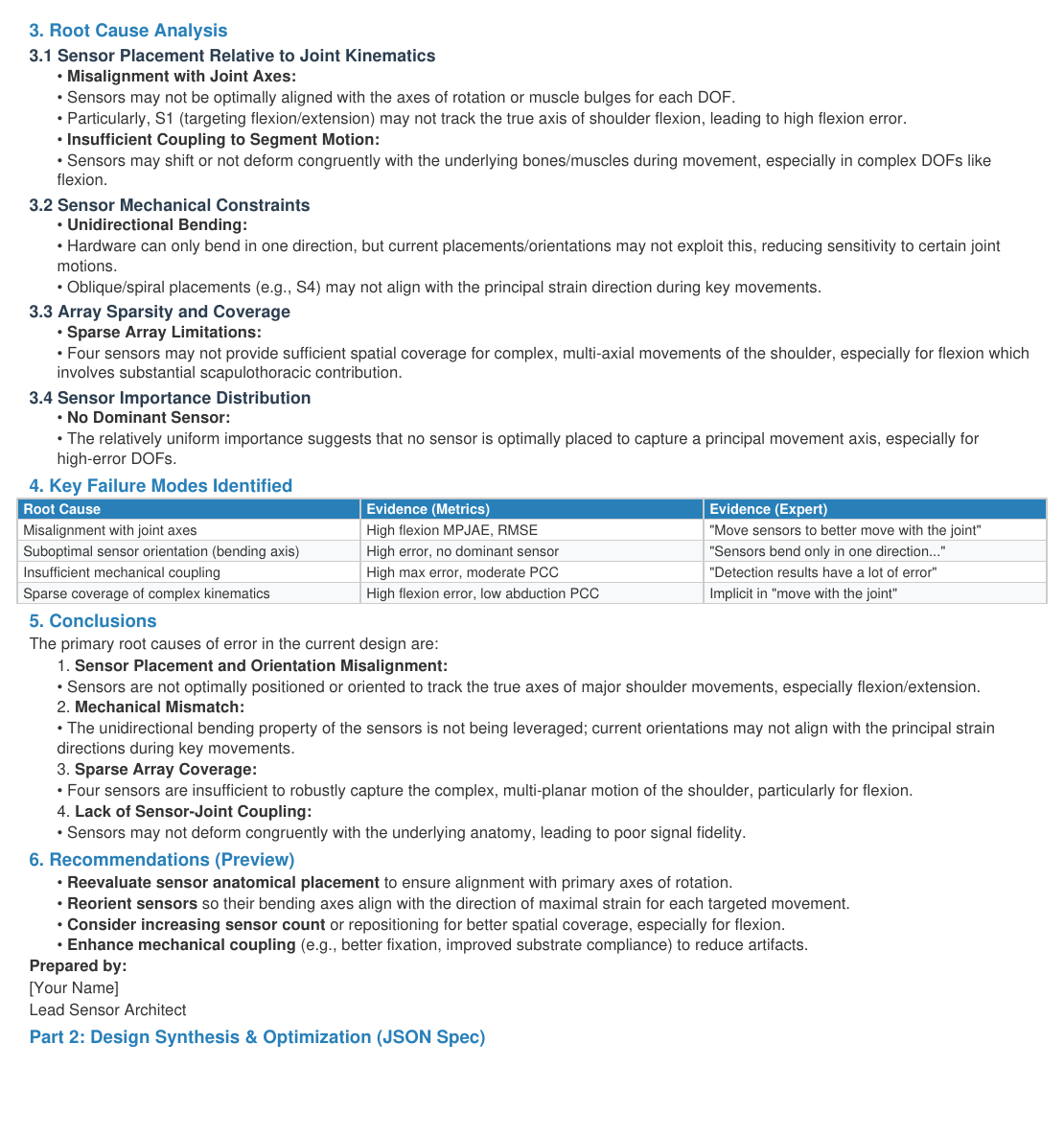}
\caption{LLM output for Designer B, Iteration 1 (Part 2).}
\label{fig:lh-llm-b-i1-p2}
\end{figure}

\begin{figure}[H]
\centering
\includegraphics[width=\textwidth]{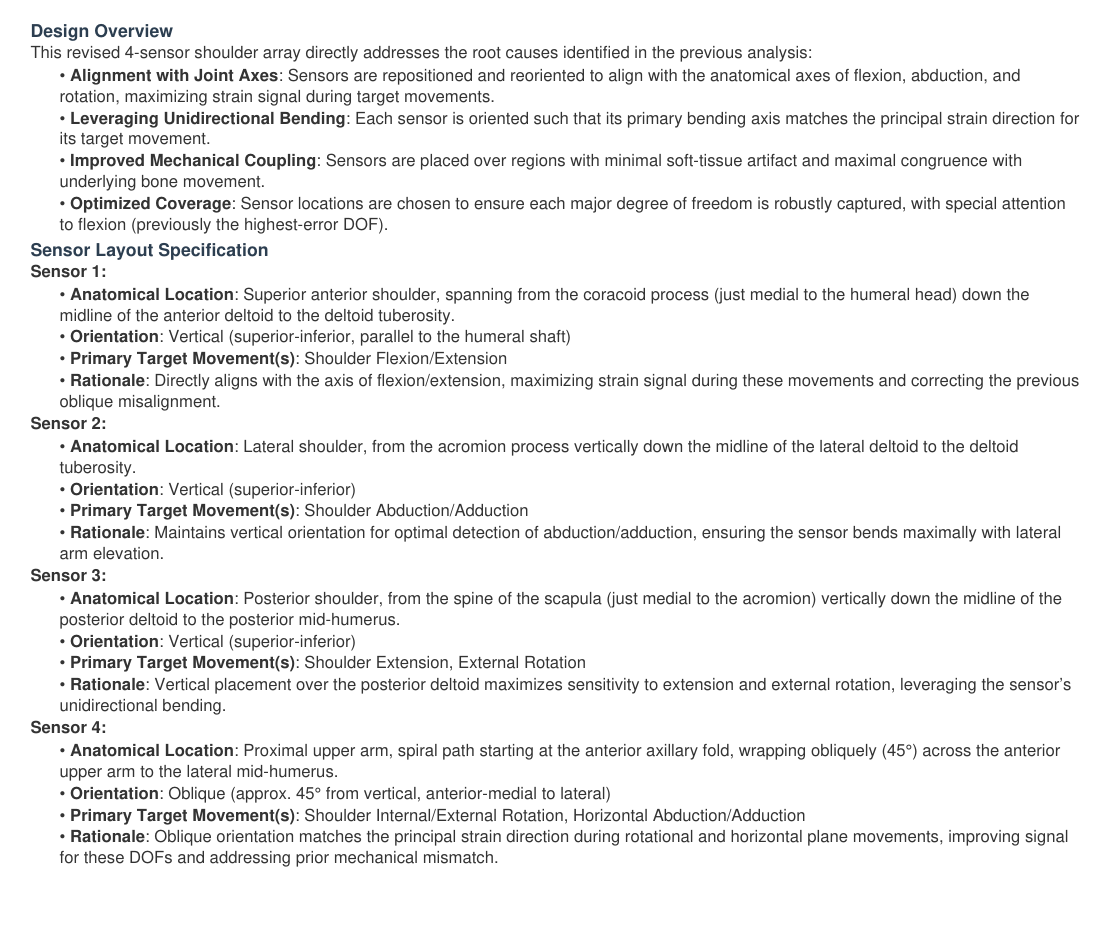}
\caption{LLM output for Designer B, Iteration 1 (Part 3).}
\label{fig:lh-llm-b-i1-p3}
\end{figure}

\begin{figure}[H]
\centering
\includegraphics[width=\textwidth]{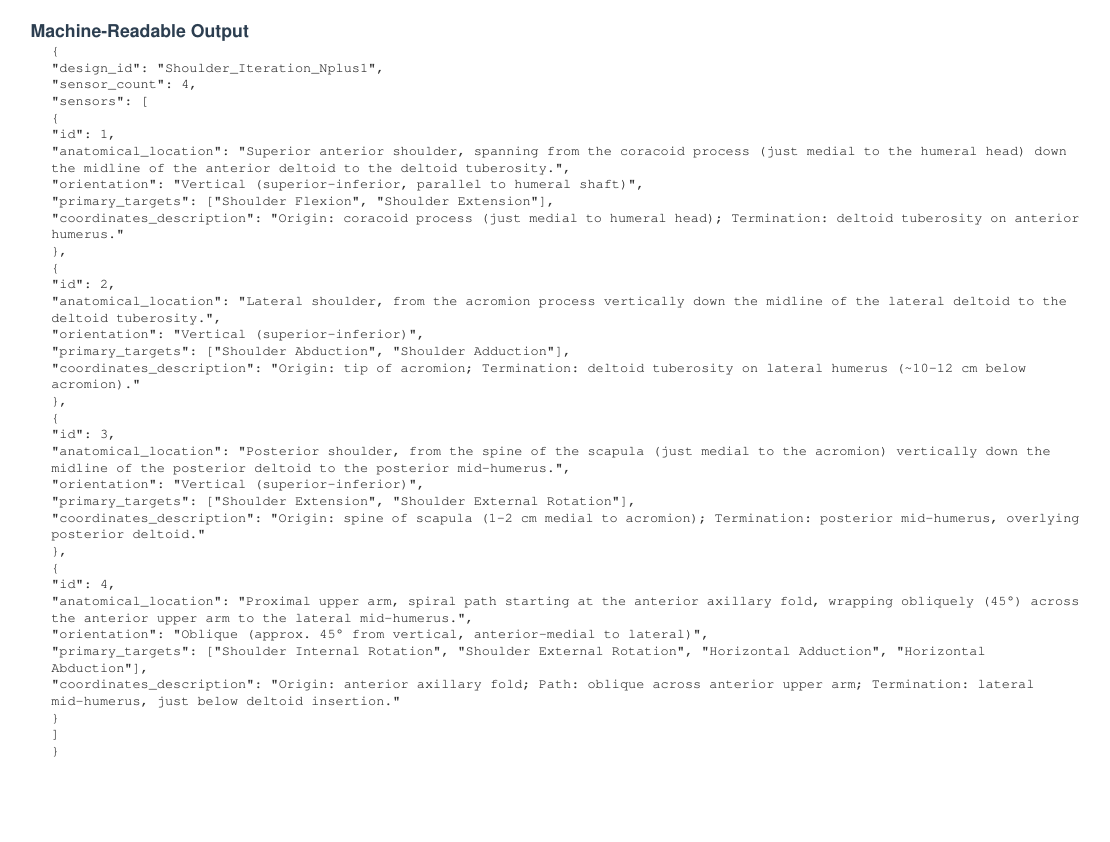}
\caption{LLM output for Designer B, Iteration 1 (Part 4).}
\label{fig:lh-llm-b-i1-p4}
\end{figure}


\begin{figure}[H]
\centering
\includegraphics[width=\textwidth]{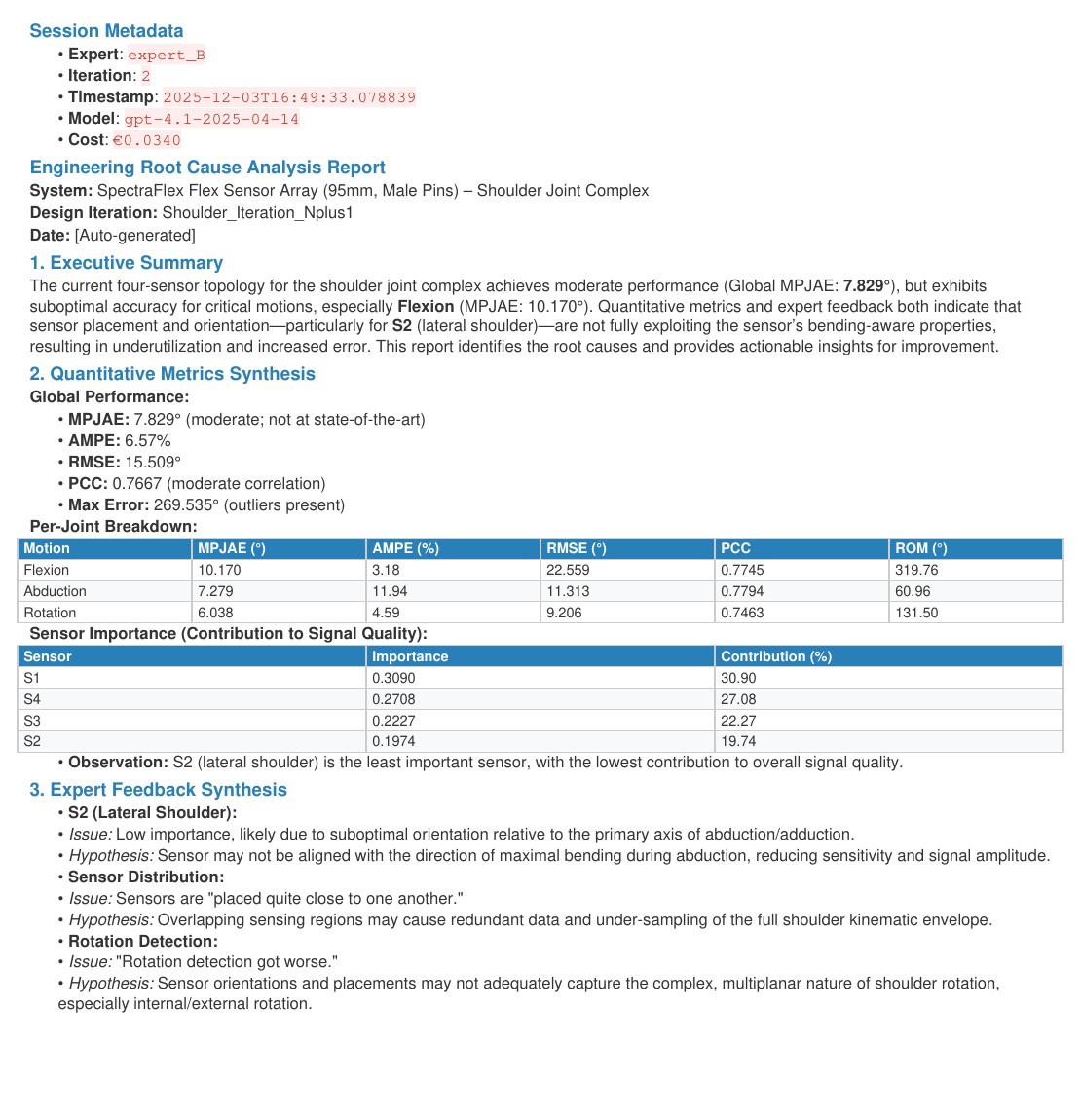}
\caption{LLM output for Designer B, Iteration 2 (Part 1).}
\label{fig:lh-llm-b-i2-p1}
\end{figure}

\begin{figure}[H]
\centering
\includegraphics[width=\textwidth]{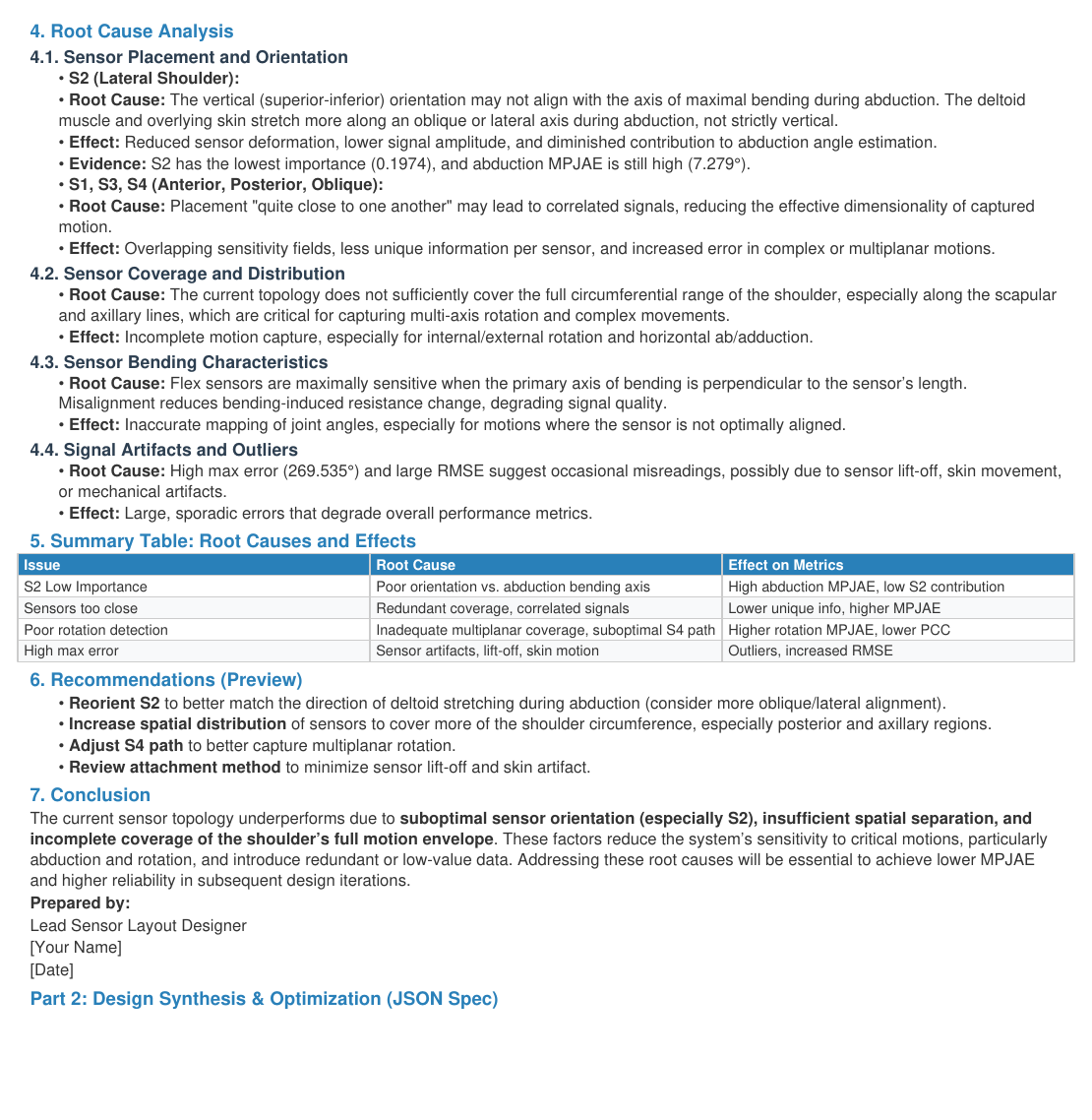}
\caption{LLM output for Designer B, Iteration 2 (Part 2).}
\label{fig:lh-llm-b-i2-p2}
\end{figure}

\begin{figure}[H]
\centering
\includegraphics[width=\textwidth]{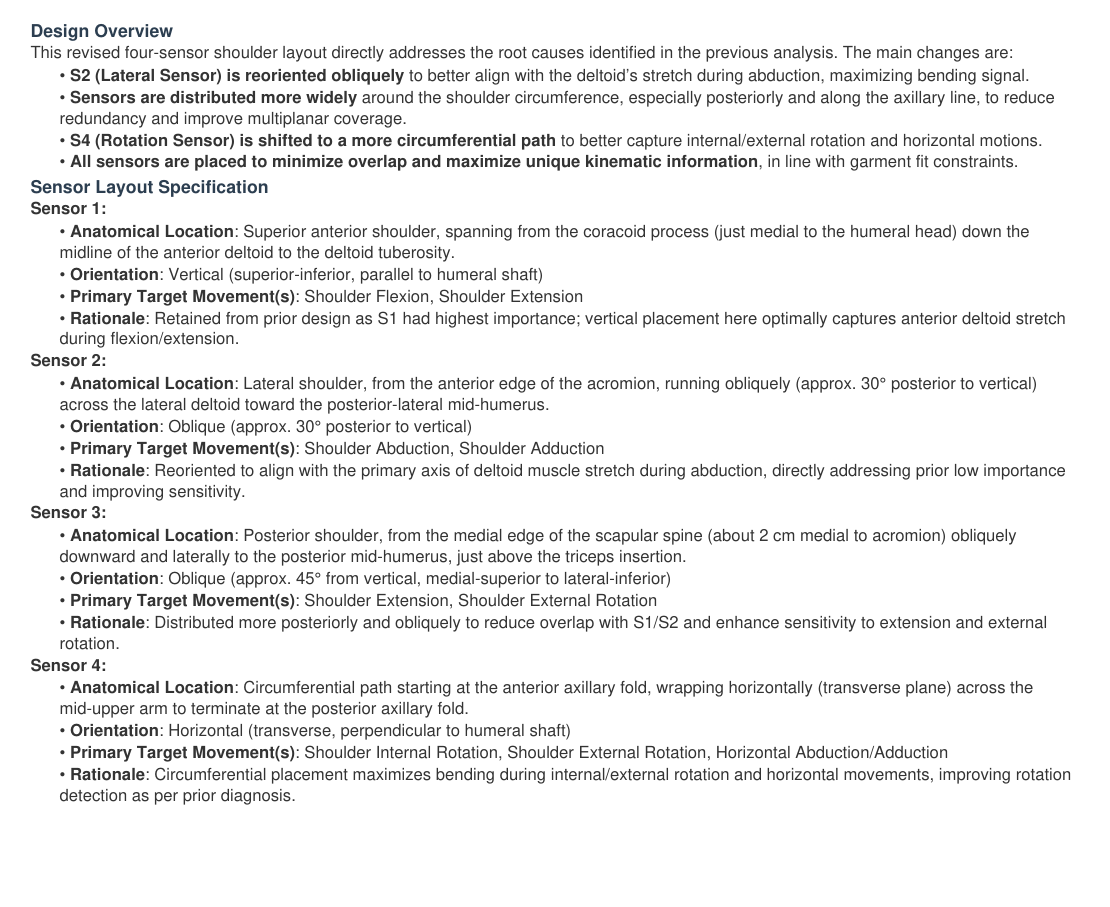}
\caption{LLM output for Designer B, Iteration 2 (Part 3).}
\label{fig:lh-llm-b-i2-p3}
\end{figure}

\begin{figure}[H]
\centering
\includegraphics[width=\textwidth]{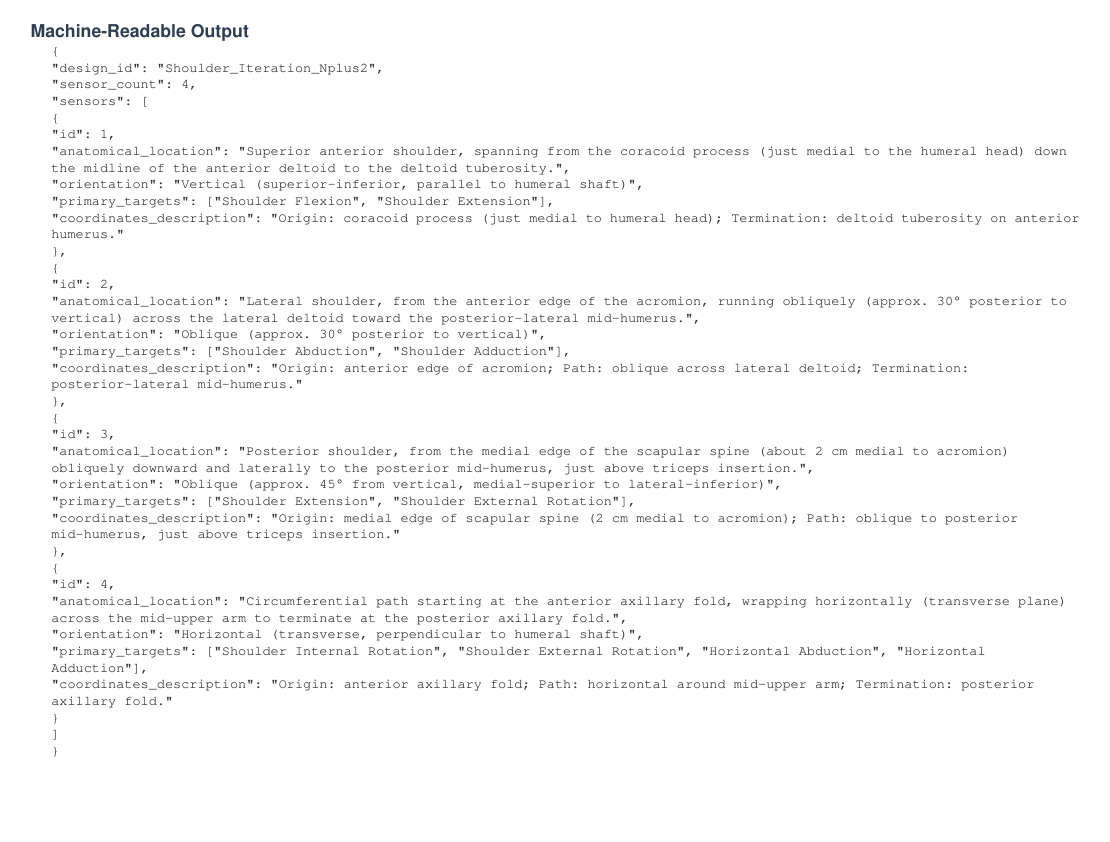}
\caption{LLM output for Designer B, Iteration 2 (Part 4).}
\label{fig:lh-llm-b-i2-p4}
\end{figure}


\begin{figure}[H]
\centering
\includegraphics[width=\textwidth]{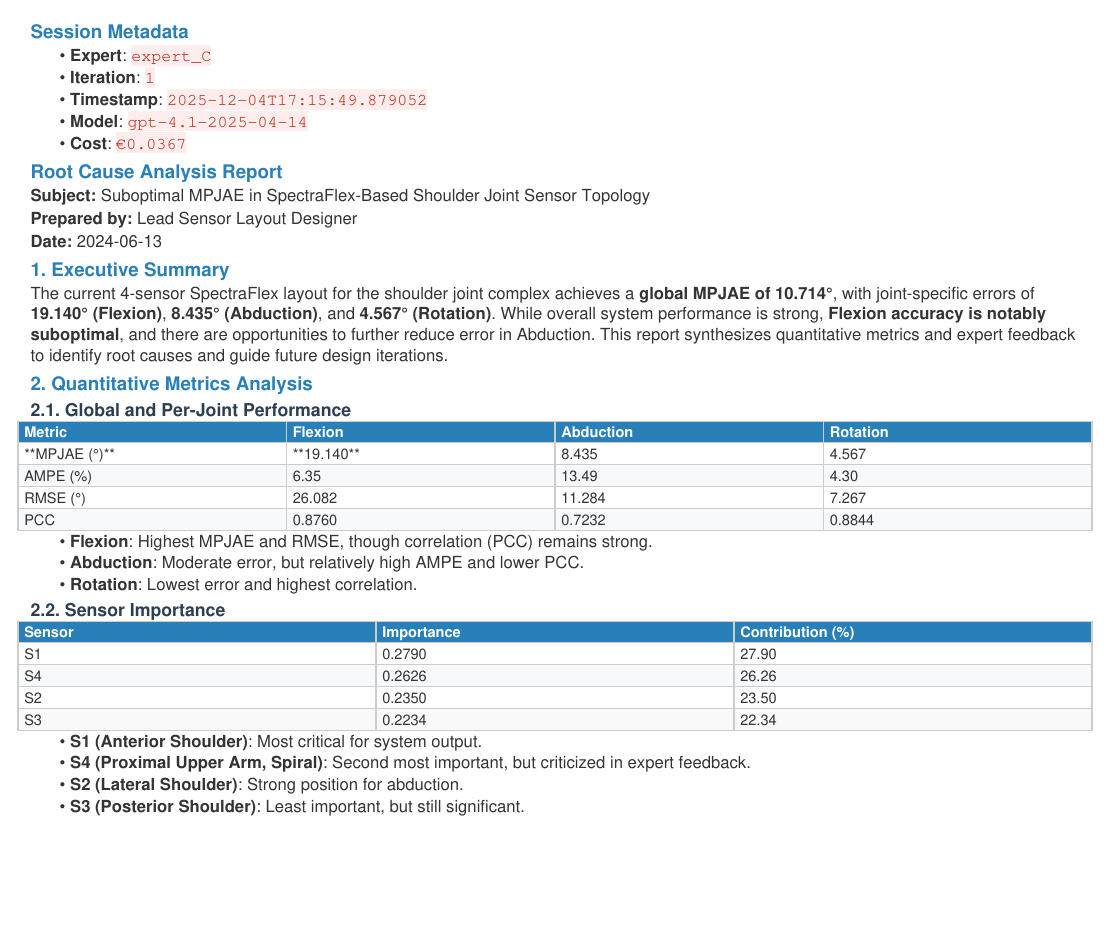}
\caption{LLM output for Designer C, Iteration 1 (Part 1).}
\label{fig:lh-llm-c-i1-p1}
\end{figure}

\begin{figure}[H]
\centering
\includegraphics[width=\textwidth]{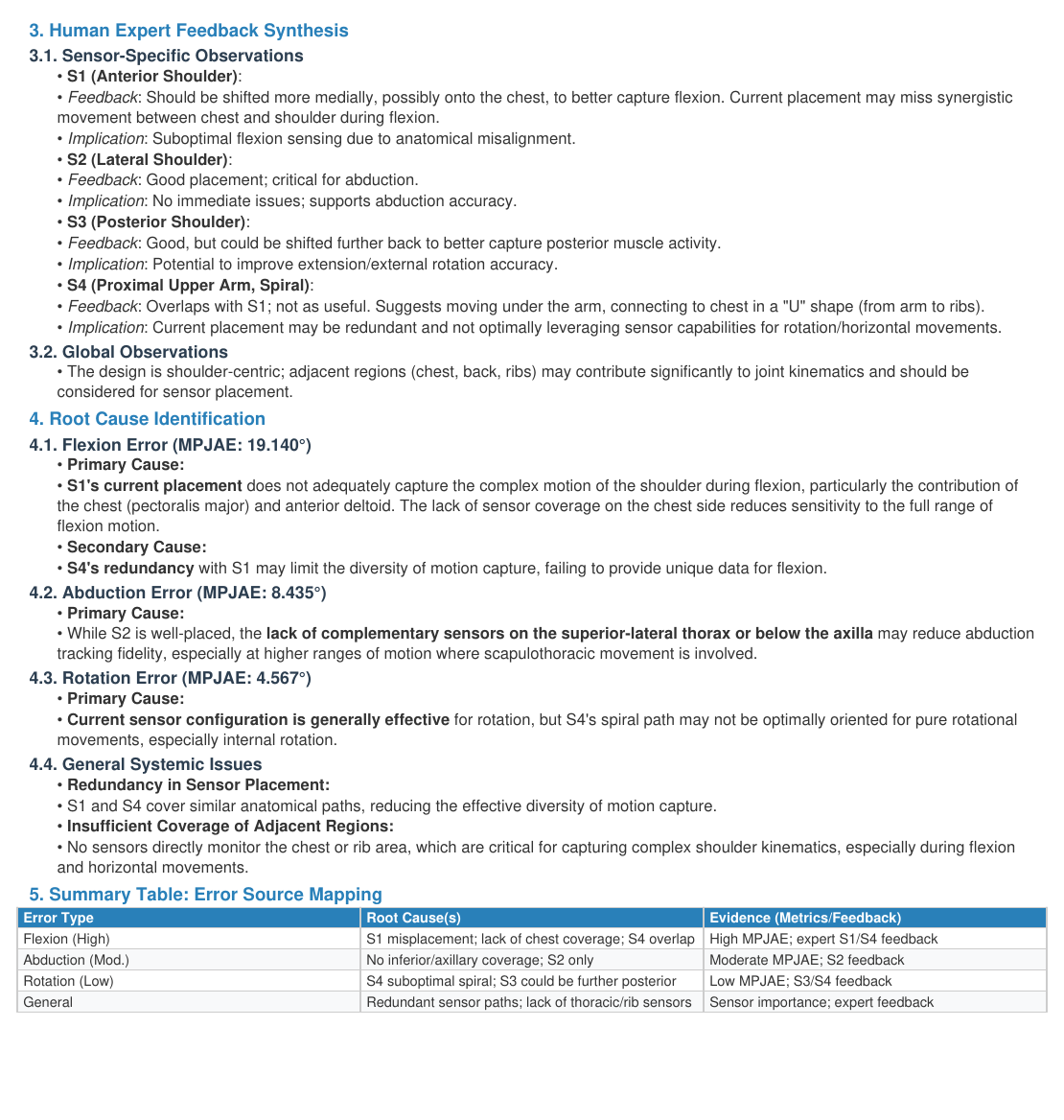}
\caption{LLM output for Designer C, Iteration 1 (Part 2).}
\label{fig:lh-llm-c-i1-p2}
\end{figure}

\begin{figure}[H]
\centering
\includegraphics[width=\textwidth]{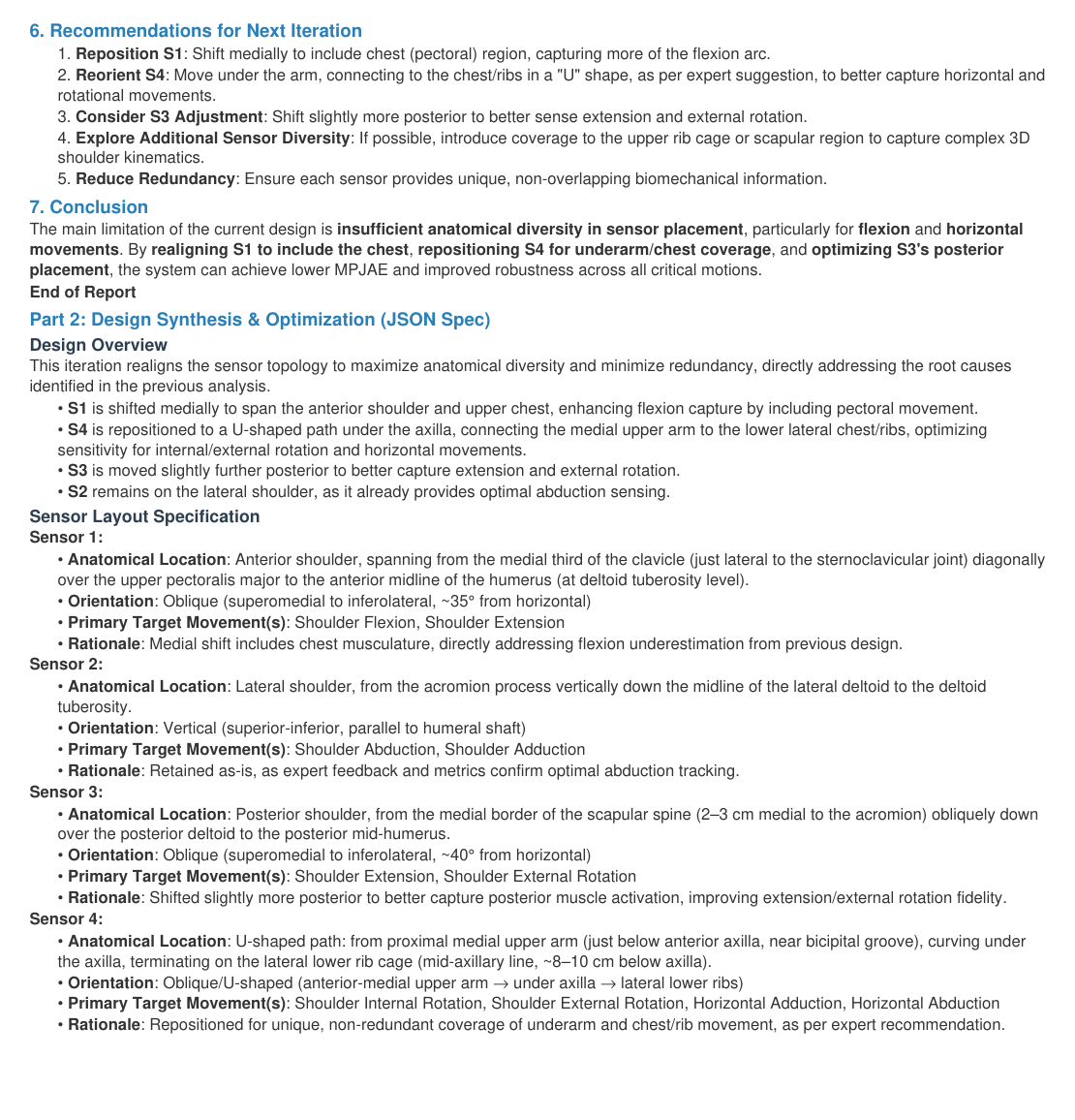}
\caption{LLM output for Designer C, Iteration 1 (Part 3).}
\label{fig:lh-llm-c-i1-p3}
\end{figure}

\begin{figure}[H]
\centering
\includegraphics[width=\textwidth]{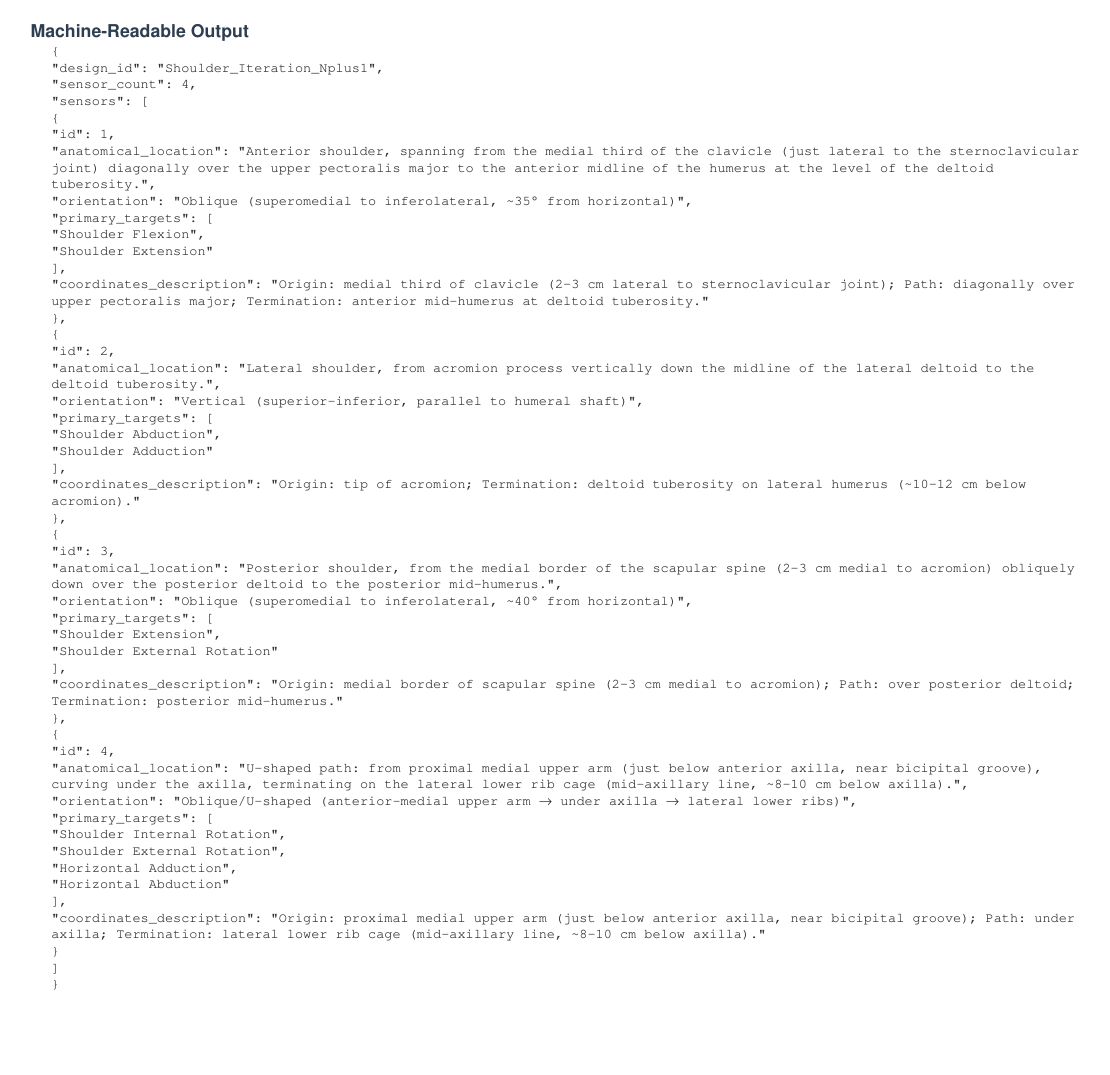}
\caption{LLM output for Designer C, Iteration 1 (Part 4).}
\label{fig:lh-llm-c-i1-p4}
\end{figure}


\begin{figure}[H]
\centering
\includegraphics[width=\textwidth]{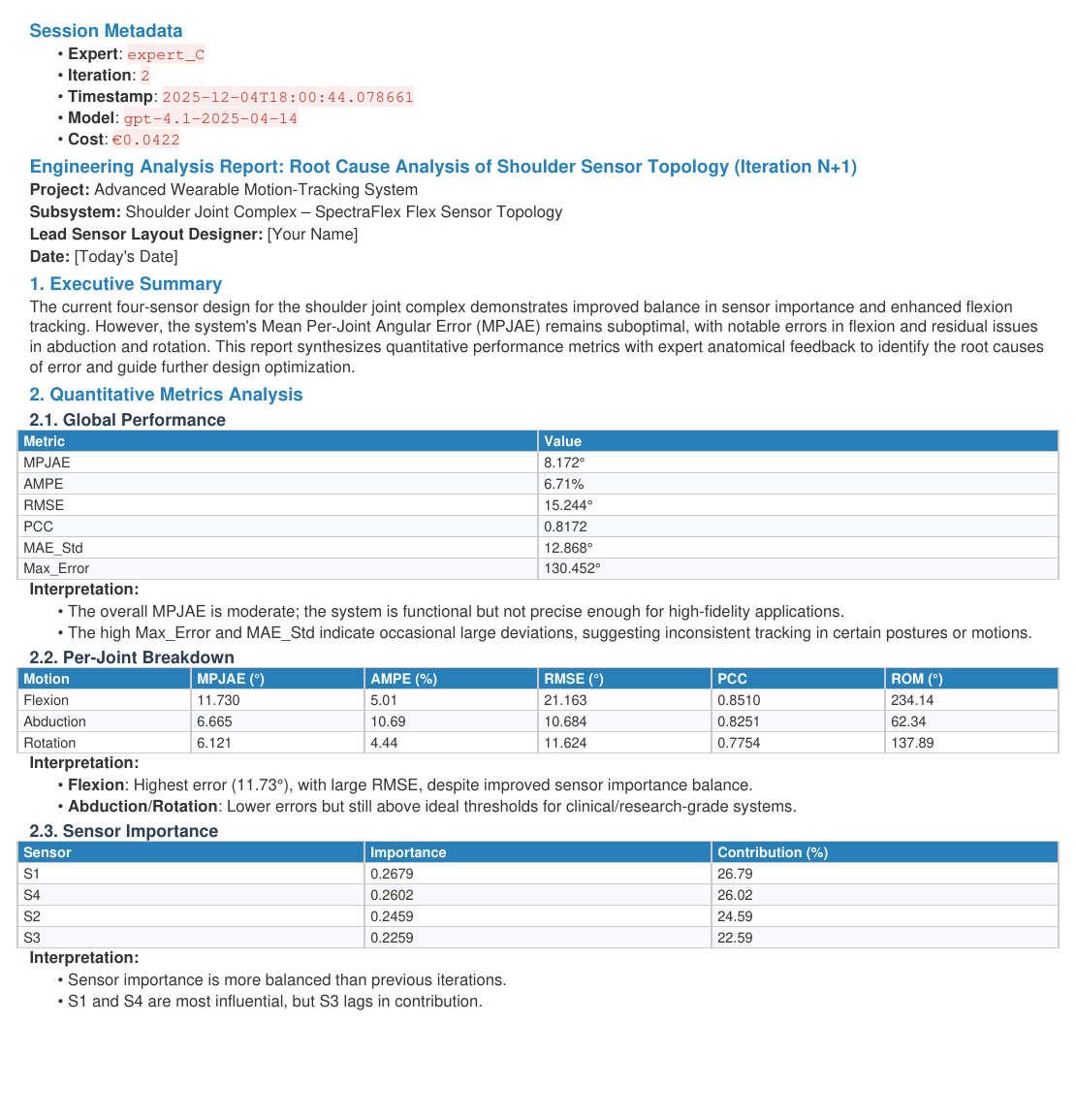}
\caption{LLM output for Designer C, Iteration 2 (Part 1).}
\label{fig:lh-llm-c-i2-p1}
\end{figure}

\begin{figure}[H]
\centering
\includegraphics[width=\textwidth]{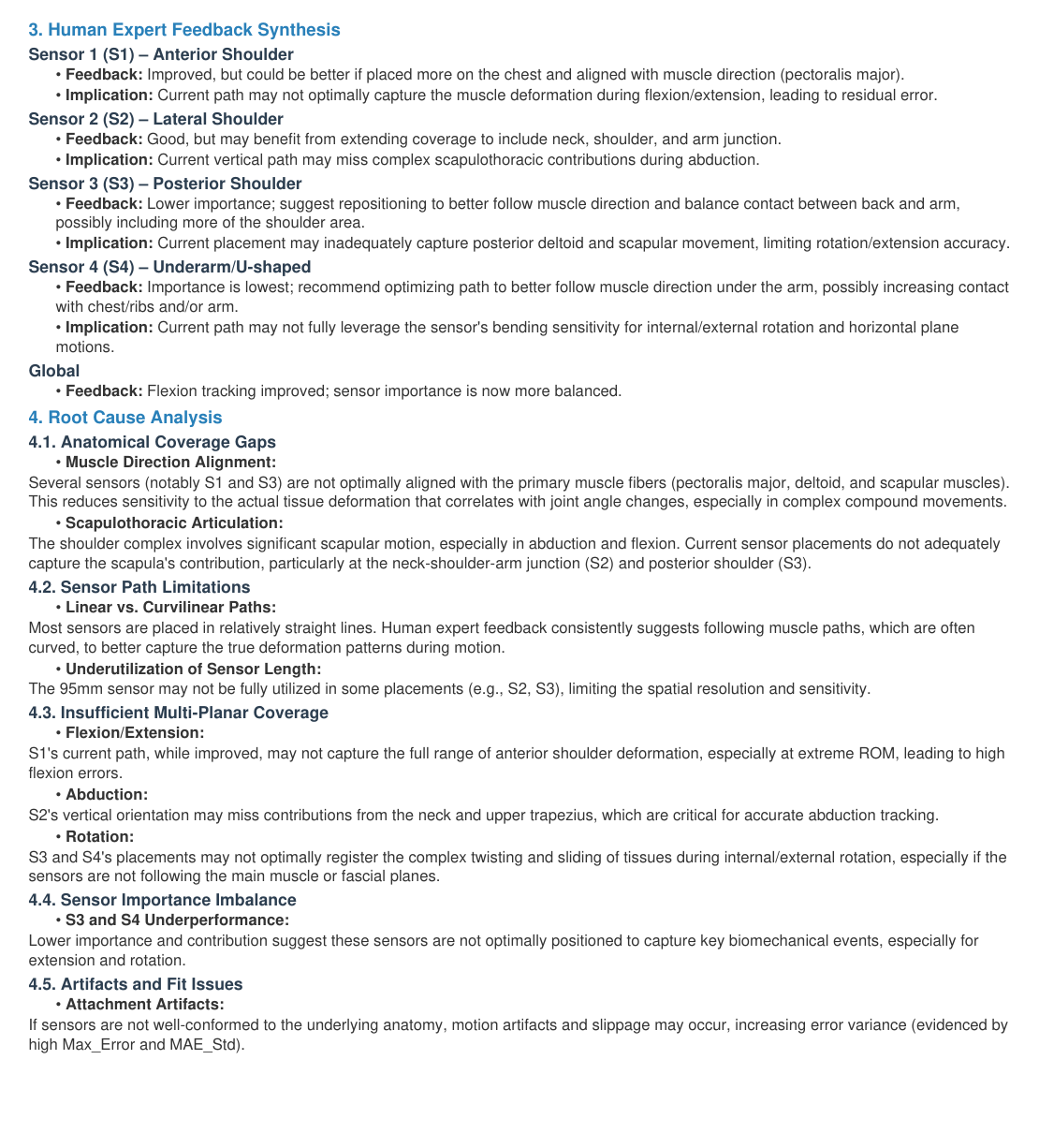}
\caption{LLM output for Designer C, Iteration 2 (Part 2).}
\label{fig:lh-llm-c-i2-p2}
\end{figure}

\begin{figure}[H]
\centering
\includegraphics[width=\textwidth]{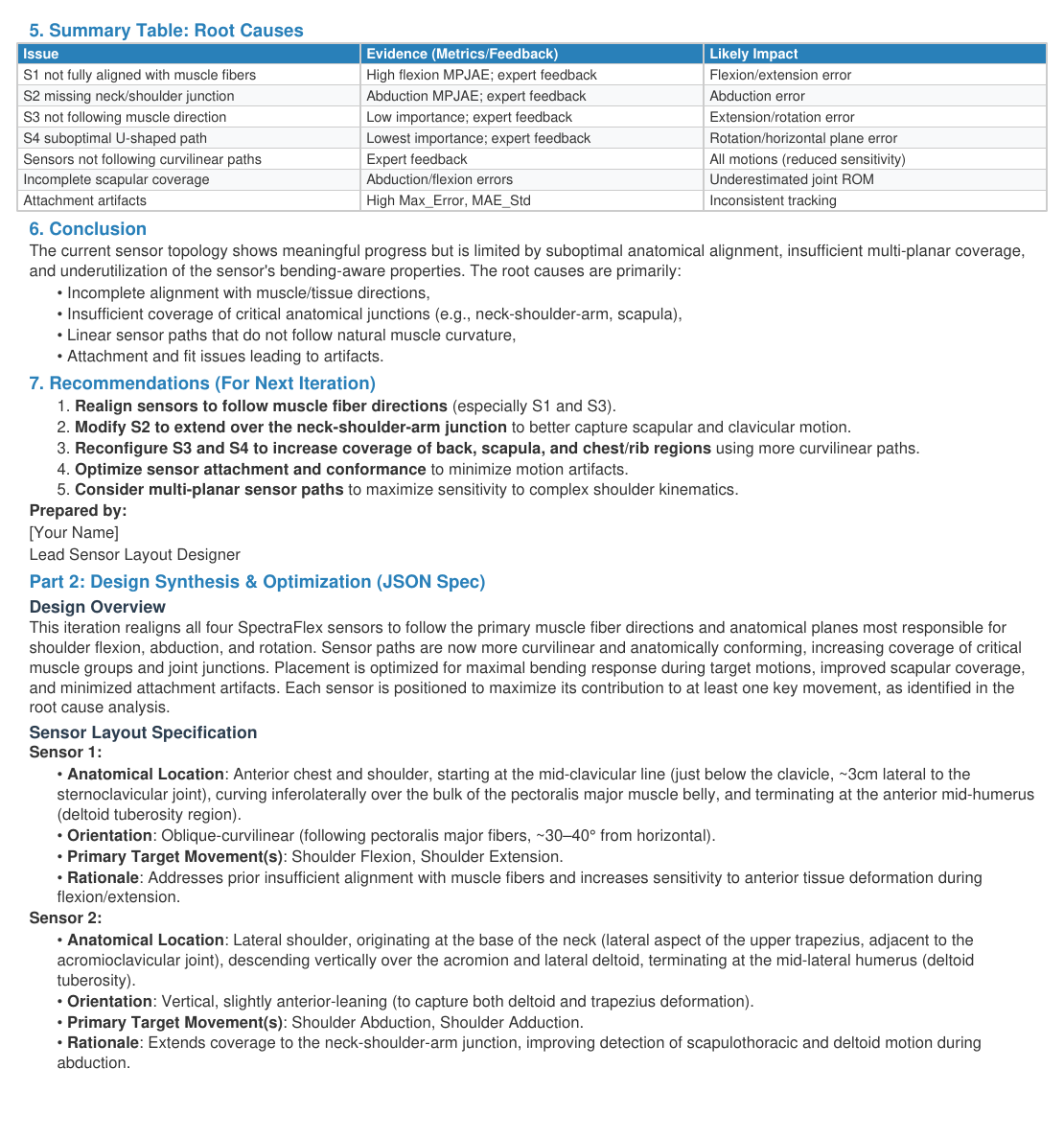}
\caption{LLM output for Designer C, Iteration 2 (Part 3).}
\label{fig:lh-llm-c-i2-p3}
\end{figure}

\begin{figure}[H]
\centering
\includegraphics[width=\textwidth]{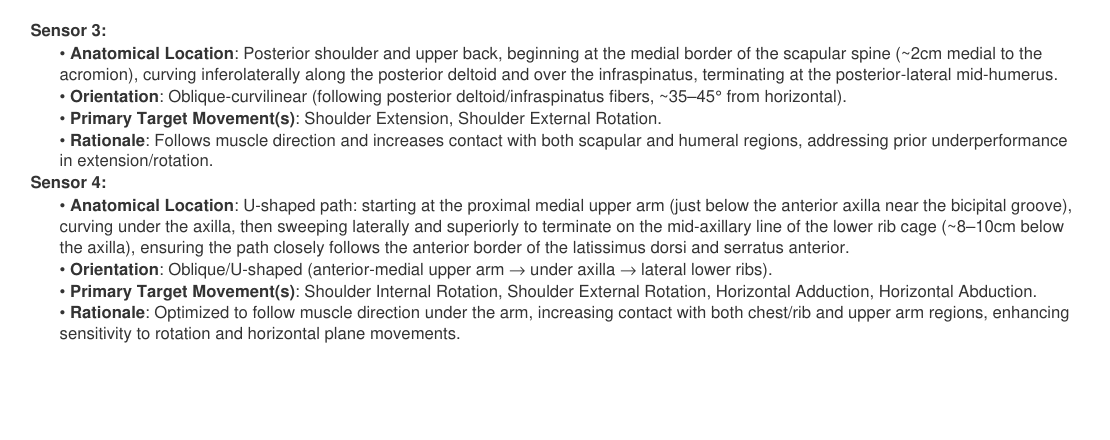}
\caption{LLM output for Designer C, Iteration 2 (Part 4).}
\label{fig:lh-llm-c-i2-p4}
\end{figure}

\begin{figure}[H]
\centering
\includegraphics[width=\textwidth]{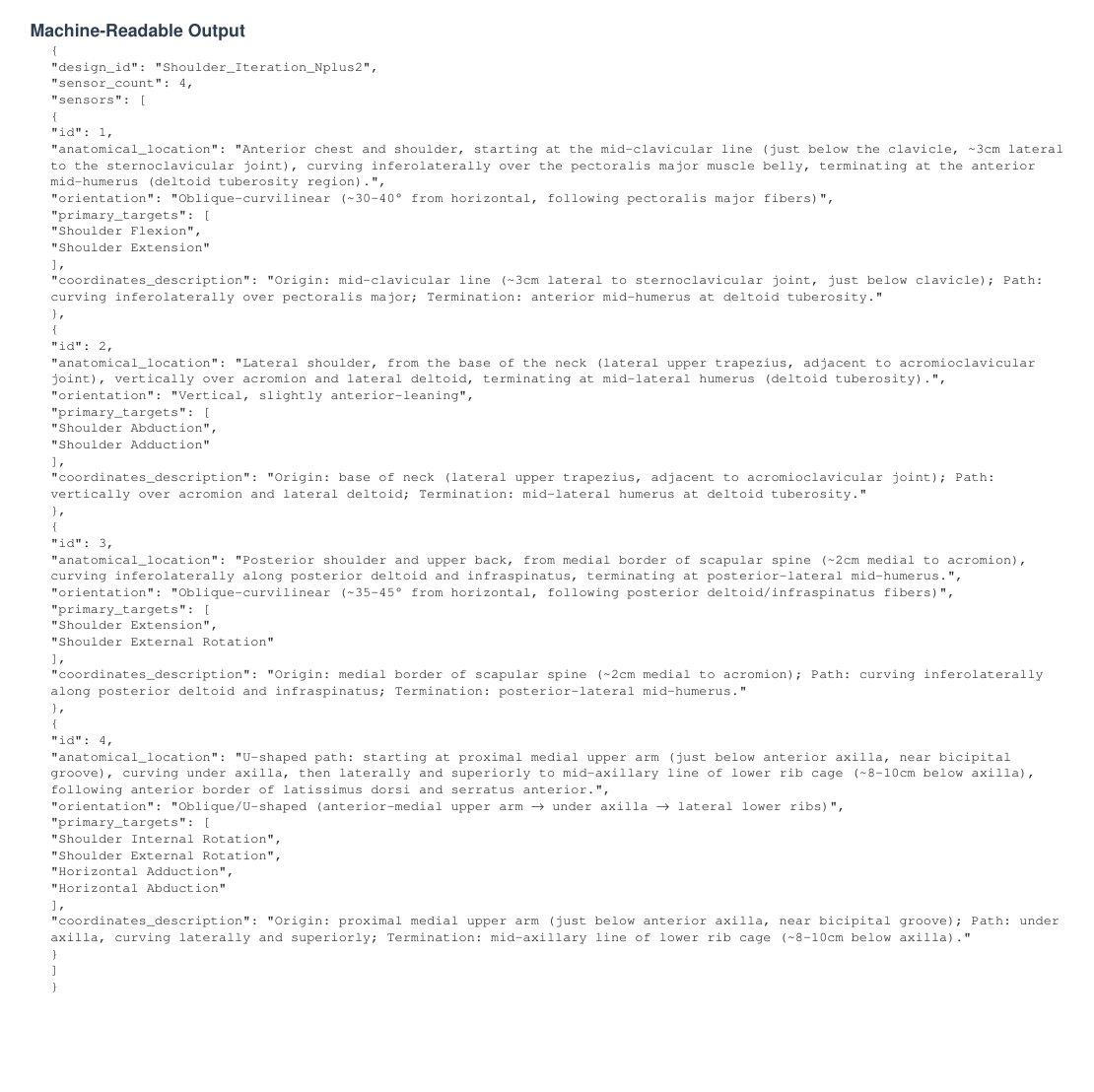}
\caption{LLM output for Designer C, Iteration 2 (Part 5).}
\label{fig:lh-llm-c-i2-p5}
\end{figure}

\end{document}